\documentclass{cernrep} 
\usepackage[export]{adjustbox}
\usepackage{texnames,bbm}
\usepackage[T1]{fontenc}
\usepackage[bookmarks, colorlinks=true, linktoc=page, pdftex, linkcolor=red, citecolor=red, urlcolor=red]{hyperref}
\def\Id{\mathbbm{1}}
\def\beq{\begin{equation}}
\def\eeq{\end{equation}}
\def\bea{\begin{eqnarray}}
\def\eea{\end{eqnarray}}
\DeclareSymbolFont{AMSb}{U}{msb}{m}{n}
\DeclareMathSymbol{\Zdouble}{\mathalpha}{AMSb}{"5A}
\DeclareMathAlphabet{\mathpzc}{OT1}{pzc}{m}{it}

\def \ve#1{\vec{#1}}
\def \vet#1{\vec{#1}^{T}}

\def\ds{\displaystyle}

\begin{document}

\title{Behind the Standard Model}
 
\author{Andrea Wulzer}

\institute{University of Padova and INFN, Padova, Italy.}

\maketitle 
 
\begin{abstract}
These lectures provide a concise introduction to the so-called ``Beyond the Standard Model'' physics, with particular emphasis on the problem of the microscopic origin of the Higgs mass term and of the Electro-Weak symmetry breaking scale in connection with Naturalness. The standard scenarios of Supersymmetry and Composite Higgs are shortly reviewed. An attempt is made to summarise the implications of the LHC run-$1$ results on what we expect to lie beyond (or behind) the Standard Model.
\end{abstract}

\begin{keywords}
Beyond the Standard Model; Supersymmetry; Composite Higgs; Naturalness.
\end{keywords}
 
\section{BSM: What For?}\label{wf}
Physics is the continuous effort towards a deeper understanding of the laws of Nature. The Standard Model (SM) theory summarises the state-of-the-art of this understanding, providing the correct description of all known fundamental particles and interactions (including Gravity) at the energy scales we have been capable to explore experimentally so far. ``Beyond the SM'' (BSM) physics aims to the next step of this understanding, namely to unveil the microscopic origin of the SM itself, of its field content, Lagrangian and parameters. From this viewpoint, the acronym ``BSM'' should better be read as ``Behind'' rather than ``Beyond'' the SM, from which the unconventional title (see however \cite{RR}) I gave to these lectures. The main focus is indeed not on new physics (beyond what predicted by the SM) per se, but on the solution of some of the mysteries associated with the microscopic theory that lie behind the SM itself. In this respect, a lack of discovery, namely a non-trivial confirmation of the SM that closes the door to BSM physics potentially associated with one of these mysteries, might be as informative as the observation of new physics.

The one described above is only one of the possible approaches to forefront research in fundamental physics. A valid alternative is to start from observations rather than from theory, in particular from those observations that cannot be accounted for by the SM, signalling the existence of new physics. What I have in mind are of course neutrino masses and oscillations and evidences of Dark Matter, Inflation and Baryogenesis. Dedicated lectures were given at this School on these topics \cite{pklect} \cite{dslect}. Even within the context of high-energy physics research, where no BSM discovery crossed our horizon yet \footnote{Still, the ongoing LHC program makes the direct exploration of the energy frontier the most promising tool of investigation we currently have to our disposal. Also, one should not forget the strong impact of Flavour physics \cite{sglect}, because of its capability of indirectly exploring very high-energy scales, on BSM physics.}, new physics searches driven by data rather than by theory are highly desirable and complementary to the study of specific signal topologies dictated by theoretical BSM scenarios. Also, we should not discard the possibility of performing theory-unbiased new physics searches in final states that appear promising because of their simplicity, of their low SM background and/or of their experimental purity. Notice however that a fully ``unbiased'' approach to new physics searches is virtually impossible. A certain degree of theory bias is unavoidably needed in order to limit the infinite variety of possible channels (or of experiments) one could search in. Even the very fact that TeV-scale reactions at the LHC are promising places to look at is in itself a theory bias, though dictated by extremely general and robust BSM considerations. Theory-unbiased or theory-driven new physics searches thus just correspond to a different gradation of BSM bias we decide to apply.

\subsection{No-Lose Theorems}
Sometimes, the quest for the microscopic origin of known particles and interactions has extremely powerful implications, leading to absolute {\emph{guarantees}} of new physics discoveries. A mathematical argument based on currently established laws of Nature, which ensures future discoveries provided the experimental conditions become favourable enough (i.e., high enough energy in the examples that follow), is what we call a ``No-Lose Theorem''. Though exceptional in the long history of science, several No-Lose Theorem could be formulated (and exploited, resulting in a number of discoveries) in the context of fundamental interaction physics over the last several decades. So many No-Lose Theorem existed, and for so long, that we got used to them, somehow forgetting their importance and their absolutely exceptional nature. They deserve a review now, after the discovery of the Higgs which prevents the formulation of new No-Lose Theorems marking the end of the age of guaranteed discoveries.

The simplest No-Lose Theorem is the one that guarantees the existence of new physics beyond (and behind) the Fermi Theory of Weak interactions. To appreciate the value of this theorem we must go back to the times when the Fermi Theory was the only experimentally established, potentially ``fundamental'', description of Weak interactions. At that times, our knowledge of the Weak force was entirely encapsulated in a four-fermions operator of energy dimension $d=6$, the Fermi interaction, with its $d=-2$ coefficient, the Fermi constant $G_F$.\footnote{Of course the Cabibbo angle was also needed in order to describe hadronic Weak processes.} The question of whether the Fermi theory can be truly fundamental or not, and correspondingly whether or not $G_F$ can be a fundamental constant of Nature, has a very sharp negative answer, schematically summarised below\\
\begin{figure}[h]
\vspace{-15pt}
  \centering
  \includegraphics[width=0.8\textwidth]{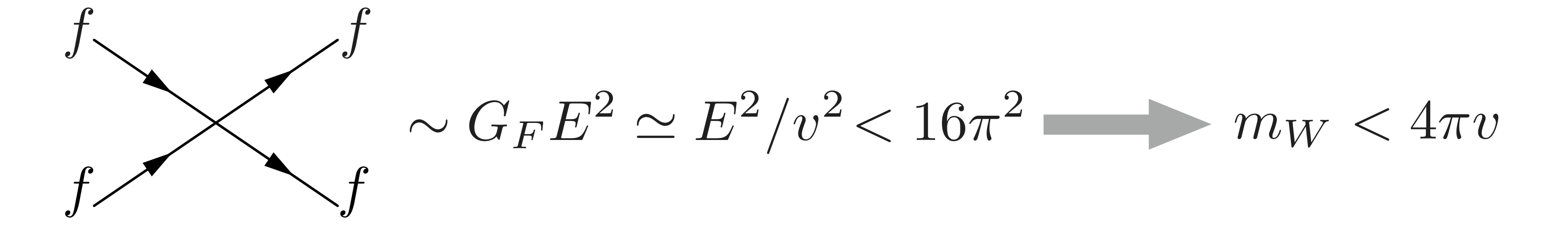} 
\vspace{-10pt}
\end{figure}

\noindent{The} point is that the four-fermions scattering amplitude grows with the square of the center-of-mass energy ``$E$'' of the reaction, a fact that trivially follows from dimensional analysis (since the amplitude is dimensionless and proportional to the $d=-2$ coupling constant $G_F$) and is intrinsically linked with the non-renormalizable nature of the Fermi Theory. But the Weak scattering amplitude becoming too large, overcoming the critical value of $16\pi^2$, means that the Weak force gets too strong to be treated as a small perturbation of the free-fields dynamics and the perturbative treatment of the theory breaks down. Of course there is nothing conceptually wrong in the Weak force entering a non-perturbative regime, the problem is that this regime cannot be described by the Fermi Theory, which is intrinsically defined in perturbation theory. Namely, the Fermi Theory does not give trustable predictions and becomes internally inconsistent as soon as the non-perturbative regime is approached. Therefore a new theory, i.e. new physics, is absolutely needed. Either in order to modify the energy behaviour of the amplitude before it reaches the non-perturbative threshold, keeping the Weak force perturbative, or to describe the new non-perturbative regime. In all cases this new, more fundamental, theory will account for the microscopic origin of the Fermi interaction and of its coupling strength $G_F$ as a low-energy effective description of the Weak force. According to the theorem, the microscopic theory must show up at an energy scale below $4\pi/\sqrt{G_F}\simeq 4\pi v$, having expressed $G_F=1/\sqrt{2}v^2$ in terms of the ElectroWeak Symmetry Breaking (EWSB) scale $v\simeq246$~GeV. We now know that the new physics beyond the Fermi Theory is the Intermediate Vector Boson (IVB) theory, which was confirmed by discovering the $W$ boson at the scale $m_W\simeq80$~GeV, far below $4\pi v$ compatibly with the theorem.

As everyone knows, well before the discovery of the $W$ discovery we already had strong indirect indications on the validity of the IVB theory and a rather precise estimate of the $W$ boson mass. These indications came from fortunate theoretical speculations and from the measurement of the Weak angle through neutrino scattering processes, and are completely unrelated with the No-Lose Theorem outlined above. Indeed, the theorem makes no assumption on, and gives no indication about, the details of the microscopic physics that lies behind the Fermi Theory. Namely, the theorem guarantees that something would have been discovered in fermion-fermion scattering, possibly not the $W$ and possibly not at a scale as low as $m_W$, even if all the theoretical speculations about the IVB theory had turned out to be radically wrong. This means in particular that if the UA$1$ and UA$2$ experiments at the CERN SPS collider had not discovered the $W$, we would have for sure continued searching for it, or for whatever new physics lies behind the Fermi theory, by the construction of higher energy machines. 

A situation like the one described above was indeed encountered in the search for the top quark, which according to a widespread belief was expected to be much lighter than $m_t\simeq 173$~GeV, where it was eventually observed. Consequently, the top discovery was expected at several lower-energy colliders, constructed before the Tevatron, which instead produced a number of negative results. However we never got discouraged and we never even considered the possibility of giving up searching for the top quark, or for some other new physics related with the bottom quark, because of a second No-Loose Theorem:\\
\begin{figure}[h]
\vspace{-15pt}
  \centering
  \includegraphics[width=0.8\textwidth]{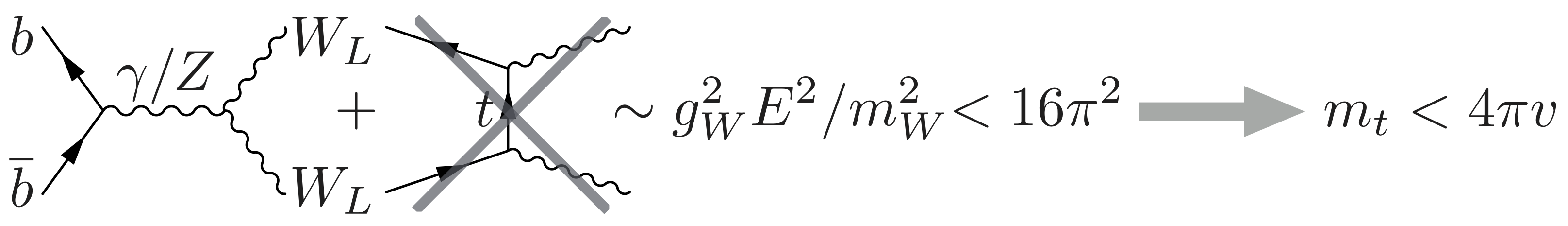} 
\vspace{-10pt}
\end{figure}

\noindent{The} theorem relies on the validity of the IVB theory and on the existence of the bottom quark with its neutral current interactions, which we consider here as experimentally established facts at the times when the top was not yet found. The observation is that the amplitude for longitudinally polarised $W$ bosons production from a $b$ $\overline{b}$ pair grows quadratically with the energy if the top quark is absent or if it is too heavy to be relevant. It is indeed the t-channel contribution from the top exchange that makes the amplitude constant at high energies in the complete SM. Perturbativity thus requires new physics at a scale below $4\pi m_W/g_W\simeq 4\pi v$, having used the relation $m_W=g_W v/2$. When interpreted in the SM, the upper bound on the new physics scale translates in the familiar perturbativity bound on the top mass, however the Theorem does not rely on the SM and on the existence of the top quark. It states that the top, or something else, must exist beyond the bottom quark in order to moderate the growth with the energy of the scattering amplitude. More physically, the Theorem says that the microscopic origin of the bottom quark (e.g., the fact that its left-handed component lives in a doublet together with the top) must reveal itself below $4\pi v$.

Another particle whose discovery was significantly ``delayed'' with respect to the expectations is the Higgs boson, which also comes with its own No-Loose Theorem:\\
\begin{figure}[h]
\vspace{-15pt}
  \centering
  \includegraphics[width=0.8\textwidth]{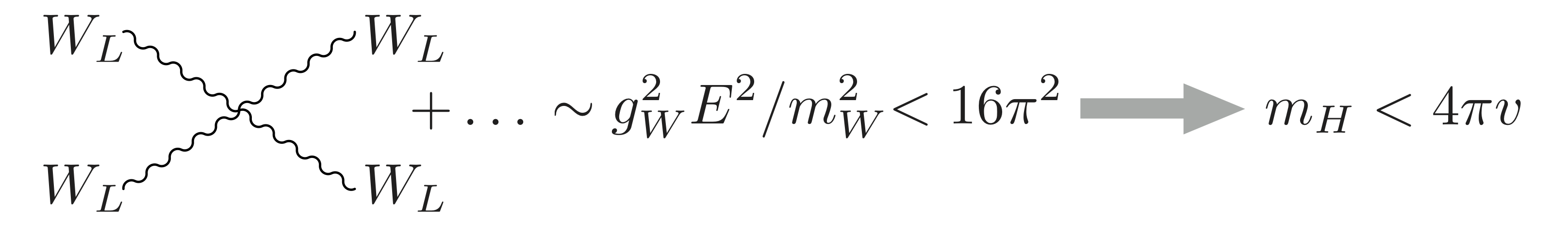} 
\vspace{-10pt}
\end{figure}

\noindent{The} growth with the energy of the longitudinally polarised $W$ bosons scattering amplitude in the IVB theory requires the presence of new particles and/or interactions, once again below the critical threshold of $4\pi v\sim3$~TeV. Given that the TeV scale is within the reach of the LHC collider, the Theorem above offered absolute guarantee of new physics discoveries at the LHC and was heavily used to motivate its construction. Now the Higgs has been found, with couplings compatible with the SM expectations, we know that it is indeed the Higgs particle the agent responsible for cancelling (at least partially, given the limited accuracy of the Higgs couplings measurements) the quadratic term in the scattering amplitude. This leaves us, as I will better explain below, with no No-Loose Theorem and thus with no guaranteed discovery to organise our future efforts in the investigation of fundamental interactions.

Each of the No-Lose theorems discussed above emerges because of the anomalous power-like growth with the energy of some scattering amplitude, a behaviour which unmistakably signals that a non-renormalizable interaction operator of energy dimension $d>4$ is present in the theory. This being the case is completely obvious for the Fermi theory, a bit less so in the two other examples. In the latter cases it requires, to be understood, somewhat technical considerations related with the Goldstone boson Equivalence Theorem \cite{Horejsi:1995jj} which go beyond the purpose of the present lectures. It suffices here to say that one given $d=6$ non-renormalizable operator, responsible for the $E^2$ growth of the scattering amplitude, can be identified for each of the $3$ No-Lose theorems above. When each theorem was ``exploited'' by discovering the associated new physics we ``got rid'' of the corresponding operator by replacing it with a more fundamental theory that explains its origin as a low-energy effective description. Having exploited all the theorems, we got rid of all the non-renormalizable operators and we are left, for the first time, with an experimentally verified renormalizable theory of electroweak and strong interactions. No new No-Lose theorems can be thus formulated in this theory, at least not as simple and powerful ones as the ones listed above.

\begin{figure}[t]
  \centering
  \includegraphics[width=0.4\textwidth]{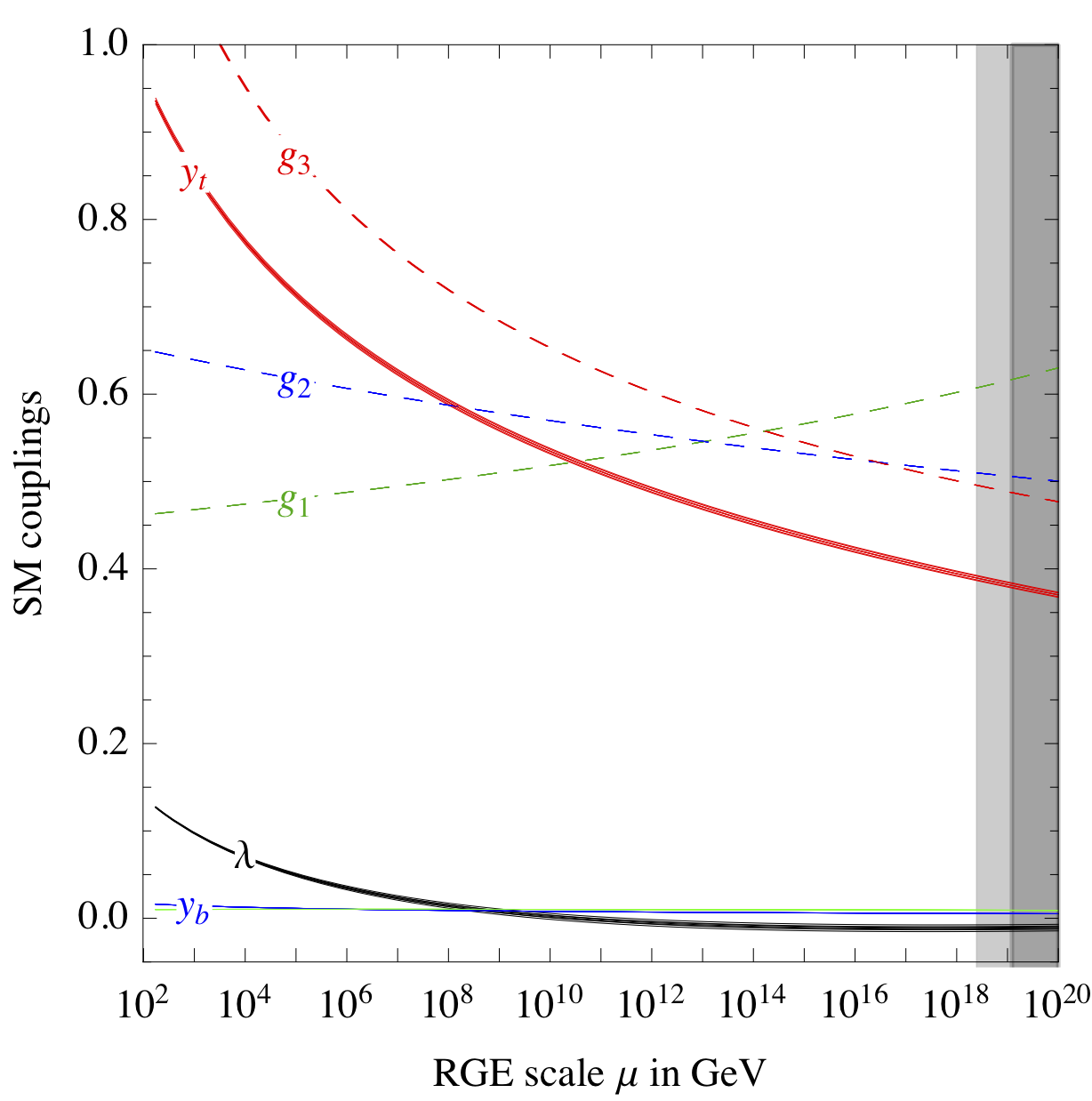}
  \caption{The RG running of the most relevant couplings of the SM, namely the three gauge couplings $g_{1,2,3}$, the top and bottom Yukawa's $y_{t,b}$ and the Higgs quartic coupling $\lambda$. See Ref.~\cite{Buttazzo:2013uya} and references therein for more details.
  \label{SM_RG}} 
\end{figure}

However the SM is not only a theory of electroweak and strong interactions. It can be (and it must be, to account for observations) extended to incorporate Gravity and the only sensible way to do so is by introducing and quantising the Einstein-Hilbert action. This produces a number of non-renormalizable interaction operators involving gravitons, giving rise to another well-known No-Lose theorem\\
\begin{figure}[h]
\vspace{-15pt}
  \centering
  \includegraphics[width=0.8\textwidth]{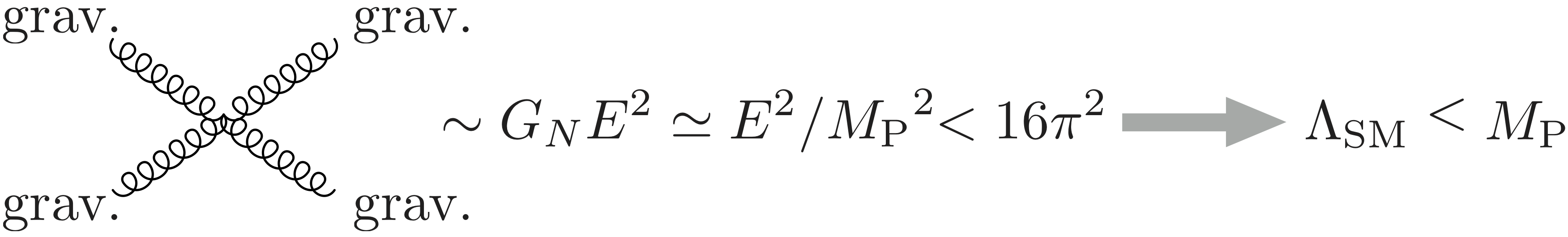} 
\vspace{-10pt}
\end{figure}

\noindent{where} $M_{\textrm{P}}\simeq10^{19}$~GeV is the Planck scale. What the theorem says is that the SM is for sure not the ``final theory'' of Nature, because it does not provide a complete description of Gravity at the quantum level. It does incorporate a description of quantum gravity that is valid and predictive at low energy but breaks down at a finite scale $\Lambda_{\textrm{{SM}}}$, which we call the ``SM cutoff''. BSM particles and interactions are present at that scale, which however can be as high as $10^{19}$~GeV. Given our technical inability to test such an enormous scale, it is unlikely that we might ever exploit this last No-Lose Theorem as a guide towards a concrete new physics discovery.

The second aspect to be discussed is that even in a renormalizable theory the scattering amplitudes can actually grow with the energy. Not with a power-law, but logarithmically, through the Renormalisation Group (RG) running of the dimensionless coupling constants of the theory. The RG evolution can make some of the couplings grow with the energy until they violate the perturbativity bound, producing a new No-Lose Theorem. Obviously this No-Lose Theorem would most likely be not as powerful as those obtainable in non-renormalizable theories because the RG evolution is logarithmically slow and thus the perturbativity violation scale is exponentially high, but still it is interesting to ask if one such a theorem exists for the SM and at which scale it points to. The answer is that perturbativity violation does not occur in the SM below the Planck mass scale, at which new physics is anyhow needed to account for gravity, as shown in fig.~\ref{SM_RG}. The only coupling that grows significantly with the energy is the one associated with the \mbox{U$(1)_Y$} gauge group, $g_1$, which however is still well below the perturbativity bound at the Planck scale. Notice that the result crucially depends on the initial conditions of the running, namely on the values of the SM parameters measured at the $100$~GeV scale. The result would have been different, and an additional No-Lose Theorem would have been produced, if that values were radically different than what we actually observed.

\begin{figure}[t]
  \centering
  \includegraphics[width=0.8\textwidth]{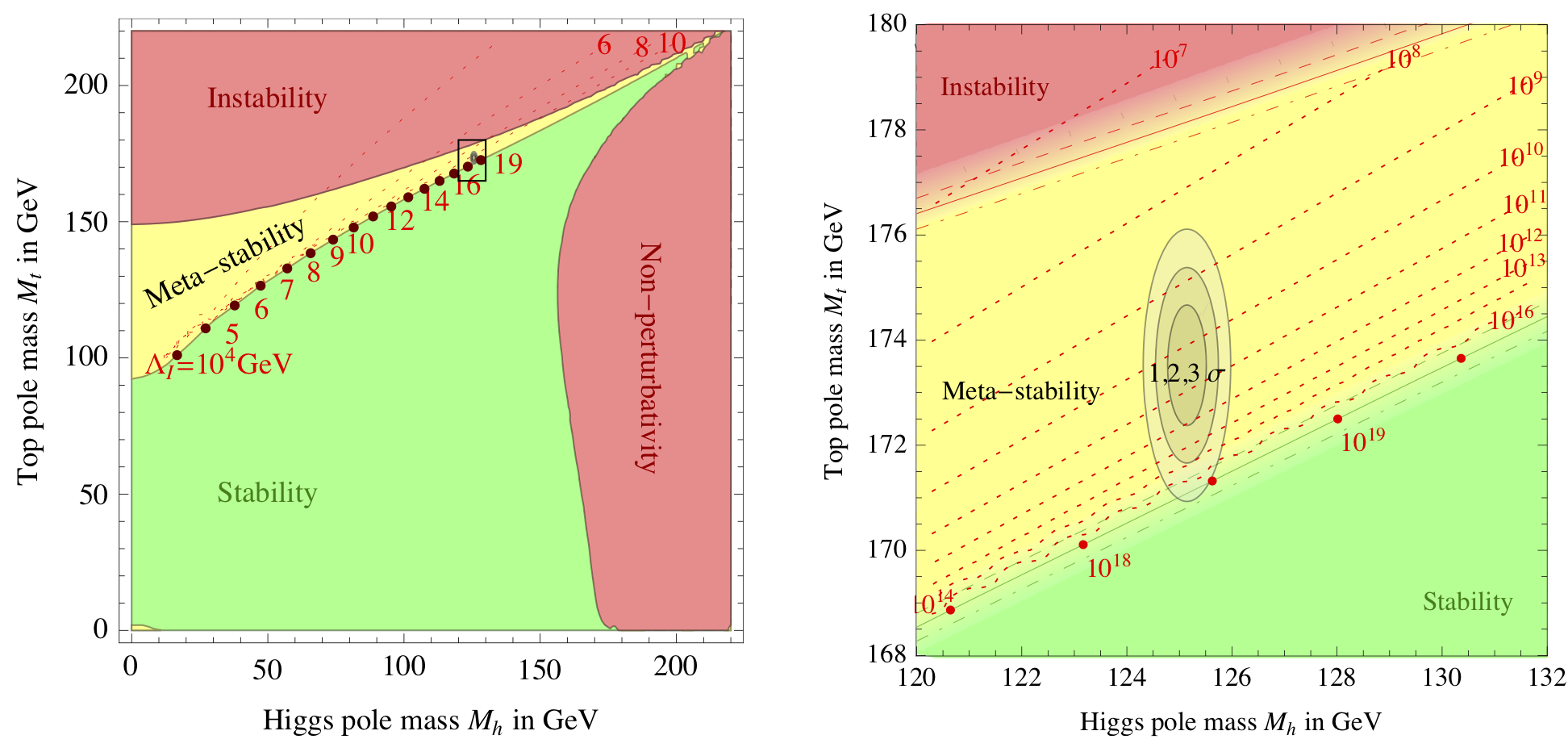}
  \caption{Stability, instability, metastability and non-perturbativity regions for the SM in the plane of the Higgs and top masses. A zoom on the experimentally viable region is displayed in the right plot, with the $1$, $2$ and $3\sigma$ regions allowed by $m_H$ and $m_t$ uncertainties. From Ref.~\cite{Buttazzo:2013uya}.
  \label{STAB}} 
\end{figure}

The vacuum stability problem \cite{Krive:1976sg} is yet another potential source of high-energy inconsistencies (and thus of No-Lose Theorems) in renormalizable theories that display, like the SM, a non-trivial structure of the vacuum state. The problem is again due to RG evolution effects, which modify the form of the Higgs potential at very high values of the Higgs field and potentially make it develop a second minimum. If the energy of this second minimum is lower than the first one, transitions can occur via quantum tunnelling from the ordinary EWSB vacuum where $v\simeq246$~GeV to an inhospitable minimum characterised by a very large vacuum expectation value (VEV) of the Higgs field. Whether this actually happens or not depends, once again, on the measured value of the SM parameters and in particular on the Higgs boson and top quark masses as displayed in fig.~\ref{STAB}. We see that our vacuum is not stable and thus it is fated to decay provided we wait long enough. However it falls in the ``meta-stability'' region of the diagram, which is where the vacuum lifetime is longer than the age of the Universe. Therefore the decay of our vacuum might not have had enough time to occur. Some people find disturbing that we live in a meta-stable vacuum. Some others \cite{Buttazzo:2013uya} find intriguing the fact that we live close (see the right panel of fig.~\ref{STAB}) to the boundary between the stability and meta-stability regions and suggest that we should measure $m_t$ better in order to be sure of how close we actually are. Anyhow what is sure (and what matters for our discussion) is that the analysis of the vacuum stability does not reveal any concrete inconsistency of the SM at high energy. Consequently, no new No-Lose Theorem is found.

\subsection{The ``SM-only'' Option}\label{SMONLY}

\begin{figure}[t]
\centering
\includegraphics[width=0.75\textwidth]{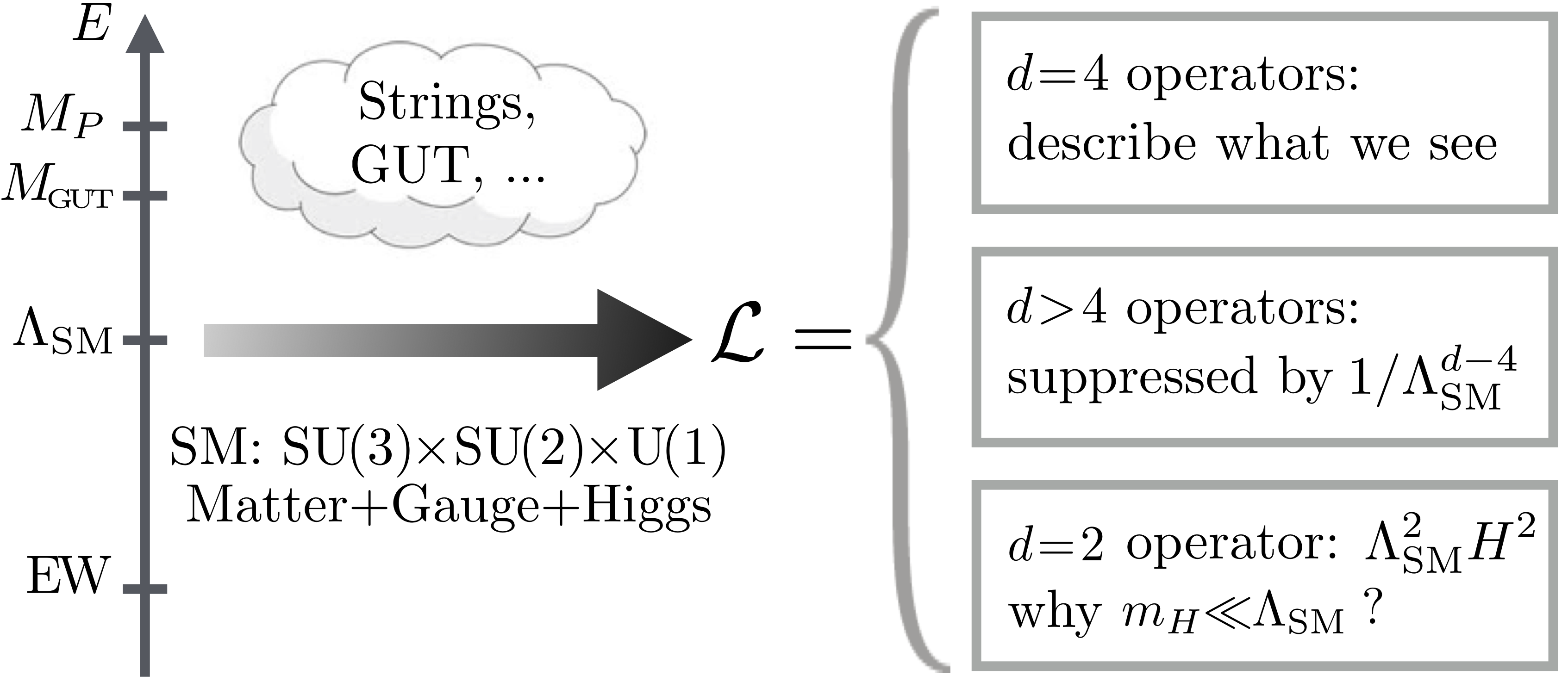}
\caption{Pictorial view of the SM as an effective field theory, with its Lagrangian generated at the scale $\Lambda_{\textrm{SM}}$.}
\label{SMEFT}
\end{figure}

Two extremely important (and in some sense contradictory) facts emerge from the previous considerations. On one hand, we know that BSM physics exists at a finite energy scale $\Lambda_{\textrm{SM}}$. This makes that the SM is necessarily an approximate low-energy description of a more fundamental theory, i.e. an Effective Field Theory (EFT) with a finite cutoff $\Lambda_{\textrm{SM}}$. On the other hand, the only upper bound on the cutoff scale is provided by the Planck mass, which is to a very good approximation equal to infinity compared with the much lower scales we are able to explore experimentally today and in any foreseeable future. We are thus led to consider the ``SM-only'' option for high-energy physics. Namely the possibility that the SM cutoff $\Lambda_{\textrm{SM}}$ (i.e., the scale of new physics) is extremely high, much above the TeV as depicted in fig.~\ref{SMEFT}. Values as high as $\Lambda_{\textrm{SM}}\sim M_{\textrm{P}}$ and $\Lambda_{\textrm{SM}}\sim10^{15}$~GeV~$\equiv{M}_{\textrm{GUT}}$ can be considered.

The SM-only option is not just a logical possibility. On the contrary, it is a predictive and phenomenologically successful scenario for high-energy physics. To appreciate its value, we look again at fig.~\ref{SMEFT}, starting from the high energy (UV) region and we ask ourselves how the SM theory emerges in the IR. As pictorially represented in the figure, we have no idea of how the theory in the UV looks like. It might be a string theory, a GUT model (for a review, see for instance Refs.~\cite{Langacker:1980js,Raby:2006sk}), or something completely different we have not yet thought about. All what we know about the UV theory is that, by assumption, its particle content reduces to the one of the SM at $\Lambda_{\textrm{SM}}$, all BSM particles being at or above that scale.\footnote{The presence of light feebly coupled BSM particles would not affect the considerations that follow.} Below $\Lambda_{\textrm{SM}}$ the UV theory thus necessarily reduces, after integrating out the heavy states, to a low-energy EFT which only describes the light SM degrees of freedom. A technically consistent description of the force carriers (gluon and EW bosons) requires invariance under the \mbox{SU$(3)_c\times$SU$(2)_L\times$U$(1)_Y$} gauge group, but apart from being gauge (and Lorentz) invariant there is not much we can tell a priori on how the SM effective Lagrangian will look like. It will consist of an infinite series of local gauge- and Lorentz-invariant operators with arbitrary energy dimension ``$d$'', constructed with the SM Matter, Gauge and Higgs fields as in fig.~\ref{SMEFT}. The coefficient of the operators must be proportional to $1/\Lambda_{\textrm{SM}}^{d-4}$ by dimensional analysis, given that $[{\mathcal{L}}]=E^4$ and $\Lambda_{\textrm{SM}}$ is the only relevant scale. This simple observation lies at the heart of the phenomenological virtues of the SM-only scenario but also, as we will see, of its main limitation.

We now classify the SM effective operators by their energy dimension and discuss their implications, starting from those with $d=4$. They describe almost all what we have seen in Nature, namely EW and strong interactions, quarks and charged leptons masses. They define a renormalizable theory and thus, together with the $d=2$ operator we will introduce later, they are present in the textbook SM Lagrangian formulated in the old times when renormalizability was taken as a fundamental principle. 

Several books have been written (see for instance Refs.~\cite{OKUN,Cheng:1985bj,Schwartz:2013pla}) on the extraordinary phenomenological success of the renormalizable SM Lagrangian in describing the enormous set of experimental data \cite{Agashe:2014kda} collected in the past decades. In a nutshell, as emphasized in Ref.~\cite{10lectures}, most of this success is due to symmetries, namely to ``accidental'' symmetries. We call ``accidental'' a symmetry that arises by accident at a given order in the operator classification, without being imposed as a principle in the construction of the theory. The renormalizable ($d\leq4$) SM Lagrangian enjoys exact (or perturbatively exact) accidental symmetries, namely baryon and lepton family number, and approximate ones such as the flavour group and custodial symmetry. For brevity, we focus here on the former symmetries, which have the most striking implications. Baryon number makes the proton absolutely stable, in accordance with the experimental limit $\Gamma_p/m_p\lesssim10^{-64}$ on the proton width over mass ratio. It is hard to imagine how we could have accounted for the proton being such a narrow resonance in the absence of a symmetry. Similarly lepton family number forbids exotic lepton decays such as $\mu\rightarrow e\gamma$, whose branching ratio is experimentally bounded at the $10^{-12}$ level. From neutrino oscillations we know that the lepton family number is actually violated, in a way that however nicely fits in the SM picture as we will see below. Clearly this is connected with the neutrino masses, which exactly vanish at $d=4$ because of the absence, in what we call here ``the SM'', of right-handed neutrino fields.

We now turn to non-renormalizable operators with $d>4$. Their coefficient is proportional to $1/\Lambda_{\textrm{SM}}^n$, with $n=d-4>0$, thus their contribution to low-energy observables is suppressed by $(E/\Lambda_{\textrm{SM}})^n$ with respect to renormalizable terms. Given that current observations are at and below the EW scale, $E\lesssim m_{\textrm{EW}}\simeq100$~GeV, their effect is extremely suppressed in the SM-only scenario where $\Lambda_{\textrm{SM}}\gg$~TeV. This could be the reason why Nature is so well described by a renormalizable theory, without renormalizability being a principle. 

Non-renormalizable operators violate the $d=4$ accidental symmetries. Lepton number stops being accidental already at $d=5$ because of the Weinberg operator \cite{Weinberg:1979sa}
\beq
\label{WOP}
\displaystyle
\frac{c}{\Lambda_{\textrm{SM}}}({\overline{\ell}}_L H^c)({{\ell}}_L^c H^c)\,,
\eeq
where $\ell_L$ denotes the lepton doublet, $\ell_L^c$ its charge conjugate, while $H$ is the Higgs doublet and $H^c = i \sigma^2 H^*$. The \mbox{SU$(2)_L$} indices are contracted within the parentheses and the spinor index between the two terms. A generic lepton flavour structure of the coefficient, leading to the breaking of lepton family number, is understood. Surprisingly enough, the Weinberg operator is the unique $d=5$ term in the SM Lagrangian. When the Higgs is set to its VEV, the Weinberg operator reduces to a Majorana mass term for the neutrinos, $m_\nu\sim{c}\,v^2/\Lambda_{\textrm{SM}}$. For $\Lambda_{\textrm{SM}}\simeq10^{14}$~GeV and order one coefficient ``$c$'' it generates neutrino masses of the correct magnitude ($m_\nu\sim0.1$~eV) and neutrino mixings that can perfectly account for all observed neutrino oscillation phenomena. Baryon number is instead still accidental at $d=5$ and its violation is postponed to $d=6$. We thus perfectly understand, qualitatively, why lepton family violation effects are ``larger'', thus easier to discover, while baryon number violation like proton decay is still unobserved. At a more quantitative level we should actually remark that the bounds on proton decay from the $d=6$ operators, with order one numerical coefficients, set a limit $\Lambda_{\textrm{SM}}\gtrsim10^{15}$~GeV that is in slight tension with what required by neutrino masses. However few orders of magnitude are not a concern here, given that there is no reason why the operator coefficient should be of order one. A suppression of the proton decay operators is actually even expected  because they involve the first family quarks and leptons, whose couplings are reduced already at the renormalizable level. Namely, it is plausible that the same mechanism that makes the first-family Yukawa couplings small also reduces proton decay, while less suppression is expected in the third family entries of the Weinberg operator coefficient that might drive the generation of the heaviest neutrino mass.

The considerations above make the SM-only option a plausible picture, which becomes particularly appealing if we set $\Lambda_{\textrm{SM}}\sim{M}_{\textrm{GUT}}$. This choice happens to coincide with the gauge coupling unification scale, but this doesn't mean that the new physics at the cutoff is necessarily a Grand Unified Theory. On the contrary, the physics at the cutoff can be very generic in this picture, the compatibility with low-energy observations being ensured by the large value of the $\Lambda_{\textrm{SM}}$ scale and not by the details of the UV theory. New physics is virtually impossible to discover directly in this scenario, but this doesn't make it completely untestable. Purely Majorana neutrino masses would be a strong indication of its validity while observing a large Dirac component would make it less appealing.

Having discussed the virtues of the SM-only scenario, we turn now to its limitations. One of those, which was already mentioned, is the hierarchy among the Yukawa couplings of the various quark and lepton flavours, which span few orders of magnitude. This tells us that the new physics at $\Lambda_{\textrm{SM}}$ cannot actually be completely generic, given that it must be capable of  generating such a hierarchy in its prediction for the Yukawa's. This limits the set of theories allowed at the cutoff but is definitely not a strong constraint. Whatever mechanism we might imagine to generate flavour hierarchies at $\Lambda_{\textrm{SM}}\sim M_{\textrm{GUT}}$, it will typically not be in contrast with observations given that the bounds on generic flavour-violating operators are  ``just'' at the $10^{8}$~GeV scale. Incorporating dark matter also requires some modification of the SM-only picture, but there are several ways in which this could be done without changing the situation dramatically. Perhaps the most appealing solution from this viewpoint is ``minimal dark matter'' \cite{Cirelli:2005uq}, a theory in which all the symmetries that are needed for phenomenological consistence are accidental. This includes not only the SM accidental symmetries, but also the additional $\Zdouble_2$ symmetry needed to keep the dark matter particle cosmologically stable. Similar considerations hold for the strong CP problem, for inflation and all other cosmological shortcomings of the SM. The latter could be addressed by light and extremely weakly-coupled new particles or by very heavy ones above the cutoff. In conclusion, none of the above-mentioned issues is powerful enough to put the basic idea of very heavy new physics scale in troubles. The only one that is capable to do so is the Naturalness (or Hierarchy) problem discussed below.\footnote{See Refs.~\cite{Barbieri:2013vca} and \cite{Giudice:2008bi} for recents essays on the Naturalness problem. The problem was first formulated in Refs.~\cite{thooftNat} and \cite{Dimopoulos:1979es,Susskind:1978ms}, however according to the latter references it was K.Wilson who first raised the issue.}

We have not yet encountered the Naturalness problem in our discussion merely because we voluntarily ignored, in our classification, the operators with $d<4$. The only such operator in the SM is the Higgs mass term, with $d=2$.\footnote{There is also the cosmological constant term, of $d=0$. It poses another Naturalness problem that I will mention later.} When studying the $d>4$ operators we concluded that their coefficient is suppressed by $1/\Lambda_{\textrm{SM}}^{d-4}$. Now we have $d=2$ and we are obliged to conclude that the operator is {\emph{enhanced}} by $\Lambda_{\textrm{SM}}^2$, {\it{i.e.}}~that the Higgs mass term reads
\beq
\label{mhUV}
\displaystyle
c\,\Lambda_{\textrm{SM}}^2 H^\dagger H\,,
\eeq
with ``$c$'' a numerical coefficient. In the SM the Higgs mass term sets the scale of EWSB and it directly controls the Higgs boson mass. Today we know that $m_H=125$~GeV and thus the mass term is $\mu^2=m_H^2/2=(89\,{\textrm{GeV}})^2$. But if  $\Lambda_{\textrm{SM}}\sim M_{\textrm{GUT}}$, what is the reason for this enormous hierarchy? Namely
\beq
\displaystyle
{\textrm{why}}\;\;\frac{\mu^2}{\Lambda_{\textrm{SM}}^2}\sim10^{-28}\lll1\;\;{\textrm{{?}}}
\nonumber
\eeq
This is the essence of the Naturalness problem.

Further considerations on the Naturalness problem and implications are postponed to the next section. However, we can already appreciate here how radically it changes our expectations on high energy physics. The SM-only picture gets sharply contradicted by the Naturalness argument since the problem is based on the same logic ({\it{i.e.}}, dimensional analysis) by which its phenomenological virtues ({\it{i.e.}}, the suppression of $d>4$ operators) were established. The new picture is that $\Lambda_{\textrm{SM}}$ is low, in the $100$~GeV to few TeV range, such that a light enough Higgs is obtained ``Naturally'', {\it{i.e.}}~in accordance with the estimate in eq.~(\ref{mhUV}). The new physics at the cutoff must now be highly non-generic, given that it cannot rely any longer on a large scale suppression of the BSM effects. To start with, baryon and lepton family number violating operators must come with a highly suppressed coefficient, which in turn requires baryon and lepton number being imposed as symmetries rather than emerging by accident. In concrete, the BSM sector must now respect these symmetries. This can occur either because it inherits them from an even more fundamental theory or because they are accidental in the BSM theory itself. Similarly, if  $\Lambda_{\textrm{SM}}\sim$~TeV flavour violation cannot be generic. Some special structure must be advocated on the BSM theory, Minimal flavour Violation (MFV) \cite{Glashow:1976nt,D'Ambrosio:2002ex} being one popular and plausible option. The limits from EW Precision Tests (EWPT) come next; they also need to be carefully addressed for TeV scale new physics. On one hand this makes Natural new physics at the TeV scale very constrained. On the other hand it gives us plenty of indications on how it should, or it should not, look like.

\subsection{The Naturalness Argument}\label{natarg}

The reader might be unsatisfied with the formulation of the Naturalness problem we gave so far. All what eq.~(\ref{mhUV}) tells us is that the numerical coefficient ``$c$'' that controls the actual value of the mass term beyond dimensional analysis should be extremely small, namely $c\sim10^{-28}$ for GUT scale new physics. Rather than pushing $\Lambda_{\textrm{SM}}$ down to the TeV scale, where all the above-mentioned constraints apply, one could consider keeping $\Lambda_{\textrm{SM}}$ high and try to invent some mechanism to explain why $c$ is small. After all, we saw that there are other coefficients that require a suppression in the SM Lagrangian, namely the light flavours Yukawa couplings. One might argue that it is hard to find a sensible theory where $c$ is small, while this is much simpler for the Yukawa's. Or that $28$ orders of magnitude are by far much more than the reduction needed in the Yukawa sector. But this would not be fully convincing and would not make full justice to the importance of the Naturalness problem.

In order to better understand Naturalness we go back to the essential message of the previous section. The SM is a low-energy effective field theory and thus the coefficients of its operators, which we regard today as fundamental input parameters, should actually be derived phenomenological parameters, to be computed one day in a more fundamental BSM theory. Things should work just like for the Fermi theory of weak interactions, where the Fermi constant $G_F$ is a fundamental input parameter that sets the strength of the weak force. We know however that the true microscopic description of the weak interactions is the IVB theory. The reason why we are sure about this is that it allows us to predict $G_F$ in terms of its microscopic parameters $g_W$ and $m_W$, in a way that agrees with the low-energy determination. What we have in mind here is merely the standard textbook formula
\beq
\displaystyle
G_F=\frac{g_W^2}{4\sqrt{2}\,m_W^2}\,,
\eeq
that allows us to carry on, operatively, the following program. Measure the microscopic parameters $g_W$ and $m_W$ at high energy; compute $G_F$; compare it with low-energy observations.\footnote{Actually $G_F$ is taken as an input parameter in actual calculations because it is better measured than $g_W$ and $m_W$, but this doesn't affect the conceptual point we are making.} Since this program succeeds we can claim that the microscopic origin of weak interaction is well-understood in terms of the IVB theory. We will now see that the Naturalness problem is an obstruction to repeating the same program for the Higgs mass and in turn for the EWSB scale.

Imagine knowing the fundamental, ``true'' theory of EWSB. It will predict the Higgs mass term $\mu^2$ or, which is the same, the physical Higgs mass $m_H^2=2\mu^2$, in terms of its own input parameters ``$p_{\textrm{true}}$'', by a formula that in full generality reads
\beq
\label{mHtrue}
\displaystyle
m_H^2=\int_0^\infty\hspace{-10pt} dE\;\frac{d m_H^2}{dE}(E;p_{\textrm{true}})\,.
\eeq
The integral over energy stands for the contributions to $m_H^2$ from all the energy scales and it extends up to infinity, or up to the very high cutoff of the ``true'' theory itself. The integrand could be localized around some specific scale or even sharply localized by a delta-function at the mass of some specific particle, corresponding to a tree-level contribution to $m_H^2$. Examples of theories with tree-level contributions are GUT \cite{Langacker:1980js,Raby:2006sk} and Supersymmetric (SUSY) models, where $m_H$ emerges from the mass terms of extended scalar sectors. The formula straightforwardly takes into account radiative contributions, which are the only ones present in the composite Higgs scenario (see sect.~\ref{CH}). Also in SUSY, as discussed in sect.~\ref{SUSY}, radiative terms have a significant impact given that the bounds on the scalar (SUSY and soft) masses that contribute at the tree-level are much milder than those on the coloured stops and gluinos that contribute radiatively. In the language of old-fashioned perturbation theory \cite{Weinberg:1995mt}, ``$E$'' should be regarded as the energy of the virtual particles that run into the diagrams through which $m_H^2$ is computed.

\begin{figure}[t]
\centering
\includegraphics[width=1\textwidth]{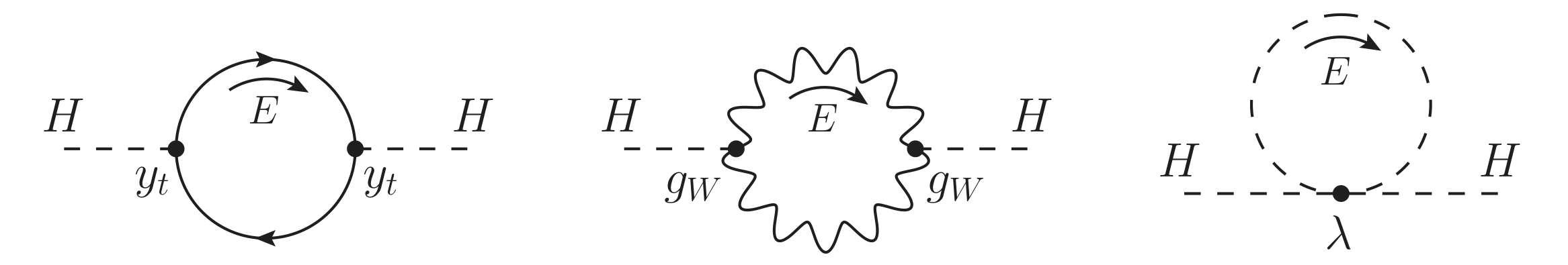}
\caption{Some representative top, gauge and Higgs boson loop diagrams that contribute to the Higgs mass.}
\label{HP}
\end{figure}

Consider now splitting the integral in two regions defined by an intermediate scale that we take just a bit below the SM cutoff. We have
\bea
\label{splitint}
\displaystyle
m_H^2&&=\int_0^{\lesssim{\Lambda_{\textrm{SM}}}}\hspace{-10pt} dE\;\frac{d m_H^2}{dE}(E;p_{\textrm{true}})+\int_{\lesssim{\Lambda_{\textrm{SM}}}}^\infty\hspace{-10pt} dE\;\frac{d m_H^2}{dE}(E;p_{\textrm{true}})\nonumber\\
&&=\delta_{\textrm{SM}}m_H^2+\delta_{\textrm{BSM}}m_H^2\,,
\eea
where $\delta_{\textrm{BSM}}m_H^2$ is a completely unknown contribution, resulting from energies at and above $\Lambda_{\textrm{SM}}$, while $\delta_{\textrm{SM}}m_H^2$ comes from virtual quanta below the cutoff, whose dynamics is by assumption well described by the SM. While there is nothing we can tell about $\delta_{\textrm{BSM}}m_H^2$ before we know what the BSM theory is, we can easily estimate $\delta_{\textrm{SM}}m_H^2$ by the diagrams in Figure~\ref{HP}, obtaining
\beq
\label{deltamh}
\delta_{\textrm{SM}}m_H^2=
\frac{3 y_t^2}{4\pi^2}\Lambda_{\textrm{SM}}^2
-
\frac{3 g_W^2}{8\pi^2}\left(\frac14+\frac1{8\cos^2\theta_W}\right)\Lambda_{\textrm{SM}}^2
-\frac{3 \lambda}{8\pi^2}\Lambda_{\textrm{SM}}^2\,,
\eeq
from, respectively, the top quark, EW bosons and Higgs loops. The idea is that we know that the BSM theory must reduce to the SM for $E<\Lambda_{\textrm{SM}}$. Therefore no matter what the physics at $\Lambda_{\textrm{SM}}$ is, its prediction for $m_H^2$ must contain the diagrams in fig~\ref{HP} and thus the terms in eq.~(\ref{deltamh}). These terms are obtained by computing $dm_H^2/dE$ from the SM diagrams and integrating it up to $\Lambda_{\textrm{SM}}$, which effectively acts as a hard momentum cutoff. The most relevant contributions come from the quadratic divergences of the diagrams, thus eq.~(\ref{deltamh}) can be poorly viewed as the ``calculation'' of quadratic divergences. Obviously quadratic divergences are unphysical in quantum field theory. They are canceled by renormalization and they are even absent in certain regularizations schemes such as dimensional regularization. However the calculation makes sense, in the spirit above, as an estimate of the low-energy contributions to $m_H^2$.

The true nature of the Naturalness problem starts now to show up. The full finite formula for $m_H^2$ obtained in the ``true'' theory receives two contributions that are completely unrelated since they emerge from separate energy scales. At least one of those, $\delta_{\textrm{SM}}m_H^2$, is for sure very large if $\Lambda_{\textrm{SM}}$ is large. The other one is thus obliged to be large as well, almost equal and with opposite sign in order to reproduce the light Higgs mass we observe. A cancellation is taking place between the two terms, which we quantify by a fine-tuning $\Delta$ of at least
\beq
\label{deltatuning}
\Delta\geq\frac{\delta_{\textrm{SM}}m_H^2}{m_H^2}
=\frac{3\, y_t^2}{4\pi^2}\left(\frac{\Lambda_{\textrm{SM}}}{m_H}\right)^2\simeq \left(\frac{\Lambda_{\textrm{SM}}}{450\,{\textrm{GeV}}}\right)^2\,.
\eeq
Only the top loop term in eq.~(\ref{deltamh}) has been retained for the estimate since the top dominates because of its large Yukawa coupling and because of color multiplicity. Notice that the one above is just a lower bound on the total amount of cancellation $\Delta$ needed to adjust $m_H$ in the true theory. The high energy contribution $\delta_{\textrm{BSM}}m_H^2$, on which we have no control, might itself be the result of a cancellation, needed to arrange for $\delta_{\textrm{BSM}}m_H^2\simeq -\delta_{\textrm{SM}}m_H^2$. Examples of this situation exist both in SUSY and in composite Higgs.

The problem is now clear. Even if we were able to write down a theory that formally predicts the Higgs mass, and even if this theory turned out to be correct we will never be able to really predict $m_H$ if $\Lambda_{\textrm{SM}}$ is much above the TeV scale, because of the cancellation. For $\Lambda_{\textrm{SM}} = M_{\textrm{GUT}}$, for instance, we have $\Delta\gtrsim10^{24}$. This means that in the ``true'' theory formula for $m_H$ a $24$ digits cancellation is taking place between two a priori unrelated terms. Each of these terms must thus be known with at least $24$ digits accuracy even if we content ourselves with an order one estimate of $m_H$. We will never achieve such an accuracy, neither in the experimental determination of the $p_{\textrm{true}}$ ``true'' theory parameters $m_H$ depends on, nor in the theoretical calculation of the Higgs mass formula. Therefore, we will never be able to repeat for $m_H$ the program we carried on for $G_F$ and we will never be able to claim we understand its microscopic origin and in turn the microscopic origin of the EWSB scale. A BSM theory with $\Lambda_{\textrm{SM}} = M_{\textrm{GUT}}$ has, in practice, the same predictive power on $m_H$ as the SM itself, where eq.~(\ref{mHtrue}) is replaced by the much simpler formula
\beq
m_H^2=m_H^2\,.
\eeq
Namely if such an high-scale BSM theory was realized in Nature $m_H$ will remain forever an input parameter like in the SM. The microscopic origin of $m_H$, if any, must necessarily come from new physics at the TeV scale, for which the fine-tuning $\Delta$ in eq.~(\ref{deltatuning}) can be reasonably small.

The Higgs mass term is the only parameter of the SM for which such an argument can be made. Consider for instance writing down the analog of eq.~(\ref{mHtrue}) for the Yukawa couplings and splitting the integral as in eq.~(\ref{splitint}). The SM contribution to the Yukawa's is small even for $\Lambda_{\textrm{SM}} = M_{\textrm{GUT}}$, because of two reasons. First, the Yukawa's are dimensionless and thus, given that there are no couplings in the SM with negative energy dimension, they do not receive quadratically divergent contributions. The quadratic divergence is replaced by a logarithmic one, with a much milder dependence on $\Lambda_{\textrm{SM}}$. Second, the Yukawa's break the flavour group of the SM. Therefore there exist selection rules (namely those of MFV) that make radiative corrections proportional to the Yukawa matrix itself. The Yukawa's, and the hierarchies among them, are thus ``radiatively stable'' in the SM (see sect.~\ref{tale} for more details). This marks the essential difference with the Higgs mass term and implies that their microscopic origin and the prediction of their values could come at any scale, even at a very high one. The same holds for all the SM parameters apart from $m_H$.

The formulation in terms of fine-tuning (\ref{deltatuning}) turns the Naturalness problem from a vague aesthetic issue to a concrete semiquantitative question. Depending on the actual value of $\Delta$ the Higgs mass can be operatively harder or easier to predict, making the problem more or less severe. If for instance $\Delta\sim10$, we will not have much troubles in overcoming a one digit cancellation once we will know and we will have experimental access to the ``true'' theory. After some work, sufficiently accurate predictions and measurements will become available and the program of predicting $m_H$ will succeed. The occurrence of a one digit cancellation will at most be reported as a curiosity in next generation particle physics books and we will eventually forget about it. A larger tuning $\Delta=1000$ will instead be impossible to overcome. The experimental exploration of the high energy frontier will tell us, through eq.~(\ref{deltatuning}), what to expect about $\Delta$. Either by discovering new physics that addresses the Naturalness problem or by pushing $\Lambda_{\textrm{SM}}$ higher and higher until no hope is left to understand the origin of the EWSB scale in the sense specified above. One way or another, a fundamental result will be obtained. 

\subsection{What if Un-Natural?}

I argued above that searching for Naturalness at the LHC is relevant regardless of the actual outcome of the experiment. Such a bold statement needs to be more extensively defended. The case of a discovery is so easy that it would not even be worth discussing. If new particles are found at the TeV scale, with properties that resemble what predicted by a Natural BSM theory such as the ones described in the following sections, Naturalness would have guided us towards the discovery of new physics. Moreover, it will provide the theoretical framework for the interpretation of the discoveries, by which the new particles will eventually find their place in a concrete BSM model. If instead nothing related with Naturalness will be found, strong limits will be set on $\Delta$ and we will be pushed towards the idea that the $m_H^2$ parameter does not have a canonical ``microscopic'' origin as previously explained. This would still qualify as a discovery: the discovery of ``Un-Naturalness''.\footnote{Deciding whether or not negative LHC results will have the last word on Naturalness is a matter of taste, to some extent, since it is unclear how much tuning we can tolerate. It also depends on how good we will be in searching for Natural new physics and consequently how strong and robust the limit on $\Delta$ will actually be. It is nevertheless undoubtable that negative LHC results will put the idea of Naturalness in serious troubles.} The profound implications of this potential discovery are discussed below.

If Un-Naturalness will be discovered, other options will have to be considered to explain the origin of the Higgs mass term. The two known possibilities are that $m_H^2$ has an ``environmental'' or a ``dynamical'' origin rather than a ``microscopic'' one, as previously assumed. A well-known parameter with environmental origin is the Gravity of Earth $g=9.8\,m/s^2$. It is the input parameter of Ballistics, a theory of great historical relevance which in Galileo's times might have been conceivably thought to be a fundamental theory of Nature. The origin of $g$ is obviously dictated by the environment in which the theory is formulated, namely by the fact that Ballistics applies to processes that occur close to the surface of Earth. Its value depends on the Earth's mass and radius and it cannot be inferred just based on the knowledge of the ``truly fundamental'' theory of Gravity (Newton's law) and of its parameters (Newton's constant). This is not the case for those parameters, such as $G_F$, with a purely microscopic origin. The dependence on the environment can help explaining the size of an environmental parameter by the so-called ``Anthropic'' argument. In fact, the value of $g=9.8\,m/s^2$ is rather peculiar. It is much larger than the one we would observe in interstellar space and much smaller than the one on the surface of a neutron star, very much like $m_H$ is much smaller than $M_{\textrm{P}}$ or $M_{\textrm{GUT}}$. However we do perfectly understand the magnitude of $g$, for the very simple reason that no ancient physicist might have lived in empty space or on a neutron star. The magnitude of $g$ must be compatible with what is needed for the development of intelligent life, otherwise no physicist would have existed and nobody would have measured it.

The Weinberg prediction of the cosmological constant \cite{Weinberg:1987dv} proceeds along similar lines. The cosmological constant operator suffers of exactly the same Naturalness problem as the Higgs mass. Provided we claim we understand gravity well enough to estimate them, radiative corrections push the cosmological constant to very high values, tens of orders of magnitude above what we knew it had to be (and was subsequently observed) in order for galaxies being able to form in the early universe. Weinberg pointed out that the most plausible value for the cosmological constant should thus be close to the maximal allowed value for the formation of galaxies because galaxies are essential for the development of intelligent life. The idea is that if many ground state configurations (a landscape of vacua) are possible in the fundamental theory, typically characterised by a very large cosmological constant but with a tail in the distribution that extends up to zero, the largest possible value compatible with galaxies formation, and thus with the very existence of the observer, will be actually observed. A similar argument can be made for the Higgs mass (see for instance Ref.~\cite{Hall:2007ja}), however it is harder in the SM to identify sharply the boundary of the anthropically allowed region of the parameter space. 

I tried here to vulgarise the mechanism of anthropic vacua selection by the example of Gravity of Earth, however the analogy is imperfect under several respects. Perhaps the most important difference is that the landscape of vacua cannot be viewed as a set of physical regions (like the interstellar space or the neutron star) separated in space, where $m_H$ or the cosmological constant assume different values. Or at least, since the other vacua live in space-time regions that are causally disconnected from us, it will be impossible to have access to them and check directly that the mechanism works.

The possibility of a ``dynamical'' origin of the Higgs mass term is quite new \cite{Graham:2015cka} and not much studied.\footnote{The word ``dynamical'' is used here in its proper sense, related with evolution in the course of time. It has nothing to do with the generation of energy scales (e.g,, the QCD confinement scale) induced by an underlying strongly-coupled theory, which is also said to be a ``dynamical'' generation mechanism.} The idea, first proposed in \cite{Abbott:1984qf} as an unsuccessful attempt to solve the cosmological constant problem, is that $m_H$ might be set by the expectation value of a new scalar field, whose value evolves during cosmological Inflation. This field is called ``relaxion'' in \cite{Graham:2015cka} because it is similar to the QCD axion needed to address the strong-CP problem and because it sets the value of $m_H$ by a dynamical relaxation mechanism. At the beginning of Inflation, the relaxion VEV is such that the Higgs mass term is large and positive, but it evolves in the course of time making the Higgs mass term decrease and eventually cross zero so that EWSB can take place. The structure of the theory is such that once a non-vanishing Higgs VEV is generated, a barrier develops in the relaxion potential and makes it stop evolving. The Higgs mass term gets thus frozen to the value which is just sufficient for an high enough barrier to form. If the theory is special enough (but not necessarily complicate), this value can be small and the Hierarchy problem can be solved.

You might find these speculations extremely interesting. Or you might believe that they have no chance to be true. Anyhow, their vey existence demonstrates how radically the discovery of Un-Naturalness would change our perspective on the physics of fundamental interactions. They show the capital importance of searching for Naturalness or Un-Naturalness at the LHC and, perhaps, at future colliders.

\section{Composite Higgs}\label{CH}

One aspect of the Naturalness problem which has not yet emerged is the fact that addressing it requires BSM physics of rather specific nature at $\Lambda_{\textrm{{SM}}}\lesssim$~TeV. Namely, it is true that any BSM scenario that Naturally explains the origin of $m_H$ is obliged to show up at the TeV by eq.~(\ref{deltatuning}), but this does not mean that the presence of generic new particles at the TeV scale would solve the Naturalness problem. Conversely, it is not true that any BSM particle we might  happen not to discover at the TeV scale would signal that the theory is fine-tuned as a naive application of eq.~(\ref{deltatuning}) would suggest. Natural BSM physics would show up through new particles (and/or, indirect effects on SM processes) of specific nature and it is only the non-discovery of these particles the one that matters for the tuning $\Delta$. Addressing this point requires studying concrete BSM solutions to the Naturalness problem.

Among the various scenarios which have been proposed to address the Naturalness problem I decided to focus on two of them: Supersymmetry and Composite Higgs. The reason for this choice is that they are representative of the only two known mechanisms which truly address the problem of the microscopic origin of $m_H$ by a well-defined high-energy picture. Alternative Natural models are often reformulations or deformations of these basic scenarios, or a combination of the two.\footnote{For instance, certain Randall-Sundrum models are reformulations of the Composite Higgs scenario with or without the Higgs being a pseudo-Nambu--Goldstone Boson (pNGB). Little Higgs (see \cite{Schmaltz:2005ky,Perelstein:2005ka} for a review) is a pNGB Higgs endowed with a special mechanism which could make it more Natural. Twin Higgs \cite{Chacko:2005pe} is an additional protection for $m_H$ which postpones the emergence of coloured particles in the spectrum. It can be applied both to the Composite Higgs and to the SUSY scenario.} You are referred to Ref.~\cite{Pomarol:2012sb} for a comprehensive overview.

\subsection{The Basic Idea}\label{TBI}

\begin{figure}[t]
  \centering
  \includegraphics[width=0.7\textwidth]{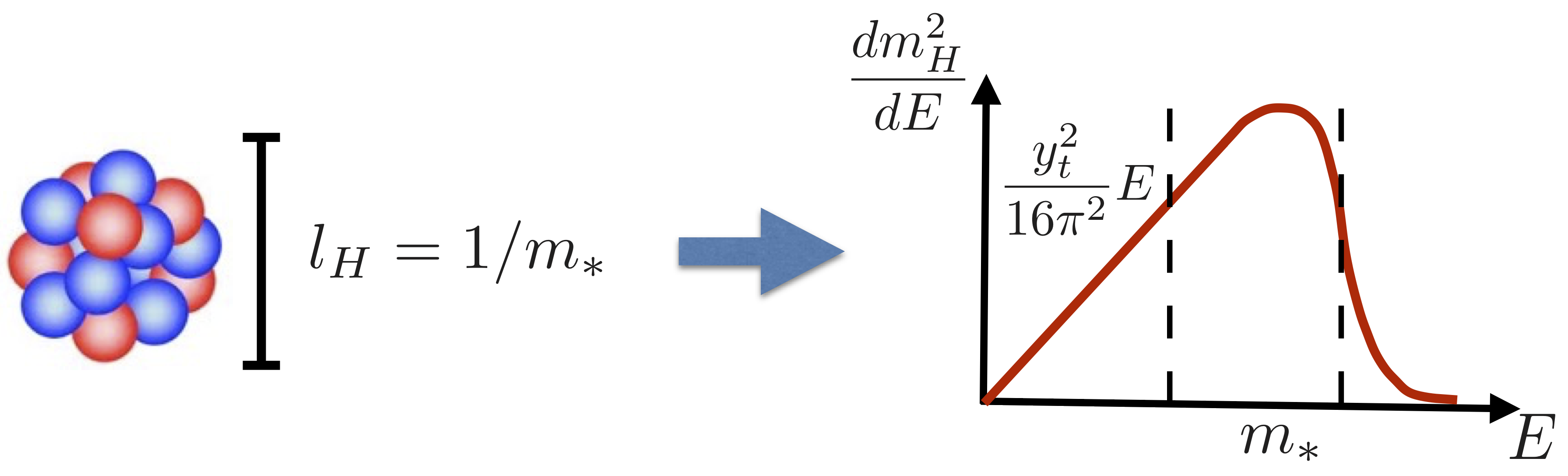} 
\caption{\label{CH_drawing}Pictorial representation of the Composite Higgs solution to the  Naturalness problem.} 
\end{figure}

The composite Higgs scenario offers a simple solution to the problem of Naturalness. Suppose that the Higgs, rather than being a point-like particle as in the SM, is instead an extended object with a finite geometric size $l_H$. We will make it so by assuming that it is the bound state of a new strong force characterised by a confinement scale $m_*=1/l_H$ of TeV order. In this new theory the $dm_H^2/dE$ integrand in the Higgs mass formula (\ref{mHtrue}), which stands for the contribution of virtual quanta with a given energy, behaves as shown in fig.~\ref{CH_drawing}. Low energy quanta have too a large wavelength to resolve the Higgs size $l_H$. Therefore the Higgs behaves like an elementary particle and the integrand grows linearly with $E$ like in the SM, resulting in a quadratic sensitivity to the upper integration limit. However this growth gets canceled by the finite size effects that start becoming visible when $E$ approaches and eventually overcomes $m_*$. Exactly like the proton when hit by a virtual photon of wavelength below the proton radius, the composite Higgs is transparent to high-energy quanta and the integrand decreases. The linear SM behaviour is thus replaced by a peak at $E\sim m_*$ followed by a steep fall. The Higgs mass generation phenomenon gets localised at $m_*=1/l_H$ and $m_H$ is insensitive to much higher energies. This latter fact is also obvious from the fact that no Higgs particle is present much above $m_*$. Therefore there exist no Higgs field and no $d=2$ Higgs mass term to worry about.

\begin{figure}[t]
\centering
\includegraphics[width=0.75\textwidth]{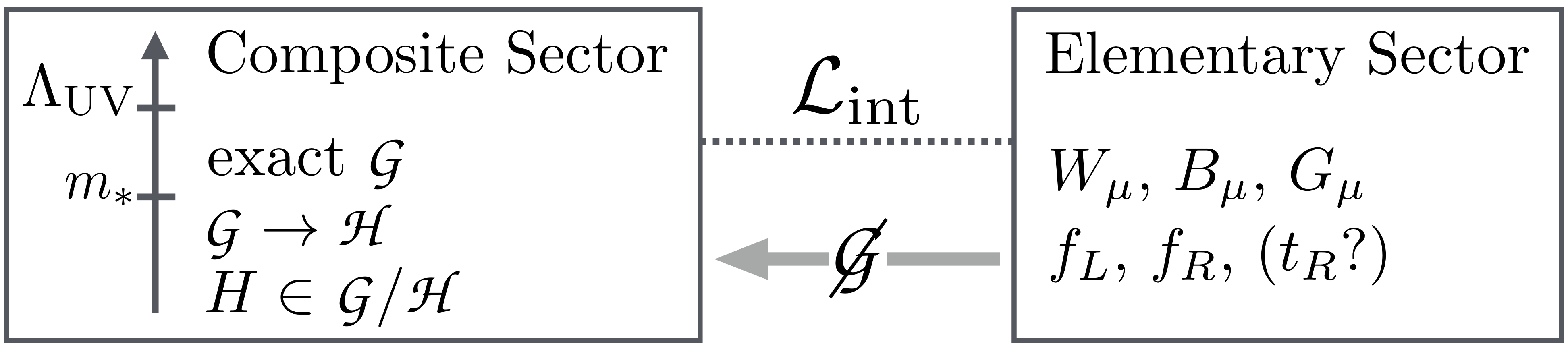}
\caption{The basic structure of the composite Higgs scenario.}
\label{elcomp}
\end{figure}

Implementing this idea in practice requires a theory with the structure in fig.~\ref{elcomp}. The three basic elements are a ``Composite Sector'' (CS), an ``Elementary Sector'' (ES) and a set of interactions ``${\mathcal{L}}_{\textrm{int}}$'' connecting the two. The Composite Sector contains the new particles and interactions that form the Higgs as a bound state and it should be viewed as analogous to the QCD theory of quarks and gluons. The CS plays the main role for the composite Higgs solution to the Naturalness problem as it gives physical origin to the Higgs compositeness scale $m_*$. In the analogy with QCD, $m_*$ corresponds to the QCD confinement scale $\Lambda_{\textrm{QCD}}$ and it is generated, again like in QCD, by the mechanism of dimensional transmutation. Thanks to this mechanism it is insensitive to other much larger scales which are present in the problem. For instance the microscopic origin of the CS itself might well be placed at $\Lambda_{\textrm{UV}}\sim{M}_{\textrm{GUT}}$, but still $m_*$ could be Naturally of TeV order, very much like $\Lambda_{\textrm{QCD}}\sim300$~MeV$\,\ll{m}_{\textrm{EW}}$ is perfectly Natural within the SM. 

The Elementary Sector contains all the particles we know, by phenomenology, cannot be composite at the TeV scale.\footnote{Those particles might be ``partially composite'', a concept that we will introduce below.} Those are basically all the SM gauge and fermion fields with the possible exception of the right-handed component of the top quark. The most relevant operators in the ES Lagrangian, namely those that are not suppressed by $1/\Lambda_{\textrm{UV}}^n$, are thus just the ordinary $d=4$ SM gauge and fermion kinetic terms and gauge interactions. Since there is no Higgs, no dangerous $d=2$ operator is present in the ES and thus the theory is perfectly Natural. Obviously the lack of a Higgs also forbids Yukawa couplings and a different mechanism will have to be in place to generate fermion masses and mixings. 

The Elementary-Composite interactions ${\mathcal{L}}_{\textrm{int}}$ consist of two classes of terms: those involving the elementary gauge fields and those involving the elementary fermionic field. The latter are responsible for fermion masses and will be discussed later. The former are instead sharply dictated by gauge invariance and read
\beq\ds
{\mathcal{L}}_{\textrm{int}}^{\textrm{gauge}}=\sum\limits_{i=1,2,3}g_i A^\mu_i J_\mu^i\,,
\label{gaugeint}
\eeq
where $i$ runs over the three \mbox{SU$(3)_c\times$SU$(2)_L\times$U$(1)_Y$} irreducible factors of the SM gauge group and $g_i$ denotes the corresponding gauge coupling. In the equation, $J_\mu^i$ represents the global current operators of the Composite Sector, namely the Noether currents associate with each of the three irreducible factors of the SM group. Notice that for this to make sense the CS must be invariant under the SM symmetries, therefore the complete global symmetry group of the CS, denoted by ``${\mathpzc{G}}$'' in fig.~\ref{elcomp}, must at least contain the SM one as a subgroup. Good reasons to make ${\mathpzc{G}}$ larger will be discussed shortly. Pushing forward the analogy with low-energy QCD and hadron physics, the ES sector is analogous to the photon plus light leptons system, whose coupling to the CS proceed through the electromagnetic gauge interaction precisely as in eq.~(\ref{gaugeint}).

The generic framework described until now has an important pitfall, which is overcame in what we nowadays properly call the ``Composite Higgs'' scenario \footnote{See \cite{Kaplan:1983fs,Kaplan:1983sm,Dugan:1984hq} for earlier references and \cite{Agashe:2004rs,Giudice:2007fh} for more recent ones.} by the fact that the Higgs is a pseudo Nambu--Goldstone Boson (pNGB). The pitfall is that if the Higgs is a generic bound state of the CS  dynamics one generically expects its mass to be of the order of the CS confinement scale $m_*$, namely $m_H\sim{m}_*$. In a sense, the point is that the mechanism of fig.~\ref{CH_drawing} does indeed solve the Naturalness problem by making the shape of $dm_H^2/dE$ localised at $m_*$ but tells us nothing about the normalisation of the $dm_H^2/dE$ function. In the absence of a special mechanism one can estimate $dm_H^2/dE\sim{m}_*$ at $E\sim{m_*}$ and the result of the integral is $m_H^2\sim{m}_*^2$. One can reach the same conclusion heuristically by exploiting the analogy with QCD and browsing one of the many PDG \cite{Agashe:2014kda} summary tables devoted to the properties of hadrons. By picking one generic (random) hadron in the list one would find that its mass is around the QCD confinement scale $\Lambda_{\textrm{QCD}}$ and that it is surrounded by many other hadrons (a bit heavier or lighter) with similar properties. The Higgs particle is instead alone in the spectrum, or at least we are pretty sure that we would have seen (directly and/or indirectly) at least some of the other particles that would come with it if $m_*$ was around $m_H\sim100$~GeV. Therefore $m_*$ must be of the TeV or multi-TeV order and some mechanism must be in place to explain why $m_H\ll{m}_*$. The problem is actually even more severe than that because the Higgs, on top of being light, is a narrow weakly coupled particle and furthermore its couplings are measured to agree with what predicted by the SM at the $10$ or $20\%$ level.\footnote{We nowadays know this directly from the LHC Higgs couplings determinations. Indirect evidences of SM-like couplings for the Higgs boson could however already be extracted from precision LEP data.} The existence of a CS resonance obeying these non-trivial properties by accident for no special underlying reason, appears extremely unlikely. The explanation of all these facts might be that the Higgs is a pNGB, namely a special CS hadron associated with the spontaneous breaking of the CS's global symmetry group ${\mathpzc{G}}$. The Higgs is said a ``pseudo'' NGB (pNGB) because ${\mathpzc{G}}$ is not an exact but an approximate symmetry. This is precisely what happens in QCD, where the $\pi$ mesons are light because they are pNGB's associated with the spontaneous breaking of the chiral group. The Higgs might be analogous to a pion, rather than to a random hadron in the PDG list.

The theory of Nambu--Goldstone Bosons works as follows. If the CS is endowed by the global group of symmetry ${\mathpzc{G}}$, it is generically expected that this group will be broken spontaneously to a subgroup ${\mathpzc{H}}\subset{\mathpzc{G}}$ by CS confinement. If this happens, the Goldstone Theorem guarantees that a set of scalar particles, exactly massless as long as ${\mathpzc{G}}$ is an exact symmetry, are present in the spectrum. The theorem says that one such massless NGB particle arises for each of the symmetry generators that are broken in the ${\mathpzc{G}}\rightarrow {\mathpzc{H}}$ pattern, namely one for each generator in ${\mathpzc{G}}$ which is not part of the unbroken ${\mathpzc{H}}$. The broken generators and the corresponding NGB's are collected in what is called the ``${\mathpzc{G}}/{\mathpzc{H}}$ coset''. If the Higgs emerges as one of those particle, which we can achieve by a judicious choices of the coset as discussed in the next section, it will be Naturally light given that its mass cannot be generated from the CS alone, which is exactly invariant under ${\mathpzc{G}}$. A non-vanishing Higgs mass requires the interplay with the ES that breaks the ${\mathpzc{G}}$ symmetry and communicates the breaking to the CS trough ${\mathcal{L}}_{\textrm{int}}$ as in fig.~\ref{elcomp}. Given that the Elementary/Composite interactions are weak and perturbative, such as the gauge couplings in eq.~(\ref{gaugeint}), a considerable gap between $m_H$ and $m_*$ is Naturally expected. 

It is important to remark that the pNGB nature of the Higgs can also explain why its couplings are close to the SM expectations. This comes from a general mechanism called ``vacuum misalignment'' discovered in Refs.~\cite{Kaplan:1983fs,Kaplan:1983sm,Dugan:1984hq}. I will illustrate how it works in the next section through an example. The picture according to which the Higgs might be the lightest state of the CS, and thus the first one in being discovered, because it is a pNGB, turns out to be rather plausible.

\subsection{The Minimal Composite Higgs Couplings}\label{MCHC}

A rigorous and complete description of the Composite Higgs (CH) scenario goes beyond the purpose of these lectures, the interested reader is referred to the extensive reviews in \cite{Contino:2010rs,Panico:2015jxa}. However most of the relevant features of CH can be illustrated by performing a specific calculation in a specific CH model, namely by computing the couplings of the Higgs to SM particles in the so-called Minimal CH Model (MCHM). Studying Higgs couplings and their possible departures from the SM expectations is one of the ways in which CH models have been and are being searched for at the LHC. Therefore the relevance of the calculation goes beyond its pedagogical value.

\begin{figure}
\centering
\includegraphics[width=0.7\textwidth]{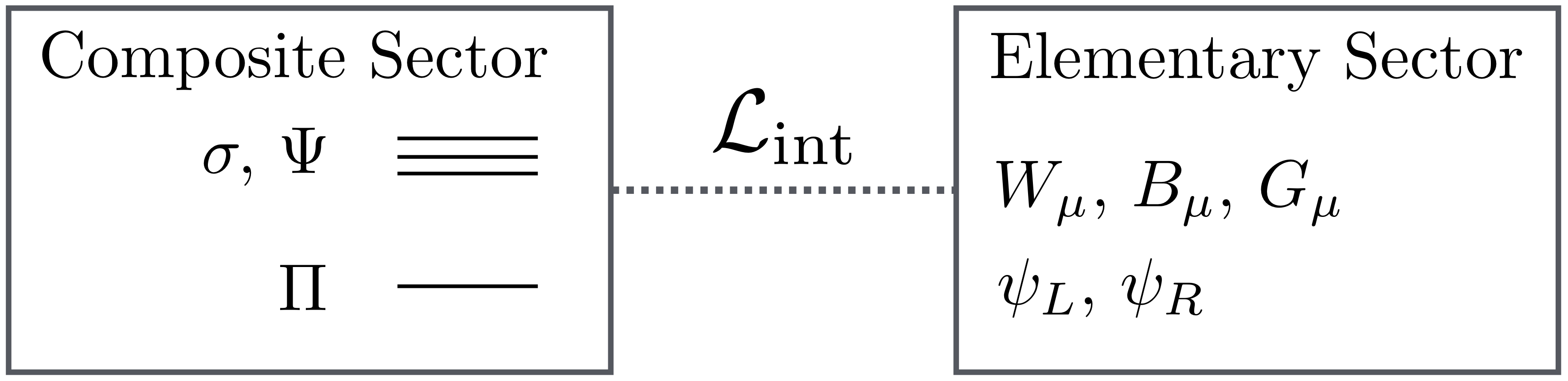}
\caption{The composite Higgs setup. The elementary SM gauge fields are the three $W$'s, the hypercharge boson $B$ and the eight QCD gluons. The elementary fermionic quark and lepton fields are collectively denoted as $\psi_L$ and $\psi_R$. The Higgs is labeled as ``$\Pi$'' (see the main text) and $\sigma$, $\Psi$ represent Composite Sector resonances.}
\label{ecomp}
\end{figure}

The MCHM \cite{Agashe:2004rs} is based on the choice ${\mathpzc{G}}={\textrm{SO}}(5)$ and ${\mathpzc{H}}={\textrm{SO}}(4)$, which delivers NGB's in the so-called ``minimal coset'' ${\textrm{SO}}(5)/{\textrm{SO}}(4)$. According to the Goldstone theorem, the number of real NGB scalar fields in this theory is $4=10-6$, equal to the number of generators in ${\textrm{SO}}(5)$ minus those in ${\textrm{SO}}(4)$. Four real scalars are just sufficient to account for the two complex components of one Higgs doublet. Therefore the ${\textrm{SO}}(5)/{\textrm{SO}}(4)$ coset delivers a single doublet, rather than an extended Higgs sector as it would be the case if larger ${\mathpzc{G}}$ and ${\mathpzc{H}}$ groups are considered. This is why it is called the minimal coset. The Goldstones, i.e. the Higgs, are the lightest particles of the CS, as shown in fig.~\ref{ecomp}. Therefore they can be  studied independently of the other hadrons of the CS (called ``resonances'') at all energies below the resonance mass scale $m_*\sim\;$~TeV. On-shell Higgs couplings are low-energy observables in this context, thus they can be computed independently of the detailed knowledge of the resonance dynamics.

\begin{figure}
\centering
\includegraphics[width=0.32\textwidth]{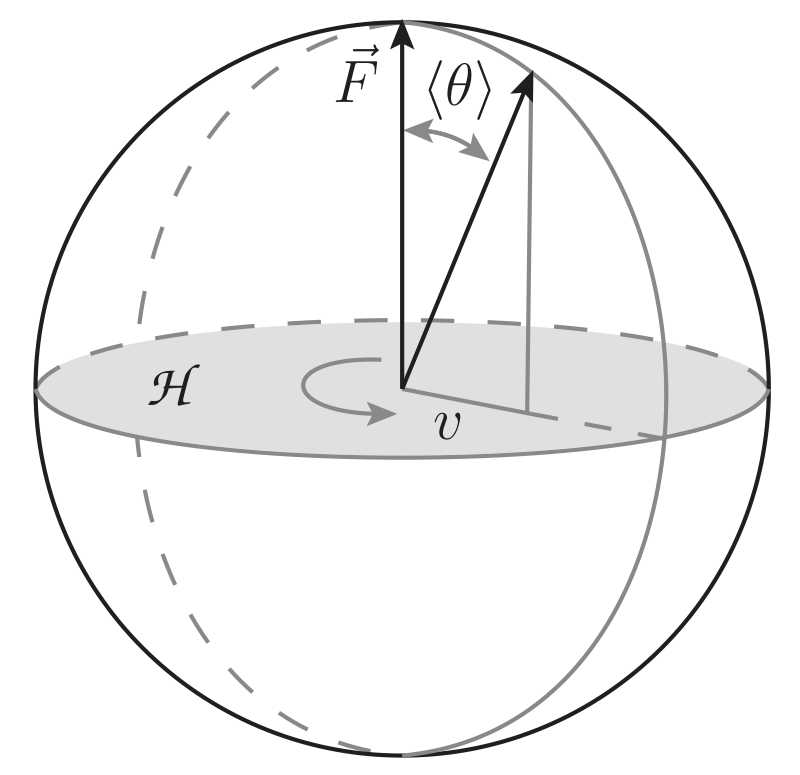}
\caption{A geometrical illustration of EWSB through vacuum misalignment, in the case of the spatial rotations group \mbox{${\mathpzc{G}}=\;$SO$(3)$} with \mbox{${\mathpzc{H}}=\;$SO$(2)$}. The \mbox{SO$(2)$} breaking from vacuum misalignment is proportional to the projection of $\vec{F}$ on the \mbox{SO$(2)$} plane, $v=f\sin\langle\theta\rangle$.}
\label{vmis}
\end{figure}

A simple model for Goldstone bosons is defined as follows. Be $\ve\Phi$ a five-components vector of real fields, on which the ${\textrm{SO}}(5)$ group acts as rotations in five dimensions, and impose on it the condition
\beq\label{constr}
\ds
\vet{\Phi}\cdot\ve{\Phi}=f^2\,.
\eeq
The constant parameter $f$ is called the ``Higgs decay constant'' because it plays in CH the same role of the pion decay constant $f_\pi$ in the low-energy theory of QCD pions. It has the dimensionality of energy and it represents the scale of ${\mathpzc{G}}\rightarrow{\mathpzc{H}}$ spontaneous breaking. The $4$ Goldstone bosons $\Pi_i$, $i=1,\ldots,4$ are introduced as the fields that parameterise the solutions to the constraint (\ref{constr}), namely 
\beq
\displaystyle
\ve\Phi=f\left[
\begin{array}{c}
\sin{\frac{\Pi}{f}}\frac{\ve{\Pi}}{\Pi}\\
\cos{\frac{\Pi}{f}}
\end{array}
\right]\,,
\label{fred5}
\eeq
where $\Pi=\sqrt{\vet\Pi\cdot\ve\Pi}$. Geometrically (see fig.~\ref{vmis}), $\ve\Phi$ lives on a sphere in the five-dimensional space and $\ve\Pi$ are the four angular variables which are needed to parametrise the sphere. Notice that the constraint (\ref{constr}) is invariant under ${\textrm{SO}}(5)$ rotations of $\ve\Phi$, therefore the theory of Goldstone Bosons we will construct out of it will respect the ${\textrm{SO}}(5)$ symmetry. A controlled and perturbative breaking of the symmetry will emerge from the coupling with SM gauge fields and fermions. 

The four $\Pi$'s are the Higgs, but this is not yet apparent because the Higgs field is typically represented as a two-components complex doublet $H=(h_u,h_d)^T$ rather than a real quadruplet. The conversion between the two notations is provided by 
\begin{equation}
\label{dh}
\vec{\Pi}=\left[\begin{matrix}\Pi_1\\ \Pi_2\\ \Pi_3\\ \Pi_4\end{matrix}\right]=
\frac1{\sqrt{2}}\left[\begin{matrix}-i\,(h_u-h_u^\dagger)\\ 
h_u+h_u^\dagger
 \\ i\,(h_d-h_d^\dagger)\\ 
h_d+h_d^\dagger
 \end{matrix}\right]\,.
\end{equation}
The deep meaning of this equation is that the unbroken group \mbox{SO$(4)$} is actually equivalent to the product of two groups, \mbox{SU$(2)_L\times$SU$(2)_R$}, where \mbox{SU$(2)_L$} is the habitual SM one and \mbox{SU$(2)_R$} is a generalisation of the SM Hypercharge \mbox{U$(1)_Y$}.\footnote{\label{cuss}This group is also called the ``custodial'' \mbox{SO$(4)_c$}. It plays a major role in BSM physics as it suppresses certain BSM effects constrained by LEP and often helps the compatibility of BSM models with data.} Namely,  \mbox{SU$(2)_R$} contains the Hypercharge, which is identified with its third generator, $Y=T_R^3$. The Higgs quadruplet $\ve\Pi$ is a $\mathbf{4}$ of \mbox{SO$(4)$}, or equivalently a $\mathbf{(2,2)}$ of \mbox{SU$(2)_L\times$SU$(2)_R$}. The $\mathbf{(2,2)}$ transforms as a $\mathbf{2_{1/2}}$ Higgs doublet under the SM \mbox{SU$(2)_L\times$U$(1)_Y$} subgroup. The conversion formula in eq.~(\ref{dh}) does depend on the convention chosen for the \mbox{SO$(4)$} generators. I thus report them for completeness
\beq
\ds
T_{L{\textrm{/}}R}^\alpha=\left[\begin{array}{cc} t_{L{\textrm{/}}R}^\alpha & 0 \\
0 & 0
\end{array}
\right]\,,\;\;\;\;\;
(t^\alpha_{L{\textrm{/}}R})_{ij} = -\frac{i}{2}\left[\varepsilon_{\alpha\beta\gamma}
\delta_i^\beta \delta_j^\gamma \pm
\left(\delta_i^\alpha \delta_j^4 - \delta_j^\alpha \delta_i^4\right)\right]\,.
\eeq
In the equation, capital $T_{L{\textrm{/}}R}^\alpha$ ($\alpha=1,2,3$) denote the $5\times5$ generators of \mbox{SO$(4)$} seen as a subgroup of \mbox{SO$(5)$}, small $t_{L{\textrm{/}}R}^\alpha$ are the habitual generators written as $4\times4$ matrices.

The Lagrangian for $\ve\Phi$, out of which the one of the Goldstones will be straightforwardly extracted, simply reads
\beq
\label{CHlag0}
\ds
{\mathcal{L}}=\frac12 D_\mu\vet\Phi\cdot D^\mu\ve\Phi\,,\;\;\;\;\;{\textrm{where}}\;\; D_\mu\ve\Phi=\left(\partial_\mu -i\, g W_{\mu}^{\alpha} T_L^\alpha - i\, g'B_\mu T_R^3\right)\ve\Phi\,.
\eeq
Notice that the couplings with the SM gauge fields $W^\alpha$ and $B$ come from the covariant derivative and  they are completely determined by the requirement of gauge invariance. This is exactly what happens when we construct the SM through the habitual gauging procedure and follows from the fact that we decided, in  eq.~(\ref{gaugeint}), to introduce the SM $W$ and $B$ as gauge fields. As a result of this fact, a very sharp prediction will be obtained for the Higgs couplings to the SM vector bosons. To compute the couplings of the physical Higgs we go to the unitary gauge 
\beq
\label{ugauge}
\displaystyle
H=\left[\begin{array}{c} 0 \\ \frac{V+h(x)}{\sqrt{2}}\end{array}\right]\,,
\eeq
and eq.~(\ref{CHlag0}) becomes
\beq
{\mathcal{L}}=\frac12\left(\partial_\mu h\right)^2 + \frac{g^2}4 f^2 \sin^2{\frac{V+h}{f}} \left(|W|^2+\frac1{2c_w^2}Z^2\right)\,,
\label{LUN}
\eeq
where $W$ and $Z$ denote the ordinary SM mass and charge eigenstate fields, $c_w$ is the cosine of the weak mixing angle defined as usual by \mbox{$\tan\theta_w=g'/g$}. The parameter $V$ denotes the VEV of the Higgs field, induced by a yet unspecified potential.

We can learn a lot on CH by looking at eq.~(\ref{LUN}). First of all, we can read the mass of the SM vector bosons
\beq
\displaystyle
m_W=c_w m_Z=\frac12 \, g f \sin\frac{V}{f}\equiv \frac12 \, g\, v\,,
\label{mwz}
\eeq
and, by comparing with the corresponding SM formulas, extract the definition of the physical EWSB scale $v\simeq 246$~GeV. We see that $v$, unlike in the SM, is not directly provided by the composite Higgs VEV, but rather it is given by
\beq\label{xval}
\ds
v=f\sin\frac{V}{f}\,.
\eeq
The geometrical reason for this equation is illustrated in fig.~\ref{vmis}. According to eq.~(\ref{fred5}), the vacuum configuration assumed by $\ve\Phi$ when the Higgs takes a VEV, call it $\langle\ve\Phi\rangle$, is a vector of norm $f$ that forms an angle $\langle\theta\rangle=\langle\Pi\rangle/f=V/f$ with the reference vector $\ve{F}=(0,0,0,0,f)^T$. The reference vector is the vacuum configuration $\ve\Phi$ would assume if the Higgs had vanishing VEV and the angle $\langle\theta\rangle$ measures how far the true VEV is from the reference vector. If $\langle\ve\Phi\rangle=\ve{F}$, the vacuum would be invariant under \mbox{SO$(4)$}, and thus in particular under the SM group which is part of \mbox{SO$(4)$}. The amount of breaking of the EW symmetry is thus measured by the transverse component of $\langle\ve\Phi\rangle$ with respect to $\ve{F}$ because it is only this component the one that makes the vacuum configuration non-invariant under the SM group. From this observation, eq.~(\ref{xval}) follows. An important property of eq.~(\ref{mwz}) that I should not forget to outline is that the $W$ and $Z$ boson masses are related by the familiar SM tree-level condition $m_W=c_w m_Z$, which is accurately established experimentally. This property is due to the unbroken \mbox{SO$(4)$} group and it furnishes one example of the ability of this ``custodial'' symmetry to suppress BSM effects as mentioned in footnote~\ref{cuss}.

Next, we can Taylor-expand eq.~(\ref{LUN}) in powers of the physical Higgs field $h(x)$ and notice that it provides  an infinite set of local interactions involving two gauge and an arbitrary number of Higgs fields. The first few terms in the expansion are
\beq\label{CHC}
\displaystyle
\frac{g^2 v^2}{4} \left(|W|^2+\frac1{2c_w^2}Z^2\right)\left[2\sqrt{1-\xi}\,\frac{h}{v} +(1-2\xi)\,\frac{h^2}{v^2} -\frac43 \xi\sqrt{1-\xi}\,\frac{h^3}{v^3}+\ldots\right]\,,
\eeq
where we traded the parameters $V$ and $f$ for the physical EWSB scale $v$ and for the parameter
\beq\label{xiset}
\ds
\xi=\frac{v^2}{f^2}=\sin^2\frac{V}{f}\leq1\,.
\eeq
$\xi$ measures how smaller the scale of EWSB scale is with respect to the scale of \mbox{SO$(5)\rightarrow\;$SO$(4)$} breaking or, equivalently, the magnitude of the misalignment angle $\langle\theta\rangle$. The capital importance of the $\xi$ parameter in CH models will become apparent by the discussion that follows. Eq.~(\ref{CHC}) contains single- and double-Higgs vertices similar to those which arise in the SM, but with modified couplings
\beq
k_V\equiv \frac{g_{hVV}^{\textrm{CH}}}{g_{hVV}^{\textrm{SM}}}=\sqrt{1-\xi}\,,\;\;\;\;\;\frac{g_{hhVV}^{\textrm{CH}}}{g_{hhVV}^{\textrm{SM}}}=1-2\xi\,.
\label{hvc}
\eeq
Also, it contains higher-dimensional vertices with more Higgs field insertions which are absent for the SM Higgs. By measuring Higgs couplings and/or (if possible) by searching for these higher-dimensional vertices we can thus test experimentally the possible composite nature of the Higgs boson.

One peculiarity of eq.~(\ref{hvc}) that you might have noticed already is that both formulas approach $1$ in the limit $\xi\rightarrow0$, meaning that both the $hVV$ and the $hhVV$ couplings reduce to the values predicted by the SM in this limit. Moreover the coupling strength of the higher-dimensional vertices in eq.~(\ref{CHC}) are proportional to $\xi$ so that they disappear for $\xi\rightarrow0$ and the same happens to all other interactions of even higher order in the Taylor series. In summary, the complete Lagrangian for the Higgs and the EW boson collapses to the one of the SM for $\xi\rightarrow0$ so that the Composite Higgs becomes effectively indistinguishable from the elementary SM Higgs in this limit. The reason for this is that the $\xi\rightarrow0$ limit is taken at fixed $v$ by sending $f\rightarrow\infty$, and $f$ is related with the typical energy scale of the Composite Sector. For $f\gg{v}$ the CS decouples from the EWSB scale while the Higgs stays light because it is a NGB. The only way in which the theory can account for this large scale separation is by turning itself, spontaneously, into the SM. Of course $\xi$ is not zero, but provided it is sufficiently small this phenomenon explains why the measured couplings of the Higgs boson are close to the SM predictions, which is a priori not trivial at all as discussed in section~\ref{TBI}. The very existence of the parameter $\xi$ and the possibility of adjusting it in order to mimic the SM predictions with arbitrary accuracy marks the essential difference between the modern CH construction and the old idea of Technicolor \cite{Weinberg:1975gm,Weinberg:1979bn,Susskind:1978ms}  (see Ref.~\cite{Lane:2002wv} for a review). Not only in Technicolor, unlike in CH, there is no structural reason to expect the presence of a light Higgs boson. There is not even a reason why this scalar, if accidentally present in the spectrum, should have couplings which are similar to the SM ones. Notice however that taking $\xi$ very small, as we will be obliged to do if the agreement with the SM will survive more precise measurement, does not come for free in CH models. I will come back to this point in the next section.

Let us now turn to the calculation of the Higgs couplings to fermions. In order to proceed we first need to specify the structure of the fermionic part of the interaction that connects the elementary and the composite sector as in fig.~\ref{ecomp}. This is taken to be similar to the gauge part in eq.~(\ref{gaugeint}), namely
\beq\label{ferint}
\ds
{\mathcal{L}}_{\textrm{int}}^{\textrm{fermion}}\sim\lambda\,\overline\psi\,{\mathcal{O}}\,,
\eeq
where $\psi$ is one of the SM fermion fields in the elementary sector, ${\mathcal{O}}$ is a composite sector local operator and $\lambda$ is a free parameter that sets the strength of the interaction. One such operator is present for each of the SM chiral fermions, each with its own coupling strength $\lambda$. Below we will mostly focus on the top quark sector, in which case the relevant SM fields are the $\psi=q_L$ doublet and the $\psi=t_R$ singlet. The similarity with eq.~(\ref{gaugeint}) consists in the fact that $\psi$ is an elementary sector field just like $A_\mu$, which is coupled linearly to an operator ${\mathcal{O}}$ made of composite sector constituents very much like $A_\mu$ couples to the composite sector current operator $J_\mu$. Linear fermion couplings of the type (\ref{ferint}) were first introduced in Ref.~\cite{Kaplan:1991dc} and are said to have the ``Partial Compositeness'' structure for a reason that I will explain in the next section.

An important difference between gauge (\ref{gaugeint}) and fermion (\ref{ferint}) interactions is that in the former case we do know perfectly what the CS operator $J$ is, while in the latter one we have to deal with an operator ${\mathcal{O}}$ of yet unspecified properties. What we know is that ${\mathcal{O}}$ must be a spin $1/2$ fermionic operator in order for equation (\ref{ferint}) to comply with Lorentz invariance and that it must be a triplet of QCD colour to respect the \mbox{SU$(3)_c$} symmetry. This latter property will have important phenomenological implications in that it obliges the CS to carry QCD colour and thus to produce coloured resonances which are easy to produce at the LHC. We also know that ${\mathcal{O}}$ must be in some multiplet of the CS global group ${\mathpzc{G}}$ but we don't know in which one. The only constraint is that the representation in which ${\mathcal{O}}$ lives must contain the SM \mbox{SU$(2)_L\times$U$(1)_Y$} group representation of the corresponding $\psi$ fermion, in order for eq.~(\ref{ferint}) not to break the EW group. Few options (focussing on reasonably small multiplets) exist to solve this constraint and for each option the calculation of Higgs couplings might produce a different result. Unlike those with gauge bosons, Higgs couplings to fermions are thus not uniquely predicted in terms of $\xi$. 

One simple option is to make ${\mathcal{O}}$ be in the ${\mathbf{5}}$, in which case eq.~(\ref{ferint}) becomes
\beq\ds
\label{pcrew}
{\mathcal{L}}_{\textrm{int}}^{\textrm{fermion}}=\lambda_{L}\left(\overline{Q}_{L}\right)^I{\mathcal{O}}_I+\lambda_{R}\left(\overline{T}_R\right)^I{\mathcal{O}}_I\,.
\eeq
The index $I$ runs from $1$ to $5$ and it transforms in the ${\mathbf{5}}$ of ${\mathpzc{G}}={\textrm{SO}}(5)$. The capital $Q$ and $T$ fields are two quintuplets that contain the elementary $q_L=(t_L,b_L)$ and $t_R$ fermions. Explicitly, they are
\beq\ds
{\ve{Q}}_{L}=\frac1{\sqrt{2}}(-i\,b_L,\,-b_L,\,-i\,t_L,\,t_L,\,0)^T\,,\;\;\;\;\;
{\ve{T}}_R=(0,\,0,\,0,\,0,\,t_R)^T\,.
\eeq
Their form is chosen in such a way that $(t_L,b_L)$ and $t_R$ appear precisely in those components of the $Q_{L}$ and $T_R$ quintuplets that display the transformation properties of a ${\mathbf{2}}_{1/6}$ and of a ${\mathbf{1}}_{2/3}$ of the SM \mbox{SU$(2)_L\times$U$(1)_Y$} subgroup. In short, the form of the embeddings is fixed by the requirement that eq.~(\ref{pcrew}) must respect the SM gauge symmetry.\footnote{I'm being quite sloppy here. In order to make the thing work one needs to enlarge the global group of the CS promoting it to ${\mathpzc{G}}={\textrm{SO}}(5)\times{\textrm{U}}(1)_X$ and to change the definition of the SM Hypercharge into $Y=T_R^3+X$, with $X$ the charge under the newly introduced ${\textrm{U}}(1)_X$ group. It is only by giving an $X$ charge of $2/3$ to ${\mathcal{O}}$ and to $Q_L$ and $T_R$ that one finds a ${\mathbf{2}}_{1/6}$ and of a ${\mathbf{1}}_{2/3}$ in the decomposition and eq.~(\ref{pcrew}) truly complies with gauge invariance.\label{u1x}}

\begin{table}
\centering
\begin{tabular}{| c  | c | c | }
\hline
\ & Top & Bottom   \\
\hline
\ & \ & \  \\[-10pt]
${\mathbf{5\oplus5}}$ & $\ds k_t=\frac{1-2\,\xi}{\sqrt{1-\xi}}\;\;\;\;\;c_2=-2\xi$   & $\ds k_b=\frac{1-2\,\xi}{\sqrt{1-\xi}}$ \\[8pt] \hline
\ & \ & \  \\[-10pt]
${\mathbf{4\oplus4}}$ & $\ds k_t=\sqrt{1-\xi}\;\;\;\;\;c_2=-\frac{\xi}2$   & $k_b={\sqrt{1-\xi}}$ \\[8pt] \hline
\ & \ & \  \\[-10pt]
${\mathbf{14\oplus1}}$ & $\ds k_t=\frac{1-2\,\xi}{\sqrt{1-\xi}}\;\;\;\;\;c_2=-2\xi$   & $\ds k_b=\frac{1-2\,\xi}{\sqrt{1-\xi}}$ \\[8pt] \hline
\end{tabular}
\caption{Kappa factor and anomalous $c_2$ coupling predictions in the top and bottom quark sector for different choices of the fermionic operators representations under the \mbox{SO$(5)$} group.
\label{tabC}}
\end{table}

Once the representation is chosen, Higgs couplings are determined by symmetries. There is indeed a unique ${\mathpzc{G}}$-invariant operator we can form with $\ve\Phi$ (i.e., the Higgs), the embeddings and no derivatives. Furthermore the coefficient of this operator is fixed by the fact that the correct top mass must be reproduced when the Higgs is set to its VEV. The operator is
\bea
\label{yukup}
\displaystyle
{\mathcal{L}}_{\textrm{Yukawa}}^{t}&&=-\frac{\sqrt{2}m_t}{\sqrt{\xi(1-\xi)}}\Phi_I\overline{{Q}}_{L}^I{T}_R=-\frac{m_t}{2} \frac{1}{\sqrt{\xi(1-\xi)}}\sin\frac{2(V+h)}{f}\overline{t} t \nonumber\\
&&=-m_t \overline{t} t - k_t \frac{m_t}{v}  h \,  \overline{t} t - c_2 \frac{ m_t}{v^2}h^2  \overline{t} t+\ldots\,.
\eea
It produces the top quark mass plus, after Taylor-expanding, a set of interactions of the physical Higgs with $t{\overline{t}}$. The first interaction is an $h$-$t{\overline{t}}$ vertex like the one we have in the SM. The second one is an exotic $hh$-$t{\overline{t}}$ coupling which is absent in the SM and could be tested in the double-Higgs production process \cite{Grober:2010yv,Contino:2012xk}. The modified single-Higgs coupling and the double-Higgs vertex read
\beq
k_t^{\mathbf{5}}\equiv \frac{g_{htt}^{\textrm{comp}}}{g_{htt}^{\textrm{SM}}}=\frac{1-2\,\xi}{\sqrt{1-\xi}}\,,\;\;\;\;\;c_2^{\mathbf{5}}=-2\xi\,,
\label{htc5}
\eeq
where the $^{\mathbf{5}}$ superscript reminds us that the prediction depends on the choice of the representation (the ${\mathbf{5}}$) for the fermionic operator ${\mathcal{O}}$. 

One proceeds in exactly the same way to generate the mass and the Yukawa coupling for the bottom quark, obtaining the bottom coupling modification $k_b$ and an anomalous $hh$-$b{\overline{b}}$ vertex, which however is weighted by the bottom mass and thus it is too small to be phenomenologically relevant. Also for the bottom, the ${\mathbf{5}}$ could be a valid representation for the corresponding ${\mathcal{O}}$ operator. Other choices like the ${\mathbf{4}}$ could be considered both for the bottom and for the top, with the results reported in table~\ref{tabC}. In the table, the notation ``${\mathbf{5\oplus5}}$'' means that the fermionic operators that couples to the left-handed doublet $q_L$ and the one that couples to the right-handed singlet ($t_R$ or $b_R$) are in the same representation, i.e. the ${\mathbf{5}}$, while their are both in the ${\mathbf{4}}$ in the ``${\mathbf{4\oplus4}}$'' case. However the two representations might be different, in spite of the fact that a single name was given for shortness to the two ${\mathcal{O}}$ operators in eq.~(\ref{pcrew}). A reasonable option is to take the doublet mixed with a ${\mathbf{14}}$ and the singlet mixed with a singlet operator. This is denoted as the ``${\mathbf{14\oplus1}}$''  case in the table. Up to caveats which is not worth discussing here, table~\ref{tabC} exhausts what are considered to be the ``most reasonable'' options for the fermionic operator representations and the corresponding predictions of Higgs couplings. Other patterns which could be worth studying are in Appendix~B of Ref.~\cite{Pomarol:2012qf}.

\subsection{Composite Higgs Signatures}

Now that the basic structure of the CH scenario has been introduced, I can start illustrating its phenomenology. Additional structural aspects that were left out from the previous discussion will be introduced when needed. The signatures of CH that have been searched for at the $8$~TeV LHC run (run-$1$) and we will keep studying at run-$2$ and possibly at future colliders are Higgs couplings modifications, vector resonances and top partners.

\subsubsection*{Higgs Couplings Modifications}

The current status of our field is that we are not sure of which kind of new physics we are looking for. This is much different from what it used to be the case when the Higgs still had to be discovered. In searching for the Higgs one could rely on one single full-fledged model (the SM) with only one at that time unknown parameter (the Higgs mass). Searching for the Higgs boson was basically equivalent to searching for the SM theory, which was capable to provide detailed and specific predictions for the expected signal to be searched for in the data. We are not anymore in this situation. Even if we focus on one given BSM hypothesis (CH, in the present case, but the same applies to SUSY, WIMP DM or whatever else), this hypothesis is not at all equivalent to a single specific model. This is why in BSM searches so much importance is given to model-independence. Namely to the fact that we should not organise our efforts around specific signatures of specific benchmark models, but rather on generic model-independent features of the scenario we aim to investigate, ideally on those features that are unmistakably present in all the models that provide specific realisations of the generic scenario.

\begin{figure}[t]
\centering
\includegraphics[width=.6\textwidth]{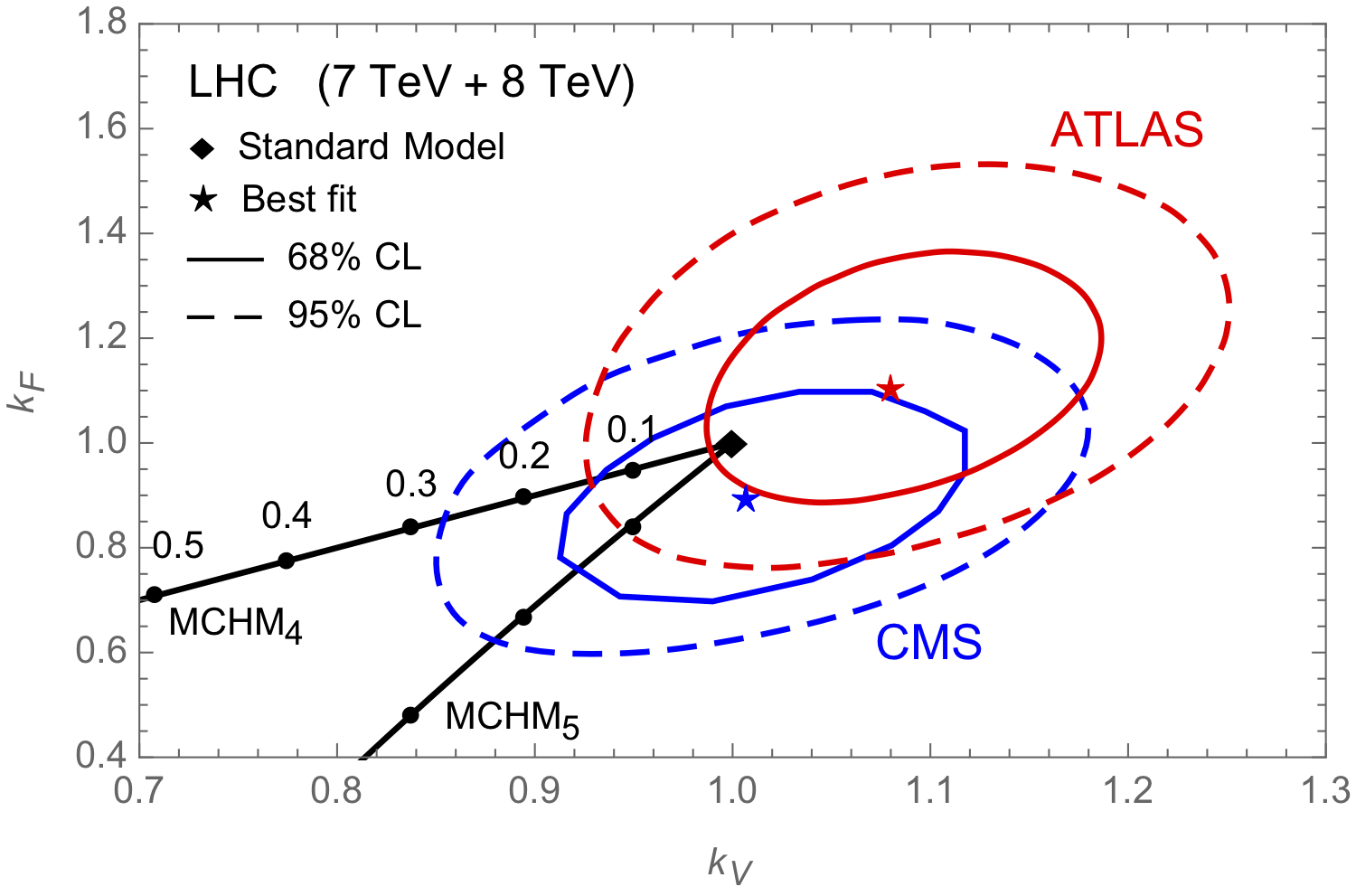}
\caption{Fit of the Higgs coupling strength to the gauge bosons ($k_V$) and fermions ($k_F$) obtained by the
ATLAS (red contours) and CMS collaborations (blue contours) from the combination of the $7$ and $8\ \mathrm{TeV}$ LHC data.
Solid black lines show the CH predictions, depending on the fermionic operators representation, at different values of $\xi$.}
\label{fig:kvkf}
\end{figure}

Model-independence is the first reason to be interested in coupling modifications in CH, given that we saw in the previous section how Higgs couplings can be universally predicted as a function of $\xi$. This prediction is independent of the detailed dynamics of the Composite Sector resonances, for which many different explicit models (with plenty of free parameters) can be written down (see e.g.~\cite{Agashe:2004rs,Panico:2011pw}). The Higgs couplings predictions in all these models are always (up to small corrections) those in eq.~(\ref{hvc}) and in table~\ref{tabC}. Higgs couplings have been measured at the LHC run-$1$ both by  ATLAS \cite{Aad:2015gba} and CMS \cite{Khachatryan:2014jba}, with the result reported in fig.~\ref{fig:kvkf} in the \mbox{$k_V$-$k_F$} plane. $k_F$ is a common rescaling factor for the SM coupling to fermions, therefore the plot assumes $k_t=k_b=k_F$. The CH predictions are also reported on the plot for different values of $\xi$. The curve labeled ``\mbox{MCHM$_{4}$}'' follows the trajectory in the second line of table~\ref{tabC}, while the ``\mbox{MCHM$_{5}$}'' one represents the first and the third lines. The resulting limit quoted by ATLAS in Ref.~\cite{Aad:2015pla} is $\xi<0.12$ in the  \mbox{MCHM$_{4}$} and $\xi<0.10$ in the  \mbox{MCHM$_{5}$} at $95\%$ CL. ATLAS limit is stronger than the CMS one because the ATLAS central value is slightly away from the SM in the opposite direction than the one predicted by CH. The resulting limit is thus stronger than the expected one. Because of this stringent bound, it is unlikely that much progress will be made with the next runs of the LHC, given that the expected limit with the full luminosity of \mbox{$300$\,fb$^{-1}$} is of around $\xi<0.1$ \cite{CMS-NOTE-2012/006,ATL-PHYS-PUB-2013-014,Dawson:2013bba}, very close to the present one. Of course if the  central value will not sit on the SM the limit could improve, but we can definitely exclude the occurrence of the discovery of a non-vanishing $\xi$.

We saw that ATLAS and CMS are doing a rather good job in studying Higgs couplings modifications due to compositeness. The study is however not fully complete, and it could be generalised in three directions. First, one can easily construct models where $\kappa_t\neq{k}_b$. It is sufficient for instance to place the fermionic operators associated with the top quark in the $\mathbf{5}$ representation while assigning those for the bottom to a $\mathbf{4}$. In this case $k_t$ will follow the prediction in the first line of table~\ref{tabC}, while $k_b$ will follow the second line. Studying this case is straightforward even if it requires going beyond the \mbox{$k_V$-$k_F$} plane. No much improvement is however expected in the compatibility of the model since $k_V$ is still the one in eq.~(\ref{hvc}) and the ATLAS preference for $k_V>1$, independently of the fermion couplings, is already sufficient to produce a limit on $\xi$ not much above $0.1$. A second direction of improvement is to study not only the modification of the Higgs vertices that exist already in the SM, but also anomalous couplings such as $hh$-$t{\overline{t}}$ in eq.~(\ref{yukup}). The latter might be visible in double-Higgs production when enough luminosity will be collected. However existing studies (see e.g.~\cite{Azatov:2015oxa}) suggest that even with the high-luminosity stage of the LHC (HL-LHC) it might be hard to reach a competitive accuracy. A third direction of improvement would be to generalise the analysis to non-minimal cosets, namely to go beyond the minimal \mbox{SO$(5)/$SO$(4)$} example we discussed here. The problem is that non-minimal cosets produce an extended Higgs sector and thus the modification of the Higgs couplings emerge from the pile-up of two effects. One has the modifications due to compositeness, which are analogous to those in eq.~(\ref{hvc}) and table~\ref{tabC}, plus further modifications due to the mixing of the Higgs boson with extra light scalar states. The former effect is easy to compute, while the latter one is hard to parametrise with a sufficient degree of generality as it depends on the properties of the extra scalars that mix with the Higgs. Furthermore, all this should be studied in correlation with the direct searches for extra scalars. A detailed phenomenological analysis of extended cosets is missing in the literature, in spite of the fact that extended cosets are not at all implausible from the view-point of model-building. The original CH model \cite{Kaplan:1983fs}, for instance, was based on an \mbox{SU$(5)/$SO$(5)$} coset, which delivers one complex and one real scalar triplet, plus one singlet, on top of the ordinary Higgs doublet. 

The second reason to be interested in Higgs couplings modification is the (almost) direct connection between the parameter $\xi$, which couplings measurements are capable to probe, and the level of fine-tuning $\Delta$ of the theory. We discussed in the previous section that for $\xi\rightarrow0$ CH models reduce to the SM, which is an eminently Un-Natural theory. It is thus expected that taking $\xi$ small might be dangerous in terms of fine-tuning. In order to illustrate how this works, let us write down the structure of the Higgs potential, as it emerges in a certain class of models and under certain approximations.\footnote{The one that follows is an approximate formula for the Higgs potential in models where the fermionic operators in the top quark sector are in the ${\mathbf{5\oplus5}}$ or in the ${\mathbf{14\oplus1}}$ configurations. The connection between the Higgs potential and the top quark sector will be explained later. Further details can be found in Chapter~3 of Ref.~\cite{Panico:2015jxa}.} It reads
\beq\label{hpot}
\ds
V[H]\simeq-\alpha f^2\sin^2\frac{H}{f}+\beta f^2\sin^4\frac{H}{f}\,,
\eeq
where the coefficients $\alpha$ and $\beta$ can be computed within explicit models (see ...) and depend on some of their free parameters. By adjusting the free parameters one can set $\alpha$ and $\beta$ in such a way that the VEV $V$ of the Higgs field (i.e., the minimum of the potential) produces our favorite value of $\xi$ through eq.~(\ref{xiset}) and also to reproduce the observed Higgs boson mass. These two constraints read, respectively
\bea\ds\label{VEVM}
\alpha&=&2\,\xi\,\beta\,,\nonumber\\
m_H^2&=&8\,\xi(1-\xi)\,\beta\,.
\eea
Both conditions might cost fine-tuning, let us however momentarily focus only on the first one. It tells us that the ``expected'' value of $\xi$ is proportional to $(\alpha)_{\textrm{expected}}/(\beta)_{\textrm{expected}}$, where by ``expected'' I mean the size of the $\alpha$ and $\beta$ coefficients that are generically encountered in the parameter space of the model. In all existing CH models, the expected magnitudes of $\alpha$ and $\beta$ either are comparable, or $\alpha$ is larger than $\beta$, making that having $\xi\ll1$ is never an expected structural feature of the model. In this situation, enforcing $\xi\ll1$ requires fine-tuning. Namely, a cancellation must take place in the prediction for $\alpha$, obtained by finely adjusting the parameters of the underlying model. This tuning is at least of order
\beq\ds\label{xitun}
\Delta=\frac{(\alpha/\beta)_{\textrm{expected}}}{\alpha/\beta}\geq\frac1{2\xi}\,.
\eeq
The above equation displays the anticipated connection between $\xi$ and the level of Un-Naturalness of the theory. The current bound $\xi<0.1$ corresponds to a not fully Natural (but still acceptably so) theory with a level of tuning $\Delta>5$.

Actually, we are not sure of the connection between $\xi$ and $\Delta$ in a fully model-independent way. In principle, it would be sufficient to find a model where $\alpha$ is structurally smaller than $\beta$ in order to avoid the tuning in the Higgs VEV and to have $\xi$ Naturally small. The problem, as mentioned above, is that no such model currently exists, but this does not mean that one could not be invented in the future. Engineer a Naturally small $\xi$ is the purpose of the Little Higgs constructions \cite{Schmaltz:2005ky,Perelstein:2005ka}, however as of now I'm not aware of any convincing and realistic model of this class.

\subsubsection*{Vector Resonances}

Searching for modified couplings of the Higgs boson is not the only way to test Higgs compositeness experimentally. Direct searches for new particles also play an important role, which will become the leading role at the LHC run-$2$ thanks to the improved collider energy. The new particles to be searched for are the resonances that emerge, together with the Higgs, from the Composite Sector of the theory (see fig.~\ref{ecomp}). Resonances at a scale $m_*\sim$~TeV are unmistakably present in CH, they are the ``hadrons'' of the new strong force we are obliged to postulate if we want the Higgs to be a composite object. If we are lucky and the CH scenario is realised in Nature, plenty of such resonances exist and a sort of new ``Subatomic Zoo'' is waiting to be discovered at the TeV scale.

Predicting the quantum numbers and the properties of the CS resonances is not completely straightforward. However a valid rule of thumb is that resonances are associated with the operators of the CS. Namely, for each resonance it should be possible to identify at least one CS operator that is capable to excite it from the vacuum. The first set of operators we encountered are the global currents ``$J$'' in eq.~(\ref{gaugeint}), associated to a set of resonances ``$\rho$" through the equation
\beq\label{op_par}
\ds
\langle\rho|J|0\rangle\neq0\,.
\eeq
The currents are bosonic operators that transform as vectors of the Lorentz group, therefore we expect $\rho$ to be a spin-$1$ vector particle in order for eq.~(\ref{op_par}) to comply with Lorentz symmetry.\footnote{$\rho$ cannot have spin greater than $1$ because a Lorentz vector operator cannot have a non-vanishing matrix element between the vacuum and a high-spin particle. Massless scalars can instead be excited from the vacuum by a conserved current if it is associated with a spontaneously broken generator. These scalars are nothing but the NGB's of the theory we already discussed extensively.} The analogous hadrons in QCD are the $\rho$ mesons, the $\omega$ and the $a_1$, each associated with one of the global currents of the chiral group. Eq.~(\ref{op_par}) also tells us the quantum numbers of $\rho$ under the SM group. If for instance ${\mathpzc{G}}={\textrm{SO}}(5)$, the global current $J$ is in the Adjoint $\mathbf{10}$ representation of the group, which decomposes in a $\mathbf{3}_0$, plus a $\mathbf{1}_0$, plus a $\mathbf{1}_1$ and a $\mathbf{2}_{1/2}$ of the SM \mbox{SU$(2)_L\times$U$(1)_Y$} subgroup (i.e., a $\mathbf{(3,1)\oplus{{(1,3)}}\oplus{(2,2)}}$ of \mbox{SO$(4)$}). $\rho$ particles in all these representations are thus expected, plus one further $\mathbf{1}_0$ because ${\mathpzc{G}}={\textrm{SO}}(5)$ actually needs to be enlarged to ${\textrm{SO}}(5)\times{\textrm{U}}(1)_X$ (see Footnote~\ref{u1x}) in order to incorporate SM fermion masses into the theory. The existence of vectors with these quantum numbers is confirmed by explicit models. A first study of their phenomenology in the context of holographic realisations of the CH scenario was performed in Ref.s~\cite{Agashe:2007ki,Agashe:2008jb,Agashe:2009bb}. Other interesting particles of this class are coloured spin-$1$ vectors, the so-called ``KK-gluons'' \cite{Agashe:2006hk}. KK-gluons emerge because the CS (see section~\ref{MCHC}) needs to carry QCD colour and thus it contains an extra \mbox{SU$(3)_c$} group of symmetry on top of the ``electroweak'' ${\textrm{SO}}(5)\times{\textrm{U}}(1)_X$ factors. This produces extra global current operators and their corresponding particles in the octet of the QCD group.

All particles above are worth searching for, however here I will focus, for definiteness, on vector resonances in the $\mathbf{3}_0$ triplet, the so-called Heavy Vector Triplet (HVT) \cite{Pappadopulo:2014qza}. The reason for this choice is that HVT's display a quite simple phenomenology, still varied enough and promising in terms of mass-reach. Furthermore, the $\mathbf{3}_0$ vectors are associated with the global currents of the SM \mbox{SU$(2)_L$} subgroup of the CS symmetry group. The existence of such subgroup is absolutely unavoidable in CH models, independently of whether or not we stick to the minimal coset or even of whether the Higgs is a pNGB or not. HVT's thus unmistakably emerge in all models where a strong dynamics is involved in the mechanism responsible for EWSB. This includes old-fashioned Technicolor, in which these particles are also present and are known as ``techni-rho'' mesons.

Characterising the HVT phenomenology requires a little digression on how we do expect, in general, Composite Sector particles to be coupled among themselves and with the gauge and fermionic fields in the Elementary Sector. This expectation can be encapsulated (see Ref.~\cite{Giudice:2007fh} and Ch.~3 of \cite{Panico:2015jxa}) in a ``power-counting rule'', namely a formula that tells us the expected size of the interaction vertices or, which is the same, of the interaction operators in the Lagrangian. The rule is based on the idea that the CS is characterised by one typical mass scale $m_*$ (the confinement scale) and by one typical coupling strength parameter ``$g_*$''. It is thus said to be a ``$1$ Scale $1$ Coupling'' (\mbox{$1$S$1$C}) power-counting. The parameter $g_*$ represent the typical magnitude of the interaction vertices involving CS particles, among which the Higgs. It can thus be expressed in terms of the Higgs decay constant $f$ and defined as
\beq\label{gsf}
g_*=\frac{m_*}{f}\,.
\eeq
The coupling $g_*$ can easily be very large, close to the absolute maximal value $g_*\sim4\pi$ a coupling strength parameter can assume. It is for instance very large in real-world QCD, where it can be identified with the $\rho$ meson coupling $g_\rho\simeq6$. It can however be smaller if the underlying strongly-interacting theory is characterised by a large number of colours $N_c$. For instance, $g_*\sim4\pi/\sqrt{N_c}\rightarrow0$ in the large-$N_c$ limit of QCD. We are thus entitled to consider values of $g_*$ anywhere from $0$ to $4\pi$, however basic phenomenological consistency of the CH scenario requires it to be above around $y_t\simeq1$. Therefore in what follows we will take $g_*\in[1,4\pi]$.

On top of $g_*$, the other couplings that are present in the theory are the SM gauge couplings ``$g$'' in eq.~(\ref{gaugeint}) and the fermionic interactions ``$\lambda$'' in eq.~(\ref{pcrew}). They control the strength of those interactions of the Elementary Sector fields (gauge and fermions, respectively) that are generated by the CS dynamics, such as for instance their couplings with the Higgs and with the CS resonances. The complete power-counting formula, which takes care both of CS particles self-couplings and of Elementary/Composite interactions, reads
\beq\label{1s1c}
\ds
{\mathcal{L}}=\frac{m_*^4}{g_*^2}{\widehat{\mathcal{L}}}\left[
\frac{\partial}{m_*},\frac{g_*H}{{m_*}},\frac{g_*\sigma}{m_*},\frac{g_*\Psi}{m_*^{3/2}},\frac{g\cdot A_\mu}{m_*},\frac{\lambda\cdot\psi}{m_*^{3/2}}
\right]\,,
\eeq
where ${\widehat{\mathcal{L}}}$ is a dimensionless polynomial function with order one coefficients. In the equation, $\sigma$ represents a bosonic CS resonance, such as a spin-$1$ particle like the $\rho$'s we aim to study, while $\Psi$ denotes a fermionic resonance such as the Top Partners we will discuss in the next section. The different power of $m_*$ in the denominator simply follows from the different energy dimensionality ($1$ and $3/2$) of bosonic and and fermionic fields. The fields $A_\mu$ and $\psi$ collectively denote the ES sector gauge and fermions, each entering in the power-counting formula with its own ``$g$'' and ``$\lambda$'' coupling. For instance $A_\mu=W_\mu^\alpha$ couples through the weak coupling $g$ while the QCD gluons, $A_\mu=G_\mu^a$, couples through the strong coupling $g_S$. Similarly the third family $q_L$ doublet couples through the $\lambda_L$ parameter in eq.~(\ref{pcrew}) and $t_R$ couples with strength $\lambda_R$. Notice that light generation quarks and leptons couple with their own strengths, which are typically much smaller than $\lambda_L$ and $\lambda_R$ because their role in the theory is to generate the light fermions Yukawa's rather than the large top Yukawa coupling. An estimate of light generation couplings is postponed to the next section, since they will turn out to be very small we are entitled to neglect them in what follows.

Let us now turn to HVT phenomenology. Since $g_*$ is the largest coupling in the theory, the strongest vertices of $\rho$ are those that only involve CS particles and no ES degrees of freedom. Among those we have a coupling with the Higgs field
\beq\label{C_H}
\ds
g_*c_H\rho_\mu^ai\,H^\dagger\tau^a{\overset{\text{\scriptsize$\leftrightarrow$}}{D}}^\mu H\,,
\eeq
where $\rho_\mu^{a=1,2,3}$ denotes the components of the triplet, $\tau^a=\sigma^a/2$ are the ${\textrm{SU}}(2)_L$ generators and the double arrow denotes the covariant derivative acting on the right minus the one acting on the left. The coefficient of the operator has been estimated with eq.~(\ref{1s1c}) up to an unknown order one parameter $c_H$. The one in eq.~(\ref{C_H}) is the unique gauge-invariant operator involving the $\rho$ and two Higgs fields that cannot be eliminated by the equations of motions. It produces couplings of $\rho$ with all the four real components of the Higgs doublet which correspond to the physical Higgs boson plus the three longitudinal polarisation components of the SM $W^\pm$ and $Z$ massive vector bosons.\footnote{\label{ET}The correspondence between longitudinally polarised vector bosons and the so-called ``unphysical'' components of the Higgs field (i.e., the charged $h_u$ and the imaginary part of the neutral $h_d$ component of the doublet) is ensured by the Equivalence Theorem \cite{Horejsi:1995jj}. It holds at energies much above the vector boson masses, which is an excellent approximation for our purposes. In practice the theorem says that the Feynman amplitudes with longitudinal vector bosons on the external legs can be equivalently be computed as the amplitude for the corresponding scalar fields.} The operator thus mediates the decay of $\rho$ to different combinations of vector bosons and Higgs final states, with decay widths 
\beq\label{bdec}
\ds
\Gamma_{\rho_0\rightarrow W^+W^-}\simeq\Gamma_{\rho_0\rightarrow Zh}\simeq\Gamma_{\rho_\pm\rightarrow W^\pm Z}\simeq\Gamma_{\rho_\pm\rightarrow W^\pm h}\simeq\frac{g_*^2 c_H^2 m_\rho}{192\pi}\,.
\eeq
With obvious notation, $\rho_0$ and $\rho_\pm$ respectively denote the electrically neutral and charged $\rho$'s, obtained as linear combinations of the $\rho^a$ triplet components. Neutral and charged resonances are approximately degenerate in mass because of the ${\textrm{SU}}(2)_L$ symmetry. Their common mass is denoted as $m_\rho$.

The second term to be considered is the one responsible for the interaction of $\rho$ with light quarks and leptons. Notice that such an interaction cannot occur directly with an operator involving light elementary fermionic fields because we argued above that the insertion of such fields in the Lagrangian (\ref{1s1c}) costs very small $\lambda$'s that make the resulting vertices negligible. However what we can do is to write, compatibly with gauge invariance, an operator that mixes $\rho$ with the elementary $W$ boson field. Since the $W$ couples to quarks and leptons just like in the SM, this $\rho$-$W$ mixing eventually generates the interaction we are looking for. In accordance with the power-counting (\ref{1s1c}), the mixing and the resulting interaction reads
\beq\label{C_F}
\ds
\frac{g}{g_*} c_F W_{\mu\nu}^a D^\mu \rho^\nu_a\,,\;\;{\Longrightarrow}\;\;\; \frac{g^2}{g_*} c_F \rho^\mu_a J_\mu^a\,,
\eeq
with $c_F$ an unknown order one parameter. In the equation, $J_\mu^a=\overline{f}_L\gamma_\mu\tau^af_L$ denotes the ordinary ${\textrm{SU}}(2)_L$ current, namely the one to which $W_\mu^a$ couples in the SM. Since the interaction emerges from the mixing with the $W$, this is precisely the structure we should have expected for the $\rho$ coupling. The scaling of the coefficient is also easily understood. The power-counting formula predicts a $g/g_*$ for the $\rho$ mixing with $W$, while the $W$ coupling with fermions gives an extra power of $g$. The result is the rather peculiar $g^2/g_*$ factor, which makes that the $\rho$ coupling with fermions decreases when $g_*$ increases and the CS becomes more and more strongly coupled. The opposite behaviour is observed for the coupling to bosons in eq.~(\ref{C_H}). The translation between the mixing and the interaction operator reported in eq.~(\ref{C_F}) is obtained by performing by a field redefinition, namely by shifting the $W$ field by an amount proportional to $\rho$ in such a way that the mixing cancels and the interaction is generated. However this technicality should not obscure the fact that the coupling physically emerges from the mixing with the $W$. 

The mixing in eq.~(\ref{C_F}) is responsible for $\rho$ decays to quarks and to leptons. Leptonic decays are particularly important because searches in $l^+l^-$ and $l\nu$ final states (with $l=e,\mu$) are extremely sensitive to the presence of resonances. These decays are controlled by one parameter only, $c_F$, therefore the processes $\rho_\pm\to l^\pm\nu$ and $\rho_0\to l^+l^-$ (and the decays to quarks as well) are universally related very much like we saw for the bosonic channels in eq.~(\ref{bdec}). The widths are
\beq\label{ldec}
\ds
\Gamma_{\rho_\pm\rightarrow l^\pm\nu}\simeq2\,\Gamma_{\rho_0\rightarrow l^+l^-}\simeq\left(\frac{g^2 c_F}{g_*}\right)^2\frac{m_\rho}{48\pi}\,.
\eeq
Notice the presence of $g_*^2$ in the denominator. Together with the $g_*^2$ factor in the numerator of the bosonic decay widths (\ref{bdec}), it makes the relative branching fraction between leptons and bosons scales like $1/g_*^4$, which is a strong suppression in the large $g_*$ limit. In this limit, leptonic final states are suppressed and the $\rho$ is better seen in diboson channels in spite of the fact that the reach in terms of cross-section is much better for the leptonic than for the diboson searches. Eq.~(\ref{C_F}) is also responsible for $\rho$ Drell-Yan production from a quark anti-quark pair.\footnote{The $\rho$ can also be produced in vector boson fusion (VBF) through the $c_H$ operator (\ref{C_H}), however the VBF rate is too small to be relevant, at least at the current stage of the LHC.}  The relative magnitude of the $\rho_\pm$ and $\rho_0$ couplings to quarks are fixed and thus the $\rho_\pm$ and $\rho_0$ relative production rate is entirely determined by the parton luminosities. For $m_\rho\sim$~TeV, $\sigma(\rho_\pm)\simeq2\,\sigma(\rho_0)$ at the LHC. The absolute normalisation of the cross section is of course also easily computed, depending on the parameter $c_F$. Together with the partial widths (\ref{bdec}) and (\ref{ldec}) (plus the analogous formula for the decay to quarks), and assuming that no other decay channel is present, cross sections times branching ratios can be computed for all the channels of interest in terms of two free parameters only, $c_H$ and $c_F$. \footnote{This is not necessarily accurate for the channels involving third family quarks. The large $\lambda$ coupling of the third family produced extra contribution to the vertex that can easily overcome the one from the mixing in eq.~(\ref{C_F}). This enhances $\rho_0\rightarrow{tt}$ and $\rho_\pm\rightarrow{tb}$ making them promising search channels \cite{Liu:2015hxi}. Composite HVT's might also dominantly decay to other composite sector particles like the fermionic top partners \cite{Bini:2011zb}, if kinematically allowed. These decays can also be searched for.} Or better, if not willing to assume a fixed $g_*$, in terms of the parameter combinations $c_Hg_*$ and $c_Fg^2/g_*$, which are those that appear in the vertices. This provides a synthetic approximate description of the HVT phenomenology  which allows for a comprehensive experimental investigation of the HVT signal \cite{Pappadopulo:2014qza}. Notice that the production rate scales like $1/g_*^2$, again due to the $1/g_*$ suppression of the vertex (\ref{C_F}), and the HVT's become more and more elusive in the strong coupling regime.

\begin{figure}[t]
  \centering
  \includegraphics[width=1\textwidth]{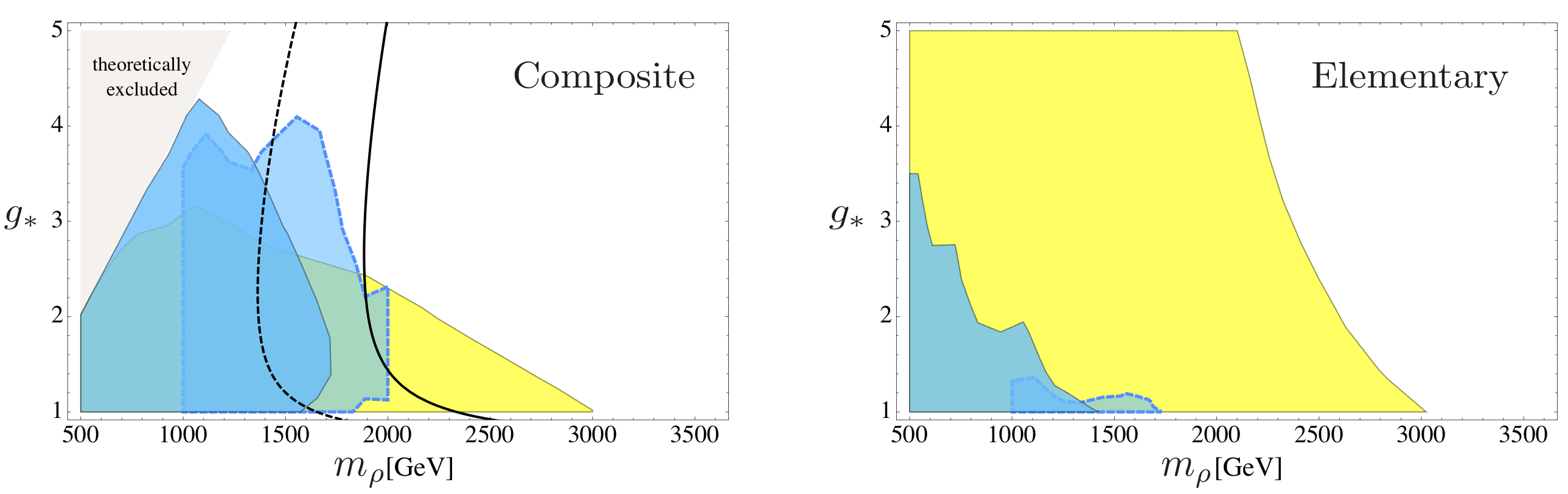} 
\caption{\label{HVT1}Run-$1$ limits on HVT's from leptonic (yellow) and bosonic (blue) searches. HVT's of the ``composite'' type, namely with properties that comply with the expectations of the CH scenario, are shown on the left panel. The case of an ``elementary'' model, namely the $W'$ of Ref.~\cite{Barger:1980ix} is displayed on the right. Black curves are limits from EWPT. See \cite{Pappadopulo:2014qza} for details.} 
\end{figure}

\begin{figure}[t]
  \centering
  \includegraphics[width=0.35\textwidth]{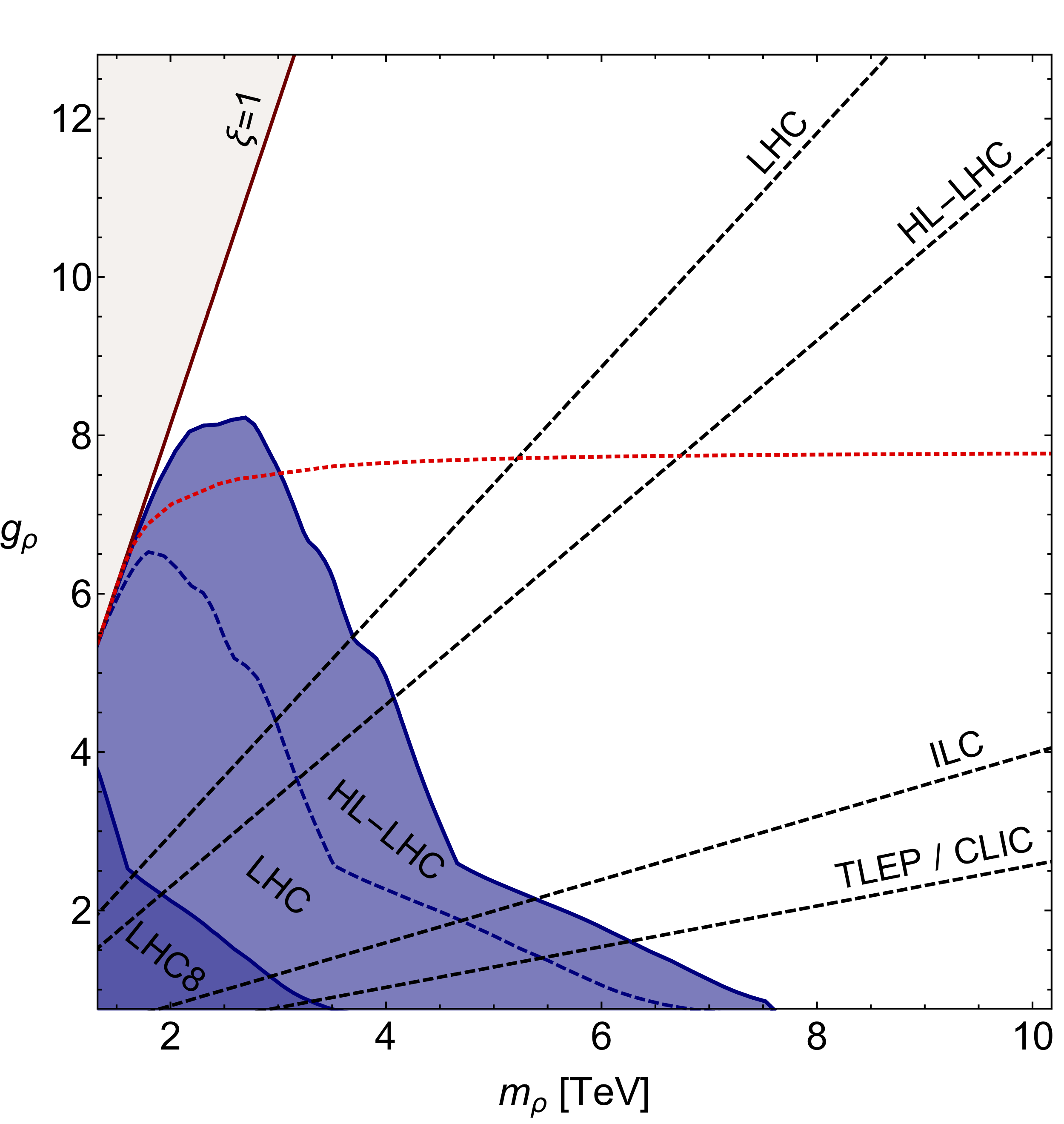} \hspace{40pt}
  \includegraphics[width=0.35\textwidth]{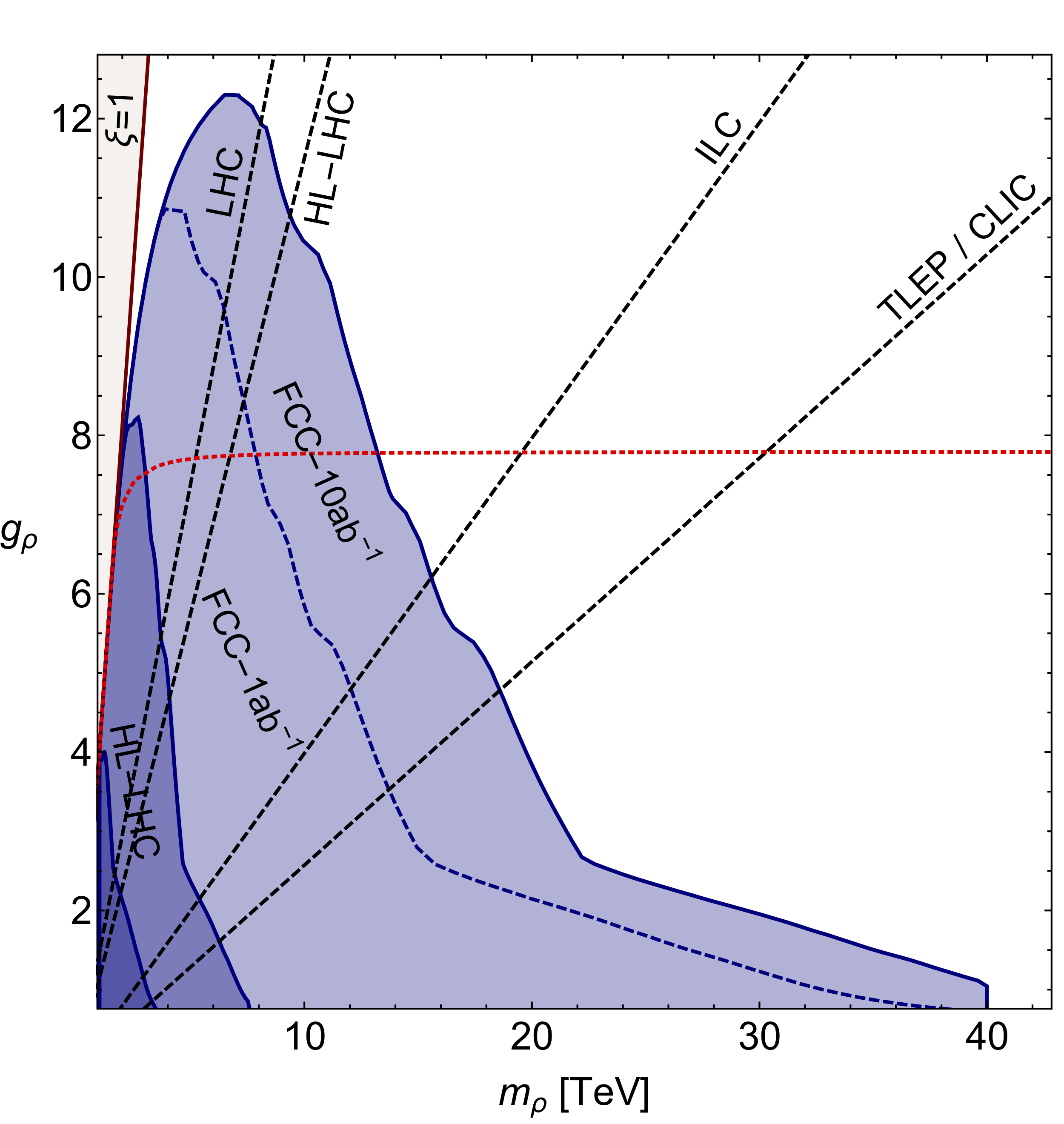} 
\caption{\label{HVT2}Expected exclusion limits on composite HVT's compared with indirect constraints from Higgs coupling measurements. From Ref.~\cite{Thamm:2015zwa}.} 
\end{figure}

The left panel of figure~\ref{HVT1} gives an idea of current limits on HVT from the negative searches performed at the LHC run-$1$. The figure assumes $c_H=c_F=1$, which leaves $g_*$ as the only free parameter. The bound is thus simply reported as an excluded region in the mass versus coupling plane. The yellow region is excluded by resonance searches in leptonic final states (specifically, $l\nu$) while two diboson searches are reported in blue (see \cite{Pappadopulo:2014qza} for details). The behaviour is the expected one. Namely, the mass reach deteriorates at large $g_*$ because of the suppression of the production rate and the one in the leptonic channel deteriorates much faster than the diboson ones because of the suppression of the leptonic branching ratio. Diboson searches thus become competitive and overcome the leptonic sensitivity for $g_*\gtrsim3$. This behaviour is peculiar of HVT's with a composite origin, as apparent from the right panel of the figure where the bounds are shown for an ``elementary'' HVT such as those encountered in $W'$ models. Elementary HVT's are massive vector bosons emerging from an underlying gauge theory, therefore all their couplings emerge as gauge interactions and thus there is no way in which the coupling to vector bosons can scale differently with $g_*$ than the one to fermions. The branching ratios to leptons and bosons thus remain comparable even at large $g_*$ and the diboson channels never win in terms of mass-reach. Overall, we see that current limits are rather poor in the composite case. Resonance as low as $2$ or $3$~TeV, perfectly compatible with Naturalness and with EWPT limits (reported in black in the figure), are still allowed for a reasonable $g_*$ of order $3$. A priori $g_*$ could be even larger than that, making composite HVT's virtually invisible, however a moderate value is suggested by other kind of considerations. The left panel of figure~\ref{HVT2} shows how much the next runs of the LHC could improve the limits, both in the high mass and in the high coupling directions. The plot is based on an approximate extrapolation of current bounds to the $14$~TeV LHC \cite{Thamm:2015zwa} and assumes a total luminosity of \mbox{$300$~fb$^{-1}$}. The HL-LHC reach, with \mbox{$3$~ab$^{-1}$}, is also reported, and the exercise is repeated in the right panel of the figure for an hypothetical future $100$~TeV collider. The dashed straight lines in the plot represent indirect limits from the Higgs coupling measurements described in the previous section. The logic is that the resonance mass $m_\rho$ is expectedly comparable with the CS confinement scale $m_*$. If we take them exactly equal we can use eq.~(\ref{gsf}) to compute $f$, and in turn $\xi$ (\ref{xiset}), on the $(m_\rho,g_*)$ plane. Lines are shown for $\xi=0.1$, $\xi=0.08$, $\xi=0.01$ and $\xi=0.004$, corresponding to the reach of the LHC, of the HL-LHC, of ILC and TLEP/CLIC future colliders (see references in \cite{Thamm:2015zwa}). This shows the complementarity of direct and indirect searches of the Composite Higgs scenario.

\subsubsection*{Top Partners}

Top partners are the Composite Sector resonances associated with the fermionic operator ${\mathcal{O}}$ introduced in eq.~(\ref{ferint}) to couple the third family $q_L=(t_L,b_L)$ doublet and the singlet $t_R$ with the CS. Similarly to what we saw for vectors in eq.~(\ref{op_par}), top partners quantum numbers can be extracted from the relation
\beq\label{op_par_TP}
\ds
\langle\Psi|{\mathcal{O}}|0\rangle\neq0\,.
\eeq
Since ${\mathcal{O}}$ is a Lorentz Dirac spinor, $\Psi$ must be a spin $1/2$ particle in order to be excited by  ${\mathcal{O}}$ from the vacuum. Also, ${\mathcal{O}}$ is in the triplet of the QCD colour group and thus $\Psi$ must also be coloured as I anticipated in section~\ref{MCHC}. Finally, top partners are CS resonances and as such their mass must be large, of order $m_*\sim$~TeV, barring special suppression mechanisms which we have no reason to expect a priori. The large top partners mass comes directly from the CS and it is unrelated with the occurrence of EWSB. Unlike quarks and leptons, top partners masses would be present in the theory even if the EW gauge symmetry was unbroken, meaning that top partners must be endowed with a perfectly gauge-invariant Dirac mass term. This requires top partners to be ``vector-like'' fermions, i.e. to come as complete Dirac fields with their left- and the right-handed components transforming in the same way under the gauge group. Coloured particles of this sort are said to be Vector-Like Quarks (VLQ's). Top partners are VLQ's of specific type and with specific properties.\footnote{VLQ's are somehow similar to a fourth family of quarks, but they are also radically different in that their vector-like mass allows them to be at the TeV scale without need of huge Yukawa couplings. Unlike a fourth family, VLQ's are not excluded by the measurement of the Higgs production rate from gluons. See \cite{Azatov:2011qy} for an analysis of the (moderate) impact of CH top partners on Higgs physics.}

Top partners gauge quantum numbers can also be extracted from eq.~(\ref{op_par_TP}). The result depends on the representation of ${\mathcal{O}}$ under the CS global group ${\textrm{SO}}(5)$, which is a priori ambiguous as I explained in section~\ref{MCHC}. However any valid representation of ${\textrm{SO}}(5)$, or actually any valid representation of any CS group ${\mathcal{G}}$ we might decide to deal with, going beyond the minimal choice ${\mathcal{G}}={\textrm{SO}}(5)$, must contain at least one SM doublet with $1/6$ Hypercharge and one singlet with Hypecharge $2/3$. The reason is of course that eq.~(\ref{ferint}) must comply with gauge invariance and thus some of the components of ${\mathcal{O}}$ must have the same gauge quantum numbers as those of the SM $q_L$ and $t_R$ fields. Top partners are thus at least one $(T,B)$ doublet and one ${\widetilde{T}}$ singlet, plus extra states that possibly emerge from the decomposition of ${\mathcal{O}}$. Among those, one extra doublet with exotic Hypercharge of $7/6$ is often present, producing one additional top partners doublet $(X_{2/3},X_{5/3})$ with electric charge $2/3$ and $5/3$, respectively. It is possible to show that all choices of the ${\mathcal{O}}$ representation for which the extra doublet is absent, such as the ${\bf{4}}$ we mentioned in section~\ref{MCHC}, are typically in serious phenomenological troubles because of unacceptably large modifications of the $Zb{\overline{b}}$ coupling \cite{Contino:2006qr,Agashe:2006at}. We thus have good reasons to expect the presence of the extra top partners doublet and thus good reasons to search for it.

\begin{figure}[t]
  \centering
  \includegraphics[width=0.8\textwidth]{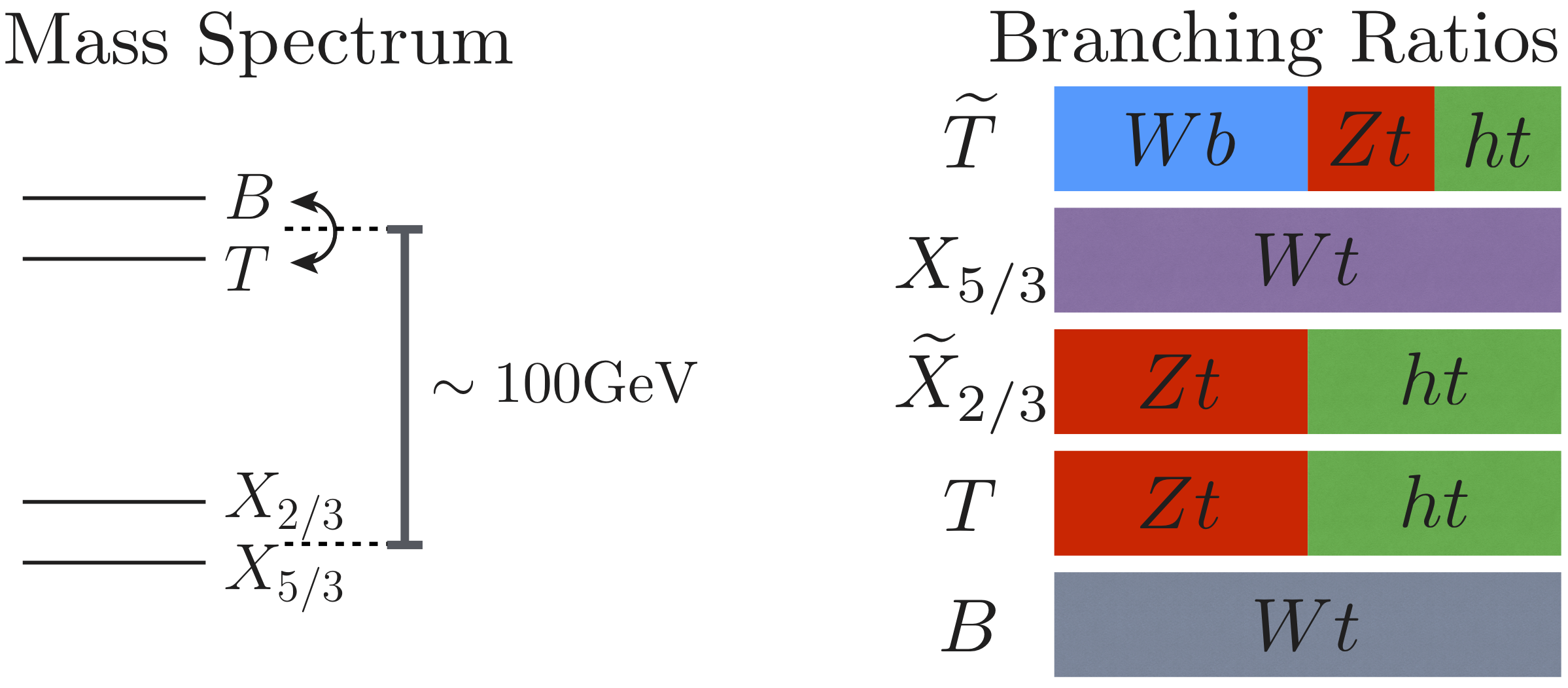} 
\caption{\label{TPS}Typical top partners mass spectrum and decay branching ratios.} 
\end{figure}

Similarly to what we saw above for vector resonances, top partners phenomenology can be characterised by employing symmetry, which constrain the structure of their interactions, and power-counting (\ref{1s1c}), which sets the expected strengths of the different couplings. The characterisation is slightly more complicate than the one for vectors, mainly because the whole symmetry structure of the theory must be taken into account and not just the SM gauge group. This includes the \mbox{SO$(4)$} unbroken group of the CS and even the full non-linearly realised \mbox{SO$(5)$} which takes care of the pNGB nature of the Higgs. The analysis produces relatively sharp predictions \cite{Matsedonskyi:2012ym,DeSimone:2012fs} of the top partners mass spectrum, decay and production processes. As shown in figure~\ref{TPS}, particles within the two doublets are essentially degenerate, but also the two doublets are quite close in mass, with a splitting between them of around $100$~GeV. The exotic Hypercharge doublet is always the lightest of the two. This spectrum is due to the fact that the two doublets emerge as a single \mbox{SO$(4)$} quadruplet and by the peculiar way in which the \mbox{SO$(4)$} symmetry is broken by the pNGB Higgs VEV. The ${\widetilde{T}}$ singlet can have any mass, significantly below or above (or close to) the two doublets. The top partners decay branching ratios are approximately universal, as shown in the right panel of the figure. This feature is not peculiar of top partners, it holds for any VLQ with a mass much above the EW scale and follows from considerations related with the Equivalence Theorem similar to those that led us to eq.~(\ref{bdec}) for vector resonances.

\begin{figure}[t]
  \centering
  \includegraphics[width=0.8\textwidth]{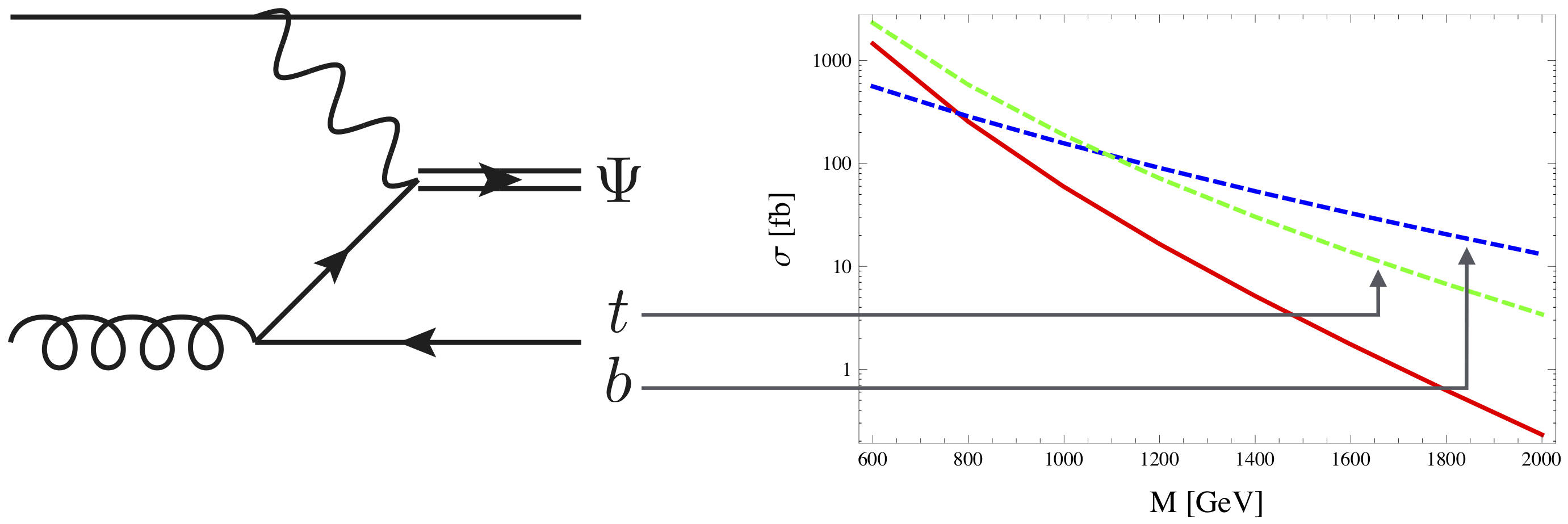} 
\caption{\label{TPSP}Top partners production cross-section for typical values of the single-production coupling at the $14$~TeV LHC. Pair production is shown as a continuous red line.} 
\end{figure}

Top partners are colour triplets, thus they are produced in pair by QCD interactions at a fixed and predictable rate as a function of their mass. Since the branching ratios are also known, negative searches for top partners pair production allow to set sharp mass limits, of around $800$ or $900$~GeV at the LHC run-$1$. The run-$2$ reach in terms of exclusions is around $1.2$ or $1.5$~TeV, and it is unlikely it will ever overcome $1.7$~TeV even when the full luminosity of the HL-LHC will be available in many years from now (see \cite{Matsedonskyi:2014mna,Matsedonskyi:2015dns} and references therein). The reach could however be extended up to around $2$~TeV by exploiting another sizeable production mechanism top partners are found to possess, namely single production (see figure~\ref{TPSP}) in association with a top or with a bottom plus a forward jet from the splitting of an EW boson out of a quark line. Single production emerges from a vertex with schematic form
\beq\label{spv}
\ds
{\mathcal{L}}_{\textrm{single}}\sim{\lambda_{L{\textrm{/}}R}}{\overline\Psi}Hq_{L{\textrm{/}}R}\,,
\eeq
with $q=t$ or $q=b$. The vertex couples top partners with third family quarks and the Higgs, and its power-counting estimate (\ref{1s1c}) is rather sizeable because it is controlled by third family $\lambda_{L{\textrm{/}}R}$ couplings. The Equivalence Theorem relates as usual the Higgs field components to longitudinally polarised EW bosons (see Footnote~\ref{ET}), therefore the operator produces single-production vertices like the one in figure~\ref{TPSP}. These vertices are of course also responsible for Top Partners decays. Single production cross-section, as the figure shows, is favoured at high mass by the steeply falling parton luminosities and readily starts to dominate over pair production. The mass-limit one can set for single production is not as sharp as the one from pair production because the reach crucially depends on the magnitude of the interaction vertex (\ref{spv}), which is not fully predicted. The above-mentioned expected reach ($\sim2$~TeV \cite{Matsedonskyi:2014mna,Matsedonskyi:2015dns}) is based on a conservative estimate of the single production coupling strength.

Top partners are arguably the most important CH signatures to be searched for in the forthcoming LHC runs, in spite of the fact that the mass-reach is not great if compared with the one on vectors that can easily overcome $3$~TeV by exploiting the complementarity between direct and indirect searches as in figure~\ref{HVT2}. The point is that a $3$ or even $5$~TeV bound on vectors would not be as problematic for the CH scenario as a $2$~TeV bound on top partners. Conversely, we don't have a strong theoretical preference for vectors below $3$ or $5$~TeV, or at least not such a strong one as we have for top partners below $2$~TeV. Of course all resonance masses are set by the same scale, $m_*$, therefore we expect them to be comparable but a factor of two hierarchy between vectors and top partners is perfectly conceivable. What makes top partners special is that it is their mass the one that actually enters in the fine-tuning formula in eq.~(\ref{deltatuning}), not the mass of vectors or of other CS resonances. Namely, the statement which I will now justify is that the generic estimate of fine-tuning in eq.~(\ref{deltatuning}) specialises, in the case of the CH scenario, to $\Lambda_{\textrm{SM}}=M_\Psi$. Top partners at $2$~TeV would thus cost a tuning well above ten.

\begin{figure}[t]
  \centering
  \raisebox{15pt}{\includegraphics[width=0.3\textwidth]{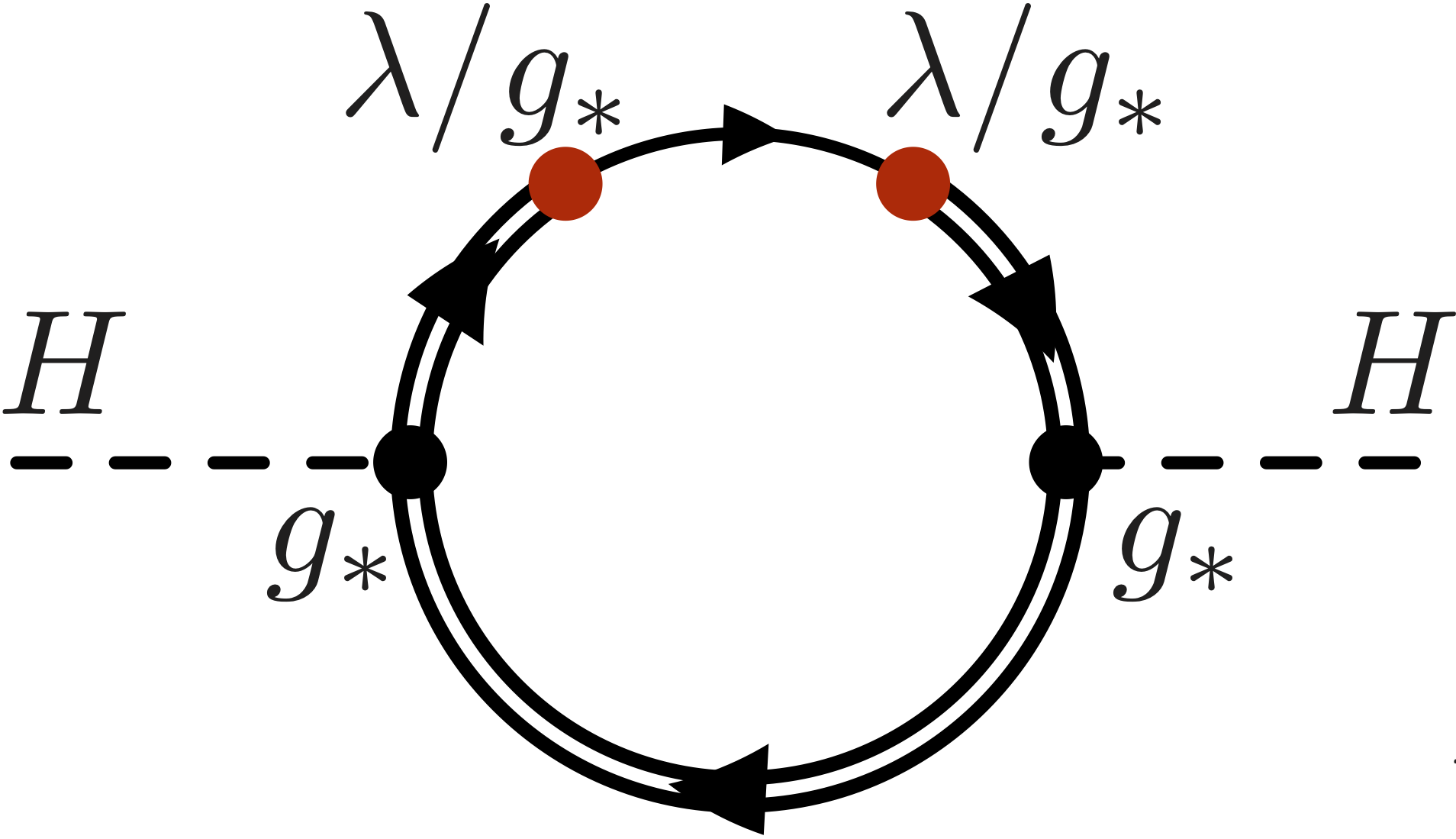}} \ \hspace{70pt} \
    \includegraphics[width=0.2\textwidth]{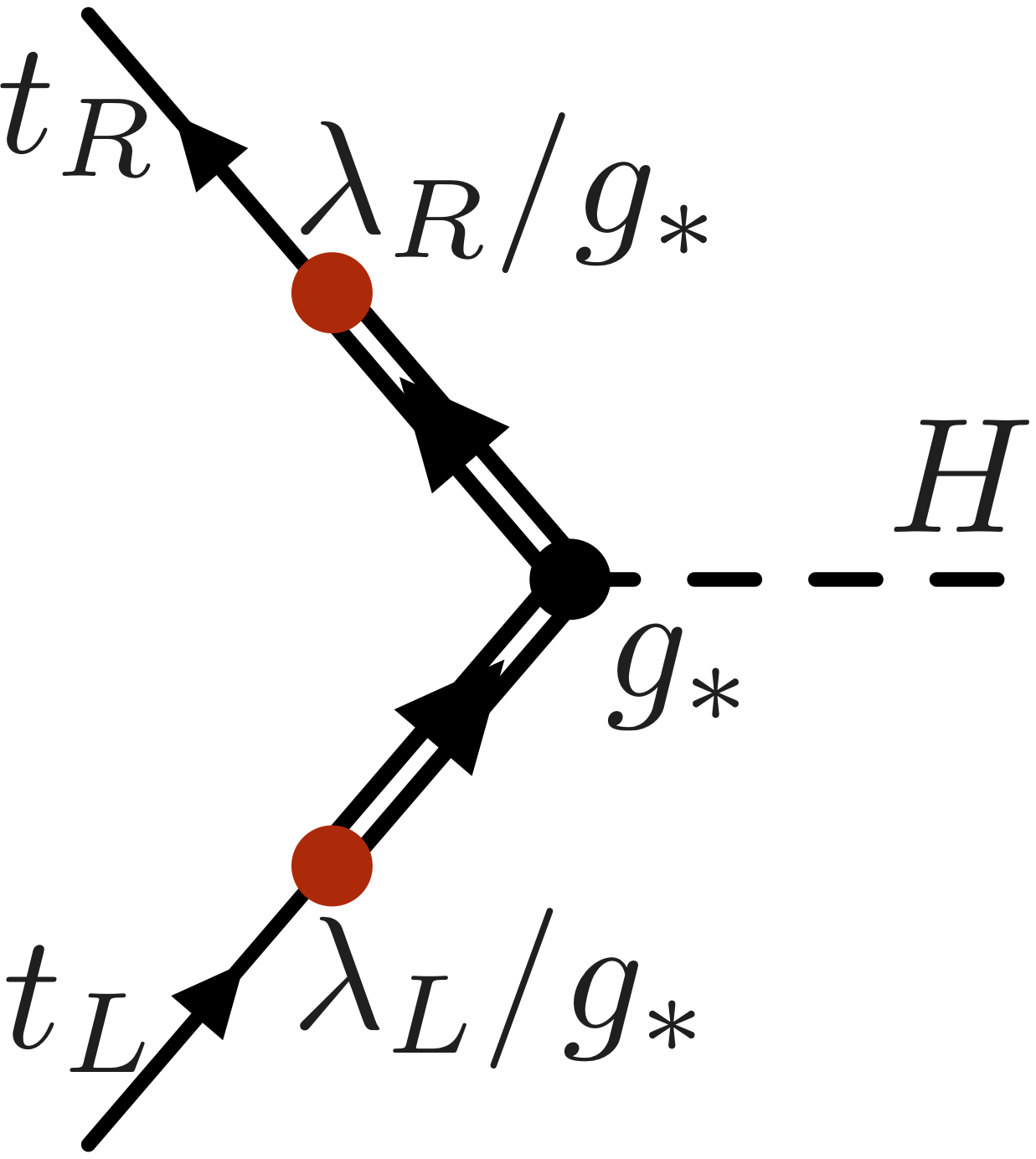} 
\caption{\label{potyuk}Left panel: one representative diagram contributing to the Higgs mass. The Higgs-top partners vertex is a purely CS interaction and thus it has been estimated as $g_*$. The insertion of the mixing weights as in eq.~(\ref{mix}). Right panel: the generation of the top Yukawa coupling through mixing.} 
\end{figure}

The connection between top partners and fine-tuning is due to the fact that top quark loops (see section~\ref{HP} and in particular figure~\ref{natarg}) are the dominant term in the low-energy contribution to the Higgs mass which is at the origin of the fine-tuning problem, and top partners are strongly coupled with the top quark. An example of such coupling is the single production operator in eq.~(\ref{spv}). Another relevant interaction is the top/top partners mixing of the form \footnote{Order one coefficients, which of course are be predicted by the power-counting formula (\ref{1s1c}), are understood in both terms.}
\beq\label{mix}
\ds
{\mathcal{L}}_{\textrm{mix}}\sim\frac{\lambda_{L}}{g_*}m_*\,{\overline{T}}\,t_{L}+\frac{\lambda_{R}}{g_*}m_*\,{\overline{\widetilde{T}}}\,t_{R}\,,
\eeq
and analogously for the $b_L$ mixing with the $B$. In explicit model it is only the mixing term above which is actually generated (in the appropriate field basis) and all the other quarks interactions such as those in eq.~(\ref{spv}) emerge after diagonalization. The mixing can be used to construct loop diagrams like the one in the left panel of figure~\ref{potyuk}, involving the exchange of a virtual top and a top partner. These diagrams generate a mass for the Higgs, of order
\beq\label{mhtop}
\ds
m_H^2\sim a_L\frac{\lambda_L^2}{16\pi^2}M_\Psi^2+a_R\frac{\lambda_R^2}{16\pi^2}M_\Psi^2\,,
\eeq
where the two terms stand respectively for the exchange of a virtual $t_L$ and $t_R$. The order one numerical coefficients $a_L$ and $a_R$ are calculable in explicit CH models (see e.g. \cite{Matsedonskyi:2012ym}) and, depending on the model's microscopic parameters, can assume any sign. The estimate of $m_H$ has been performed by counting the powers of $\lambda$ and $g_*$, reported in figure~\ref{potyuk}, multiplying by the loop factor $1/16\pi^2$ and by two powers of the top partners mass $M_\Psi$ because of dimensionality. This is quite right in spite of the fact that the diagram is still logarithmically divergent because the log only produces order one numerical coefficients which is not worth retaining in our rough estimate.\footnote{This would not be the case if a parametrically large separation was present between $M_\Psi$ and the confinement scale $m_*$ at which the loop is naturally cut off. We assume a factor of a few separation at most, which does not qualifies as parametrically large and thus the estimate is correct.}

Eq.~(\ref{mhtop}) requires some clarification. As I explained at length in the previous sections, the fact that the Higgs is a NGB prevents the generation of its mass as long as the Goldstone symmetry, i.e. the group ${\mathcal{G}}$, is an exact symmetry of the theory. Since the CS is exactly invariant under ${\mathcal{G}}$, no contribution to $m_H$ can come from the CS alone. In our language this contribution would be a tree-level Higgs mass-term, and the fact that it is absent is the reason why to estimate $m_H$ we had to go at the loop level as in figure~\ref{natarg}. The diagrams in the figure have the chance to produce a mass because they do feel ${\mathcal{G}}$ breaking through the insertion of the top/top partner mixing. Remember that are indeed the Composite/Elementary Sector interactions the ones responsible for ${\mathcal{G}}$ breaking (see figure~\ref{elcomp}) in our construction, and the mixing is one of those interaction. Moreover the mixing is the largest of those interaction because it is associated with the generation of the largest coupling of the Higgs boson, namely the top quark Yukawa $y_t$. Other Elementary/Composite interaction such as the gauge couplings also contribute to $m_H$, producing however only small corrections to eq.~(\ref{mhtop}). This is the reason why it is the top partners mass scale $M_\Psi$, and not for instance the mass of spin one resonances, the one that controls the size of the Higgs mass.

Mixed top/top partners loops generate not only a mass-term, but a full potential for the Higgs field. The potential has the form of eq.~(\ref{hpot}), with an $\alpha$ parameter 
\beq\ds
\alpha\sim a_L\lambda_L^2\frac{N_cM_\Psi^2}{16\pi^2} + a_R\lambda_R^2\frac{N_cM_\Psi^2}{16\pi^2} \,.
\eeq
This estimate is slightly more accurate than the one in eq.~(\ref{mhtop}), in particular it takes into account the number of colours $N_c=3$, but it scales in the same way with the parameters. The physical mass of the Higgs boson, obtained by combining the two lines of eq.~(\ref{VEVM}), thus reads
\beq\ds
m_H^2=4(1-\xi)\alpha\sim a_L\lambda_L^2\frac{N_cM_\Psi^2}{4\pi^2} + a_R\lambda_R^2\frac{N_cM_\Psi^2}{4\pi^2}\,.
\eeq
If $M_\Psi$ is large, obtaining the correct Higgs mass $m_H=125$~GeV requires a cancellation between the two terms, obtained by choosing the fundamental parameters of the models such that $a_L$ is almost equal and opposite to $a_R$. This means a fine-tuning 
\beq
\label{tuntpp}
\Delta = \frac{3\, \lambda^2}{4\pi^2}\left(\frac{M_\Psi}{m_H}\right)^2\simeq \lambda^2 \left(\frac{M_\Psi}{450\,{\textrm{GeV}}}\right)^2\,,
\eeq
having assumed $\lambda_L\simeq\lambda_R\equiv\lambda$, which is the configuration that minimises the required amount of tuning.  The equation clearly illustrates that light top partners are needed for a Natural (low-tuning) CH model.

Our estimate closely resembles the general formula (\ref{deltatuning}) with $\Lambda_{\textrm{SM}}=M_\Psi$, apart from the prefactor $\lambda^2$ that is replaced by $y_t^2$ in eq.~(\ref{deltatuning}). In order to see that the two formulas match we should relate $\lambda$ with the top Yukawa coupling, by proceeding as follows. The top/top partners mass-mixing (\ref{mix}) makes that the two chirality components of the physical top quark, which is massless before EWSB is taken into account, are a quantum mechanical superimposition of Elementary and Composite degrees of freedom
\bea\label{part_comp}
\ds
|t_L^{\textrm{phys.}}\rangle=\cos\phi_L|t_L^{\textrm{Elem.}}\rangle+\sin\phi_L|T_L^{\textrm{Comp.}}\rangle\,,\nonumber\\
|t_R^{\textrm{phys.}}\rangle=\cos\phi_R|t_R^{\textrm{Elem.}}\rangle+\sin\phi_R|{\widetilde{T}}_R^{\textrm{Comp.}}\rangle\,,
\eea
with $\sin\phi_L\simeq\lambda_L/g_*$ and $\sin\phi_R\simeq\lambda_R/g_*$. A similar formula holds for the $b_L$. This comes from diagonalising the mass-matrix of the top/top partners system, which consists of the mass-mixing (\ref{mix}) plus the vector-like mass-term $M_\Psi$ for the partners. For the estimate we took $m_*=M_\Psi$ in eq.~(\ref{mix}), consistently with what implicitly done in the estimate of the $m_H$. Eq.~(\ref{part_comp}) shows, in the first place, why we call ``Partial Compositeness'' \cite{Kaplan:1991dc} the mechanism (\ref{ferint}) we are using to couple ES fermions with the CS: it is because it produces physical particles that are partially made of Composite degrees of freedom. Second, the formula allows us to estimate the top Yukawa generated by mixing as in the right panel of figure~\ref{potyuk}, obtaining
\beq\label{yukest}
\ds
y_t=\sin\phi_L\sin\phi_R\,g_*\,,\;\;\;\;\;
\raisebox{8pt}{$
\overset{\lambda_L=\lambda_R=\lambda}{\raisebox{-12pt}{\resizebox{20pt}{!}{$\Rightarrow$}}}$}
\;\;\;\;\;\lambda=\sqrt{y_tg_*}\,.
\eeq
But we said that $g_*$ has to be large, at least above $y_t\simeq1$, therefore the above equation tells us $\lambda>y_t$ and eq.~(\ref{tuntpp}) can be turned into a lower bound
\beq
\label{tpptf}
\Delta > \frac{3\, y_t^2}{4\pi^2}\left(\frac{M_\Psi}{m_H}\right)^2\simeq \left(\frac{M_\Psi}{450\,{\textrm{GeV}}}\right)^2\,,
\eeq
identical to eq.~(\ref{deltatuning}). 

Notice that the estimate of the Yukawa couplings can be carried on for the light quarks (including the bottom) and leptons in exactly the same way as for the top, producing expressions for the corresponding $\lambda$ parameters which are identical to eq.~(\ref{yukest}) aside from the fact that the light quarks and leptons Yukawas, rather than $y_t$, are involved. Light generation $\lambda$'s are thus very suppressed and this is why we could systematically ignore them. Correspondingly, light fermions compositeness fraction $\sin\phi\sim\lambda/g_*$ are very small. Light fermions are thus almost entirely elementary particles, with a tiny composite component which is however essential to generate their Yukawa's and masses.\footnote{There are however exceptions to this rule. On one hand, it is possible to make largely composite one of the light quarks chirality components recovering the small Yukawa by giving very very small compositeness to the other one. This helps in evading flavour constraints \cite{Redi:2011zi,Barbieri:2012uh} and produces interesting LHC signatures related with the fermionic partners of the light quarks \cite{Redi:2013eaa,Delaunay:2013pwa}. On the other hand, it is possible to avoid partial compositeness altogether for the light fermions \cite{Matsedonskyi:2014iha,Panico:2016ull} and obtain their mass by bilinear interactions.}

In summary, the importance of top partners stems from their connection with tuning in eq.~(\ref{tpptf}). Not finding them at the LHC below $2$~TeV would cost more tuning than what negative searches of Higgs couplings modifications (whose reach is $\xi<0.1$) would imply through eq.~(\ref{xitun}).\footnote{Eq.~(\ref{tpptf}) does not supersede eq.~(\ref{xitun}). The two equations estimate tuning from different sources, namely the one from the Higgs VEV and from the Higgs mass, respectively. Therefore the maximum of the two expressions should be taken for a complete estimate of $\Delta$.} Vector resonances are mildly connected with tuning, therefore even a multi-TeV bound on their mass would not be competitive in terms of fine-tuning reach. Top partners searches are so important because their capable to put the very idea of Naturalness in serious troubles, at least in the Composite Higgs framework. We will see in the next chapter that a similar role is played in Supersymmetry by the stops. It is also important to keep in mind that top partners might very well be discovered at the LHC run-$2$. Current bounds are below $1$~TeV and thus their impact on tuning is modest, well below ten and comparable with the one from coupling measurements. The interesting mass region is the one from $1$ to $2$~TeV in which we are about to enter.

\section{Supersymmetry}\label{SUSY}

Supersymmetry (SUSY) is probably the most intensively studied theoretical subject of the last $30$ or $40$ years. Its applications range from string theory and supergravity down to collider phenomenology, with digressions on \mbox{AdS/CFT} correspondence and holography, dualities and scattering amplitude properties. I mention this to outline that the scope of SUSY is much broader than phenomenology and to explain why theorists care about SUSY a priori, independently of its applicability to the real world on a short timescale. Plenty of excellent reviews \cite{Sohnius:1985qm,Derendinger:1990tj,Drees:1996ca,Martin:1997ns,Peskin:2008nw}, lecture notes\footnote{At least $22$ of them, counting only those produced by the CERN ESHEP school founded in $1993$.} and books \cite{WessBagger,WeinbergSUSY} have been written about SUSY, just to mention some of those that are relevant in the (relatively narrow, as I mentioned) context of SUSY phenomenology. With all this literature available, it makes no sense trying to condense a self-contained introduction to SUSY in these few pages. I will thus keep introductory material to the minimum, focusing only on few basic concepts and results that are absolutely needed for the discussion. Next, in sections~\ref{tale} and \ref{after}, I will describe SUSY phenomenology building around two specific questions which I find particularly important to address in this particular moment.

\subsection{Basics of SUSY}

Symmetries are so much important in particle physics that Coleman and Mandula in `$67$ found interesting to ask themselves what is the largest symmetry content a relativistic theory of interacting particles can posses \cite{Coleman:1967ad}. Their answer was that the largest global symmetry group is Poincar\'e, generated by the $6$ $M^{\mu\nu}$ Lorentz generators plus the $4$ $P^\mu$'s associated with space-time translations, times a generic Lie group of symmetries generated by a set of charges ${\vec{Q}}_{\textrm{B}}$. Here ``times'' means direct product, namely their result was that all the internal symmetry generators have to commute with those of the Poincar\'e group
\beq\label{comm0}
\left[P^{\mu},{\vec{Q}}_{\textrm{B}}\right]=0\,,\;\;\;\;\;\left[M^{\mu\nu},{\vec{Q}}_{\textrm{B}}\right]=0\,.
\eeq
Remember that commutators are the way in which the symmetry generators act on the other operators. The first equation thus means that the ${\vec{Q}}_{\textrm{B}}$'s are invariant under translations and the second one means that the ${\vec{Q}}_{\textrm{B}}$'s are Lorentz scalars, namely that they stay the same in any reference frame. With a modern terminology we would say that what Coleman and Mandula had in mind were ``bosonic'' generators, this is why I labeled them with the subscript ``B''. Concretely what they had in mind are generators that obey ordinary commutation relations among them, of the form $[{{Q}}_{\textrm{B}}^i,{{Q}}_{\textrm{B}}^j]=if^{ijk}{{Q}}_{\textrm{B}}^k$.

However Gol'fand and Likhtman proved that Coleman and Mandula were wrong, and in so doing they discovered SUSY \cite{Golfand:1971iw}. They pointed out that a set of $2$ symmetry generators $Q_\alpha$ ($\alpha=1,2$) exist which do not obey eq.~(\ref{comm0}), but instead
\beq\label{comm1}
\left[P^{\mu},{{Q}}_{\alpha}\right]=0\,,\;\;\;\;\;\left[M^{\mu\nu},{{Q}}_{\alpha}\right]=-(\sigma^{\mu\nu})_\alpha^{\;\;\beta}Q_\beta\,.
\eeq
The $Q$'s are still invariant under translations, and in particular under time translations (which is of course obvious if they are conserved), but they are not anymore invariant under Lorentz transformations. Under Lorentz, i.e. under commutation with $M^{\mu\nu}$, the $Q$'s transform with the matrix $\sigma^{\mu\nu}$ which is a $2\times2$ representation of the Lorentz group called the (left-handed) Weyl representation. Therefore we will say that the SUSY generators (or charges) $Q_\alpha$ form a two-components Weyl spinor under the Lorentz group. 

The reader unfamiliar with the formalism of Weyl spinors is referred to standard textbooks or to Ref.~\cite{Derendinger:1990tj} (section~4.1 and Appendix~A and B) for a concise introduction. The essential point is that Weyl spinor fields are the ``building'' blocks of the habitual Dirac fermions we normally employ to describe {spin-$1/2$} particles. Namely, one four-components Dirac spinor $\Psi$ can be decomposed as
\beq\label{deco}\ds
\Psi=\left[
\begin{array}{c}
(\psi_1)_\alpha \\[20pt]
({\overline{\psi_2}})^{\dot{\alpha}}
\end{array}
\right]
\begin{array}{c}
\includegraphics[width=0.2\textwidth]{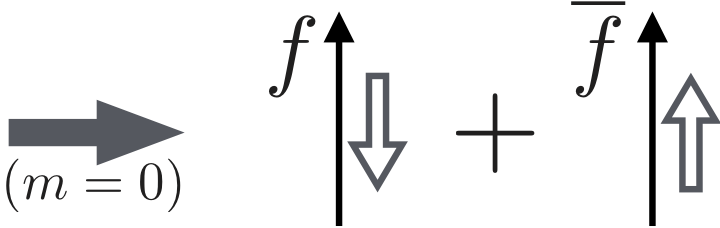}\\[0pt]
\includegraphics[width=0.2\textwidth]{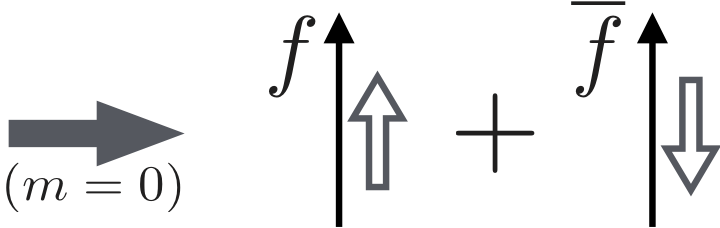}
\end{array}
\eeq

\noindent{in} terms of two two-components spinors $(\psi_1)_\alpha$ and $(\psi_2)_\alpha$ called ``left-handed'' Weyl spinors.\footnote{This assumes that the Weyl basis is chosen for the $\gamma$ matrices, otherwise the decomposition is more complicate.} As anticipated, the Lorentz generators acting on these objects are the $\sigma^{\mu\nu}$ matrices. Namely, under an infinitesimal Lorentz transformation
\beq\ds
(\delta\psi_{1,2})_\alpha=-\frac{i}2\omega_{\mu\nu}(\sigma^{\mu\nu})_{\alpha}^{\;\;\beta}\,,\;\;\;\;\sigma^{\mu\nu}=\frac{i}4(\sigma^\mu\overline\sigma^\nu-\sigma^\nu\overline\sigma^\mu)\,,
\eeq
where $\sigma^\mu=(\Id,\vec\sigma)$ and $\overline\sigma^\mu=(\Id,-\vec\sigma)$, with $\vec\sigma$ the Pauli matrices. Notice that unlike $\psi_1$, $\psi_2$ does not enter the decomposition formula directly, but rather through the object $({\overline{\psi}}_2)^{\dot{\alpha}}$ which is related to  $\psi_2$ by complex conjugation. Namely
\beq\label{LR}
\ds
({\overline{\psi_2}})^{\dot{\alpha}}=\varepsilon^{\dot\alpha\dot\beta}[(\psi_2)_\beta]^*\,,
\eeq
where $\varepsilon$ is the antisymmetric Levi-Civita tensor in two dimensions and the sum over $\beta=1,2$ is understood. We sometimes call $({\overline{\psi}}_2)^{\dot{\alpha}}$ a ``right-handed'' Weyl spinor, however the previous formula shows that there is no actual distinction between left- and right-handed spinors because one can be turned into the other by complex conjugation. What is normally done in the SUSY literature is to use only left-handed spinors to describe fermions, an habit which can be confusing for beginners. For instance, because of this convention the SM right-handed top quark is represented by a left-handed spinor with electric charge equal to $-2/3$ rather than $+2/3$ because the correspondence between left and right spinors involves complex conjugation. 

If the Dirac spinor $\Psi$ is massless, the two Weyl components are endowed with a very simple physical interpretation, pictorially reported in eq.~(\ref{deco}). $\psi_1$ corresponds to a massless fermion $f$ with helicity $h=-1/2$ plus its anti-particle $\overline{f}$ with $h=+1/2$, while $\psi_2$ is an $h=+1/2$ fermion plus an $h=-1/2$ anti-fermion. If instead the Dirac spinor is massive, namely if it is endowed with a Dirac mass term, there is no direct correspondence between Weyl spinors and physical particles because the Dirac mass mixes the two Weyl components and produces physical particles which are combinations of the two components. Still, a Weyl spinor can be in direct correspondence with a massive fermion, but only if it is a completely neutral particle, not endowed with any conserved charge or quantum number. In this case there is no way to distinguish particle from anti-particle, namely $f=\overline{f}$, and the two helicity states of each Weyl can be interpreted as the two helicity (or spin) eigenstates of a single massive fermion. A mass term given to a single Weyl spinor, which unlike the Dirac mass does not mix the two Weyl components, is called a ``Majorana'' mass. One Weyl $2$-component spinor can be equivalently representations as a $4$-component spinor called a ``Majorana spinor''. There is no physical distinction between the two representations, thus a Weyl fermion with Majorana mass is often called a Majorana fermion.

After this interlude on Weyl spinors, we return to our historical introduction to SUSY. Gol'fand and Likhtman could find a counterexample to the Coleman--Mandula theorem because Coleman and Mandula made too restrictive assumptions in their proof. Namely they assumed bosonic internal symmetry generators, characterised by  ordinary commutation relations as previously mentioned. The Gol'fand--Likhtman SUSY charges are instead fermionic generators, characterised by anti-commutation relations
\beq\label{AC}\ds
\{Q_\alpha,Q_\beta\}=0\,,\;\;\;\;\;\{Q_\alpha,{\overline{Q}}_{\dot{\beta}}\}=2(\sigma^\mu)_{\alpha\dot\beta}P_\mu,
\eeq
where $\overline{Q}$ is the conjugate of the SUSY charge \footnote{Weyl spinor indices can be raised or lowered by acting with $\varepsilon^{\alpha\beta}=\varepsilon^{\dot\alpha\dot\beta}=-\varepsilon_{\alpha\beta}=-\varepsilon_{\dot\alpha\dot\beta}$. With this convention the definition of ${\overline{Q}}_{\dot{\alpha}}$ reported below matches with eq.~(\ref{LR}).}
\beq\ds
{\overline{Q}}_{\dot{\alpha}}=[Q_\alpha]^*\,.
\eeq
SUSY charges are thus very different from the ordinary generators of internal symmetries like baryon and lepton number, isospin, etc. Unlike the latter, they do not form an algebra, specified by commutation relations, but rather what is called a ``super-algebra'', specified by relations that involve the anti-commutators. Moreover, and perhaps more importantly, SUSY generators do not commute with $M^{\mu\nu}$ (\ref{comm1}) unlike the ordinary bosonic charges (\ref{comm0}). The story ends with Haag, Lopusza\'nski and Sohnius, who had the final word on the maximal symmetry content of a relativistic theory (with massive particles) \cite{Haag:1974qh}. They found that it consists of the Poincar\'e generators, plus bosonic bosonic charges, plus a set of replicas, $Q_\alpha^i$ with $i=1,\ldots,N$, of Gol'fand--Likhtman's SUSY generators. Actually they also found other symmetries related with SUSY, called ``$R$-charges''. Extended SUSY, namely $N\neq1$, does not play an important role in phenomenology, therefore in what follows we will stick to the minimal case $N=1$. A similar consideration holds for the continuos $R$-symmetry group, aside from a discrete subgroup of it called ``$R$-parity'' which is instead very relevant and will be discussed in the next section.

SUSY-invariant theories display a number of remarkable properties, some of which can be summarised by the famous rule 
\beq\ds\label{b=f}
{\textrm{Bosons}}\;\;\raisebox{-2pt}{$\underset{\textrm{SUSY}}{\scalebox{2}{=}}$}\;\;{\textrm{Fermions}}\,.
\eeq
The rule has several meanings, the simplest one being that SUSY requires bosonic and fermionic particles with the same mass. In order to see why it is so, consider the state $|h,p\rangle$ describing a single particle with helicity $h$ and four-momentum $p$. For definiteness, we will take the particle moving along the $z$-axis so that the helicity operator coincides with the third component of the angular momentum, i.e. the $1-2$ component of the Lorentz generator, $S_3=M^{12}$. Let us now act on the state with one of the SUSY charges, $Q_\alpha$. This produces a new single-particle state, $Q_\alpha|h,p\rangle$, with the following properties
\bea\label{MH}
&&P^\mu |h,p\rangle=p^\mu |h,p\rangle\;\;\raisebox{-1pt}{\scalebox{1.5}{$\Rightarrow$}}\;\;\;P^\mu\hspace{-1pt}\left( Q_\alpha|h,p\rangle\right)=\left(Q_\alpha P^\mu+[P^\mu,Q_\alpha]\right)|h,p\rangle=p^\mu\left( Q_\alpha|h,p\rangle\right),\\[2pt]
&&\hspace{-20pt}M^{12} |h,p\rangle=h\;|h,p\rangle\hspace{2pt}\;\;\raisebox{-1pt}{\scalebox{1.5}{$\Rightarrow$}}\;\;\;M^{12}\left( Q_\alpha|h,p\rangle\right)=\left(Q_\alpha M^{12}+[M^{12},Q_\alpha]\right)|h,p\rangle=(h\mp1/2)\left( Q_\alpha|h,p\rangle\right),\nonumber
\eea
where the $-1/2$ is for $\alpha=1$ and the $+1/2$ for $\alpha=2$. The first equation tells us that the new particle has the same four-momentum as the original one, and thus in particular the same mass. It follows from the first relation in eq.~(\ref{comm1}), which states that the SUSY charges commute with the $P^\mu$ operator. This first result is of course not at all surprising. Any symmetry generator commutes with $P^\mu$ and connects among each other particles with the same mass. The second relation in eq.~(\ref{MH}) is instead peculiar of SUSY. Ordinary generators commute with $M^{12}$ and as such they connect particles with the same spin and the same helicity. The commutator of SUSY charges with $M^{12}$ is instead $[M^{12},Q_\alpha]=-1/2(\sigma^3)_\alpha^{\;\beta}Q_\beta$, as dictated by eq.~(\ref{comm1}), so that SUSY connects particles with helicity $h$ to particles with helicity $h\mp1/2$ as in eq.~(\ref{MH}). Given that it shifts the helicity by a semi-integer amount, SUSY relates bosons with fermions and thus it requires the existence of mass-degenerate multiplets containing at the same time bosonic and fermionic particles.

\begin{figure}[t]
  \centering
  \includegraphics[width=0.3\textwidth,valign=t]{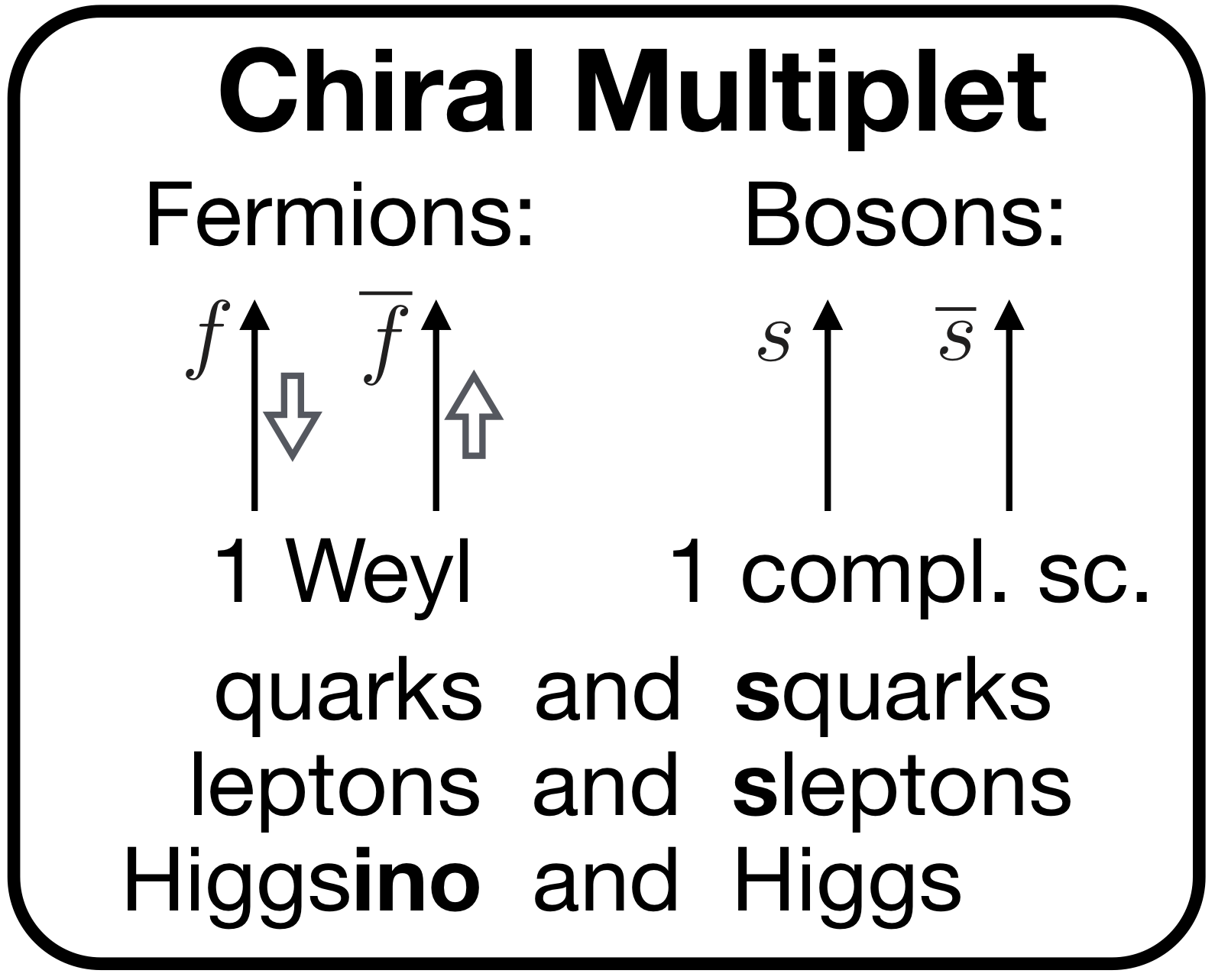}
    \includegraphics[width=0.3\textwidth,valign=t]{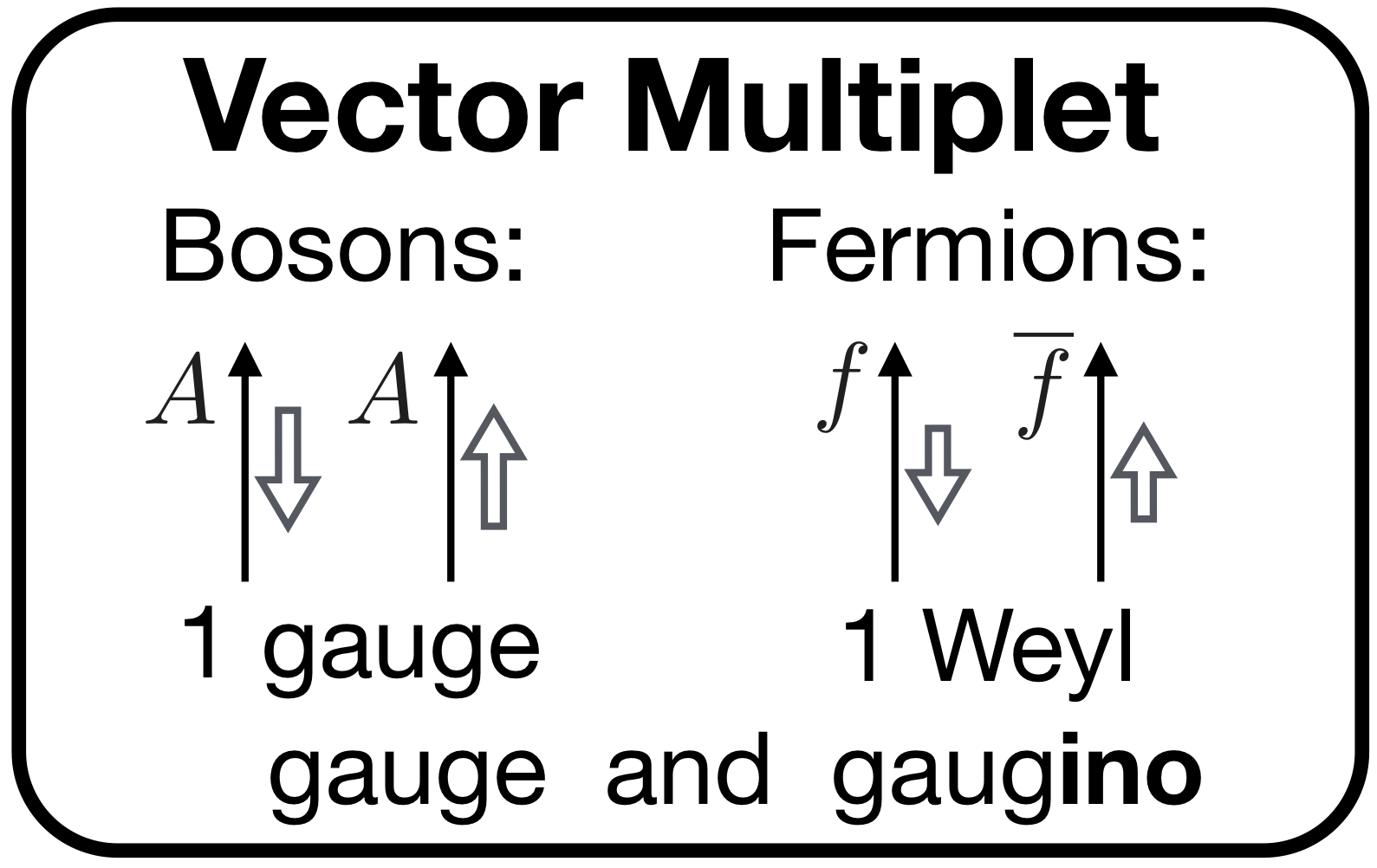}
      \includegraphics[width=0.3\textwidth,valign=t]{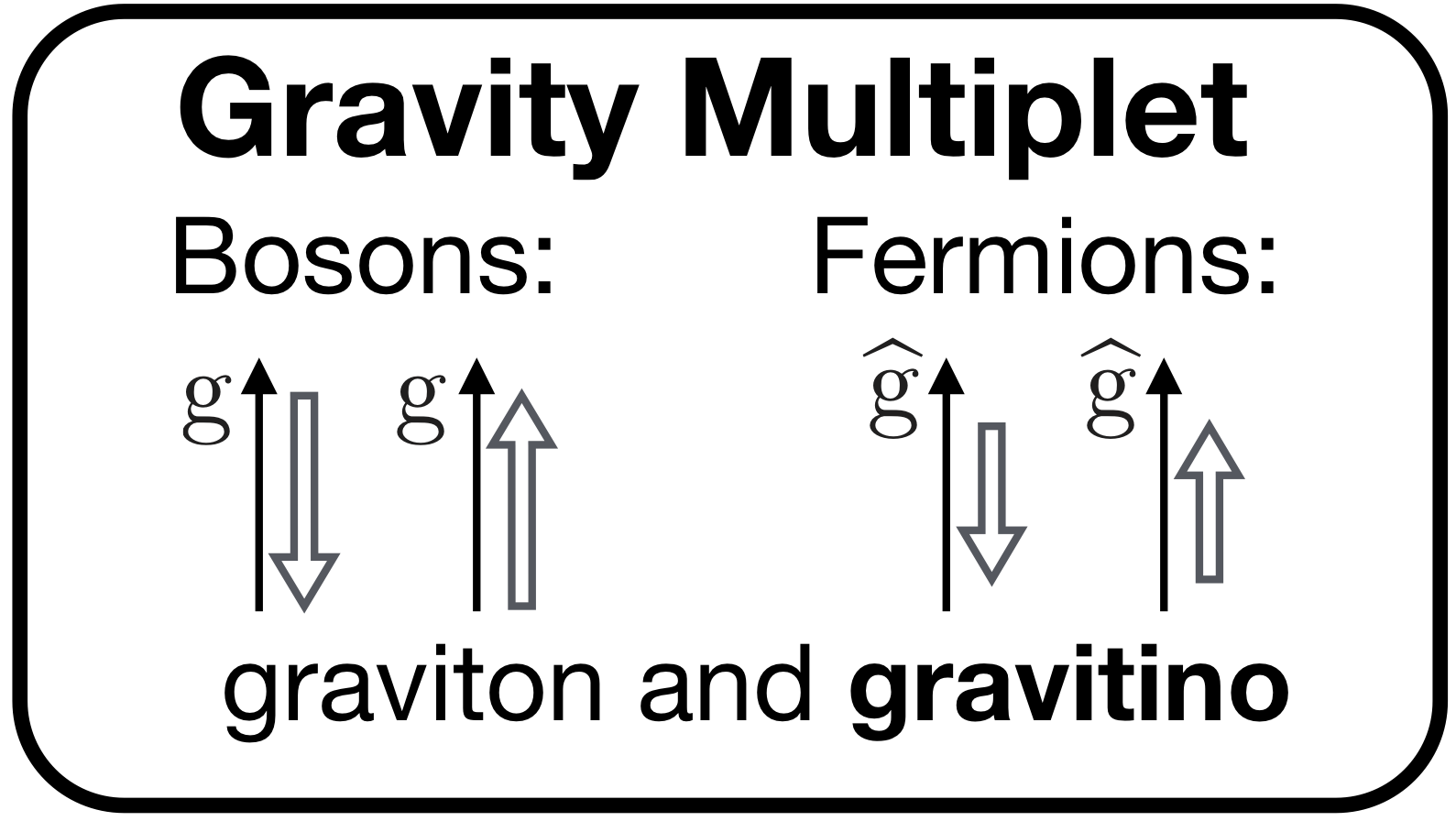}
  \caption{The $N=1$ SUSY multiplets that are relevant for phenomenology.
  \label{SUSY_mult}} 
\end{figure}

By proceeding along these lines, i.e. by repeatedly acting with $Q$ and $\overline{Q}$, one can classify the irreducible representations of $N=1$ SUSY. The relevant ones are those that contain particles of spin two at most, i.e. the chiral, vector and gravity multiplets, schematically represented in fig.~\ref{SUSY_mult}. When constructing supersymmetric extensions of the SM, chiral multiplets are used to describe the SM chiral fermions (quarks and leptons), plus the corresponding SUSY particles (squarks and sleptons). The latter are complex scalars with the same quantum numbers of the corresponding SM fermions under the SM gauge group. A chiral multiplet (actually, two of them, as we will see) also describes the SM Higgs field, plus the ``higgsinos'' superpartners, which are $2$-components fermions. Vector multiplets describe the SM gauge field (photon, gluons, $W$ and $Z$) with their partners, which are again $2$-components fermions called photino, gluinos, wino and zino. Clearly the vector multiplets describe the $W$ and $Z$ bosons, plus their super-partners, before the breaking of the EW symmetry, when they are massless. The gauge fields becoming massive require extra components taken from the Higgs multiplet, like in the SM. The graviton is part of the gravity multiplet, together with a particle of spin $3/2$, the gravitino. A proper description of the gravity multiplet and thus of the gravitino requires a supersymmetric theory of gravity, i.e. a Supergravity model. This goes far beyond the purpose of the present lectures, we will thus not consider the gravity multiplet anymore in what follows.

Notice that each of the multiplets in fig.~\ref{SUSY_mult} contains the exact same number ($2$) of bosonic and of fermionic degrees of freedom.\footnote{By ``degree of freedom'' we mean single-particles states of given helicity and quantum numbers.} If we combine them to form a SUSY theory we will thus obtain a model with the same number of bosonic and of fermionic degrees of freedom, in accordance with the general rule ``bosons $=$ fermions'' in eq.~(\ref{b=f}). Notice that if SUSY is spontaneously broken, bosons and fermions will not anymore form mass-degenerate multiplets according to fig.~\ref{SUSY_mult}, but still the total number of bosonic and of fermionic degrees of freedom in the theory will remain the same. It is interesting to remark that the validity of the ``bosons $=$ fermions'' rule crucially relies on the fact that the trivial representation, i.e. the singlet, does not exist in SUSY, unlike any other ordinary symmetry group. If it existed, it would be possible to add ``SUSY-singlet'' states (bosonic or fermionic) to the theory, violating in this way the equality of the number of bosonic and fermionic degrees of freedom. No SUSY-singlet particle exists because a singlet would be a state that is invariant under SUSY, which means that is must be annihilated both by $Q$ and by $\overline{Q}$. But since $\{Q,\overline{Q}\}\propto P_\mu$, this hypothetical SUSY singlet would be also annihilated by $P_\mu$ and thus it would have vanishing four-momentum and could not be interpreted as a particle. The only state with such properties, i.e. the only SUSY-singlet state, is a (SUSY-invariant) vacuum configuration.

Let us now turn to the problem of writing down SUSY-invariant theories. If SUSY was an ordinary (bosonic) global symmetry, this would be a trivial step to take, once the single-particle state multiplets are known. One would just introduce one field for each particle and construct a multiplet of fields that transform under the symmetry in the exact same way as the corresponding particle multiplets. Symmetric Lagrangians will eventually be obtained by constructing invariant combinations of the field multiplets. The situation is more complicate in SUSY. Constructing invariant Lagrangians requires the concept of ``auxiliary fields'' and the one of ``super-fields''. The issue comes from the  ``bosons $=$ fermions'' rule in eq.~(\ref{b=f}), which happens to hold not only for the states, but also for the fields. Namely, any set of fields that form a representation of SUSY must contain the same number of bosonic and of fermionic fields components. Consider for instance the chiral multiplet of particles. We describe its scalar degrees of freedom by one complex scalar field $\phi(x)$, which has $2$ real bosonic components, while to describe the $2$-components fermion we must use a Weyl spinor $\psi_\alpha(x)$, which amounts to $4$ real ($2$ complex) fermionic components. Purely in terms of fields, i.e. before we impose the Equations Of Motion (EOM) that reduce the number of fermionic degrees of freedom to $2$, there is a mismatch between the number of bosonic and fermionic components. This mismatch means that a SUSY multiplet cannot just contain the $\{\psi,\phi\}$ fields. One additional complex scalar field, the auxiliary field $F(x)$, is needed to match bosonic and fermionic components. The chiral multiplet is thus made of the set of fields $\{\psi,\phi,F\}$. The exact way in which the SUSY symmetry acts on this multiplet is not worth reporting here. What matters is that a consistent SUSY transformation exists and thus the problem of writing down a SUSY-invariant theory boils down, from this point on, to the one of combining these fields in order to form a SUSY-invariant Lagrangian. The super-field formalism turns out to be extremely effective for this purpose.

Before discussing super-fields, it is important to clarify the role of the auxiliary fields in the construction of SUSY theories. They are introduced in order to comply with the ``bosons $=$ fermions'' rule applied to the fields, but of course their presence cannot invalidate the rule at the particle level. Namely, auxiliary fields cannot produce extra propagating degrees of freedom, and for this being the case their Lagrangian must not contain a kinetic term. The simplest SUSY-invariant Lagrangian for a chiral multiplet indeed reads
\beq\label{simp}\ds
{\mathcal{L}}=i\overline{\psi}{\overline{\sigma}}^\mu\partial_\mu\psi-\frac{m}2(\psi\psi+\overline{\psi}\overline{\psi})+\partial_\mu\phi^\dagger\partial^\mu\phi-m(\phi F+\phi^\dagger F^\dagger)+F^\dagger F\,,
\eeq
and it is such that the dependency on the auxiliary field $F$ is purely polynomial. Consequently, the EOM for $F$ is polynomial and can be solved exactly, leading to 
\beq\ds
F=m\phi^\dagger\,.
\eeq
A field whose EOM can be uniquely solved in terms of the other fields in the theory produces no physical particles, and furthermore it can be eliminated (or, ``integrated out'') from the theory by plugging the solution into the Lagrangian. Auxiliary fields thus will not enter in the final expressions for our SUSY-invariant Lagrangian, in spite of the fact that their presence was needed in order to construct it (in the super-field formalism, at least). In the case of eq.~(\ref{simp}) we obtain
\beq\ds
{\mathcal{L}}=i\overline{\psi}{\overline{\sigma}}^\mu\partial_\mu\psi-\frac{m}2(\psi\psi+\overline{\psi}\overline{\psi})+\partial_\mu\phi^\dagger\partial^\mu\phi-m^2\phi^\dagger\phi\,,
\eeq
which is simply the Lagrangian of a free complex scalar, with mass $m$, plus a Majorana fermion (i.e., a neutral Weyl spinor endowed with a Majorana mass-term) with the same mass.

\begin{figure}[t]
  \centering
  \includegraphics[width=1\textwidth,valign=t]{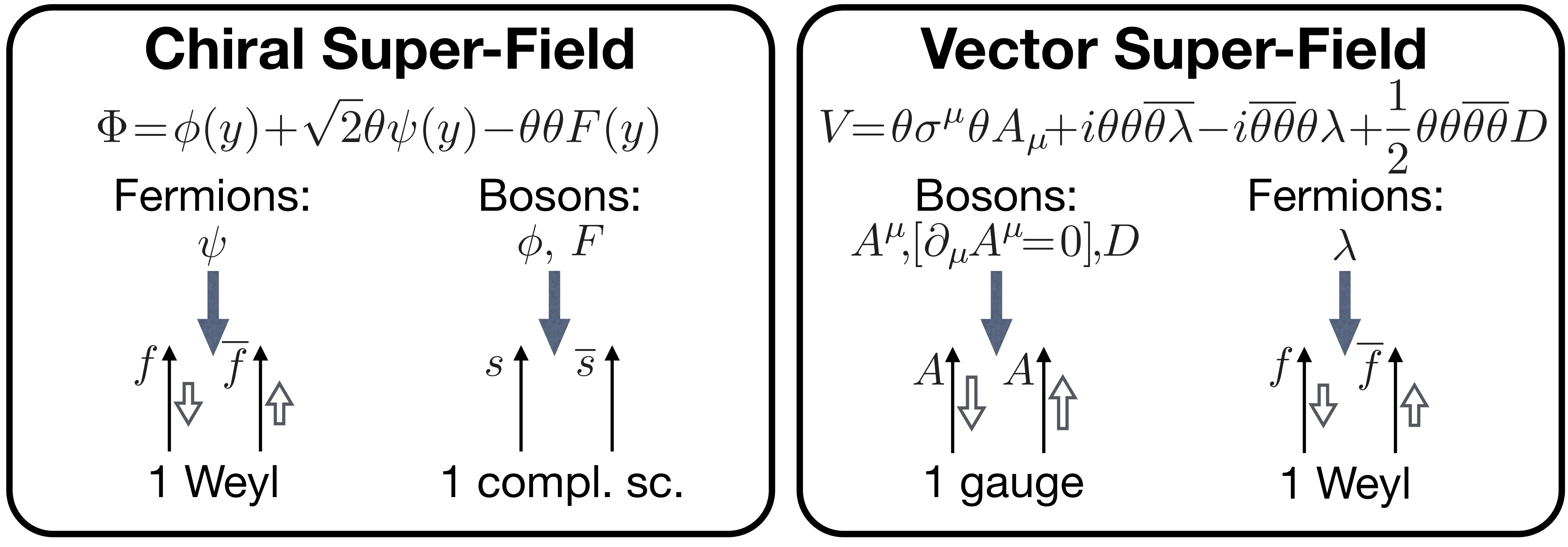}
  \caption{The chiral and vector superfields, together with the physical degrees of freedom they produce after the EOM are applied to get rid of the auxiliary fields $F$ and $D$. The variable $y$ that appears in the chiral super-field is defined as $y^\mu=x^\mu-i\theta\sigma^\mu\overline\theta$.}
  \label{SF} 
\end{figure}

All fields (including the auxiliary ones) in a SUSY multiplet can be collected in a single object, called a super-field. Super-fields can be thought as fields in an extended coordinate space (the super-space), which contains four additional ``fermionic'' coordinates $\theta^\alpha$ and $\overline{\theta}_{\dot{\alpha}}$ on top of the ordinary ``bosonic'' space-time coordinates $x^\mu$. The idea is to treat SUSY charges in analogy with the $P_\mu$ momentum operator, which acts on ordinary fields $\mathcal{F}(x)$ as a shift $x\rightarrow x+\delta x$ of the coordinates. A super-field is a function $\mathcal{F}(x,\theta,\overline\theta)$ and the SUSY charges $Q$ and $\overline{Q}$ act on it (almost) as translations $\theta\rightarrow\theta+\delta\theta$ and $\overline\theta\rightarrow\overline\theta+\delta\overline\theta$. A SUSY invariant Lagrangian \footnote{More precisely, an invariant Action since SUSY Lagrangian are often only invariant up to total derivatives} is thus constructed as a functional of the super-field that is translational-invariant in the super-space. The $\theta$ and $\overline\theta$ coordinates are however very different from the ordinary space-time ones. Rather than real numbers, they are ``Grassmann variables'', namely the product of two of them anti-commutes rather than commuting. This has several bizarre implications, among which the fact that the square of one of the $\theta$ or $\overline\theta$ components just vanishes. The most general super-field thus is not an arbitrary function of $\theta$ and $\overline\theta$, but just a fourth order (corresponding to the total number of independent components) polynomial in $\theta$ and $\overline\theta$, whose coefficients are ordinary fields in the $x$ space. Namely
\beq\label{gensf}\ds
{\mathcal{F}}(x,\theta,\overline\theta)=a(x)+b(x)\theta +c(x)\overline\theta  +d(x)\theta\theta +e(x)\overline\theta \overline\theta +f_\mu(x)\theta\sigma^\mu\overline\theta +g(x)\theta\theta\overline\theta +h(x)\overline\theta\overline\theta\theta +i(x)\overline\theta\overline\theta\theta\theta .
\eeq
The super-field is taken to be a bosonic object, therefore the fields in the decomposition that accompany even powers of $\theta$ and $\overline\theta$ (i.e., $a$, $d$, $e$, $f_\mu$ and $i$) are bosonic while the ones that come with odd powers ($b$, $c$, $g$ and $h$) are fermionic Weyl fields.

The generic super-field in eq.~(\ref{gensf}) (or, which is the same, the fields $a,\ldots,i$ it is made of) is a representation of the SUSY algebra, but it is a reducible one. Irreducible representations, corresponding the the chiral and to the vector multiplets, are restricted versions of the general super-field reported in fig.~\ref{SF}. We already discussed the auxiliary field $F$ appearing in the chiral field multiplet, we now see that it corresponds to the $\theta\theta$ component of the chiral super-field. This component is thus sometimes dubbed the ``$F$-component''. A real auxiliary field $D$ is present in the vector multiplet, together with the gauge field $A_\mu$ and the Weyl gaugino fields $\lambda_\alpha$. The auxiliary $D$ is needed because the $A_\mu$ field is taken to be in the Feynman gauge, i.e. it is subject to the condition $\partial_\mu A^\mu=0$ that reduces to three its independent components. One extra real field is thus required in order to match the $4$ real components of the gaugino field. The $D$ field is the $\theta\theta\overline\theta\overline\theta$ component of the vector super-field, which is thus called ``$D$-component''.

The rules to construct SUSY-invariant Lagrangians out of super-fields are rather simple. The first one is that (generic) super-fields, like ordinary fields, can be summed, multiplied and conjugated to produce other super-fields. Super-fields can also be derived with respect to the ordinary $x^\mu$ coordinates and also with respect to the SUSY coordinates $\theta$ and $\overline\theta$, by defining certain differential operators called ``SUSY covariant derivatives''. I will not define SUSY covariant derivatives here, the reader is referred to the literature. Chiral super-fields can also be summed and multiplied producing other chiral super-fields, but they cannot be conjugated. The conjugate of a chiral super-field is still a super-field, but not a chiral one (it is called ``anti-chiral''). The product of a chiral super-field with its conjugate is instead neither chiral nor anti-chiral. An important composite chiral super-field, which we will readily use to construct our SUSY Lagrangian, is the super-potential 
\beq\ds
W(\Phi)=a\Phi+\frac12 m \Phi^2 + \frac13 \lambda\Phi^3\,.
\eeq
It is a cubic polynomial in the chiral super-field $\Phi$, with an obvious generalisation to the case in which several super-fields $\Phi_i$ are present. The super-potential is the SUSY generalisation of the ordinary scalar potential. However unlike the latter it cannot contain the conjugate of the chiral field, $\Phi^\dagger$, otherwise it would not be a chiral super-field as previously explained. A super-potential can actually contain higher power of $\Phi$. I stopped at the third order because higher term would produce non-renormalizable interactions in the Lagrangian.

The last set of rules tells us how to extract invariant Lagrangians out of functionals (sums, products and derivatives) of super-fields. All SUSY invariants happen to be either the $D$ component (i.e., $\theta\theta\overline\theta\overline\theta$) of a generic super-field or the $F$ component (i.e., $\theta\theta$) of a chiral super-field. The most general SUSY-invariant Lagrangian for a chiral super-field (with obvious generalisation to several super-fields) is thus
\bea
&&\ds \left[\Phi^\dagger\Phi\right]_F=i\overline\psi{\overline\sigma}^\mu\partial_\mu\psi+
\partial_\mu\phi^\dagger\partial^\mu\phi+F^\dagger F\,,\nonumber\\
&&\ds \left[W(\Phi)\right]_D+\textrm{h.c.}=\left.\frac{\partial W}{\partial\Phi}\right|_\phi F
-\frac12\left.\frac{\partial^2 W}{\partial\Phi\partial\Phi}\right|_\phi\psi\psi+\textrm{h.c.}\,.
\label{lesssimp}
\eea
We see that the simple SUSY-invariant Lagrangian in eq.~(\ref{simp}) is recovered for $a=\lambda=0$ in the super-potential. Also notice that even in the more general Lagrangian in eq.~(\ref{lesssimp}) the auxiliary field $F$ does not possess a kinetic term and it can be integrated out by solving its EOM, which is just $F^\dagger=-\partial W/\partial\Phi$. This results in a potential for the scalar component $\phi$ of the chiral super-field
\beq\ds\label{fterm}
V_F(\phi)=\left|\left.\frac{\partial W}{\partial\Phi}\right|_\phi\right|^2=\left|a+m\phi+\lambda\phi^2\right|^2\,,
\eeq
which is called ``$F$-term potential''.

Similarly, one can write down the Lagrangian for the vector super-field and the interactions between the vector super-field and the chiral one. The vector super-field is the SUSY generalisation of the $A_\mu$ gauge field, therefore its interactions are dictated by gauge-invariance (plus SUSY), very much like the interaction of an ordinary gauge field. For a single vector super-field, corresponding to a \mbox{U$(1)$} gauge symmetry (the generalisation to non-abelian groups like the ones of the SM is rather straightforward) and a single chiral super-field with charge $q$ under the group, the Lagrangian consists of the two following terms
\bea
&&\ds\frac14\left[{\mathcal{W}}^\alpha {\mathcal{W}}_\alpha\right]_F+\textrm{h.c.}=
-\frac14 A_{\mu\nu} A^{\mu\nu}+
i\overline\lambda{\overline\sigma}^\mu\partial_\mu\lambda+\frac12 D^2\,,\label{gaugelag}\\
&&\ds\left[
\Phi^\dagger e^{2 q g V}\Phi
\right]_D
=D_\mu\phi^\dagger D^\mu\phi+i\overline\psi{\overline{\sigma}}^\mu D_\mu\psi 
+F^\dagger F 
-i\sqrt{2}qg\phi\overline\psi\overline\lambda +i\sqrt{2}qg\phi^\dagger\psi\lambda
-gq\phi^\dagger\phi D\,. \nonumber
\eea
The first one is simply the kinetic term for the gauge and for the gaugino fields, plus a quadratic (non-derivative, as it should) term for the auxiliary $D$.\footnote{The chiral super-fields ${\mathcal{W}}_\alpha$ are a SUSY generalisation of the field-strength in ordinary gauge theories. Their definition is not worth reporting here.} The second contains the kinetic terms of the scalar and Weyl fields in the chiral multiplet, with ``$D_\mu$'' denoting the ordinary covariant derivative with with charge $q$ and gauge coupling $g$, which produces the habitual gauge interactions. Interestingly enough, Yukawa couplings are also present involving $\phi$, $\psi$ and the gaugino $\lambda$. These are supersymmetric generalisations of the $A_\mu$ gauge interactions with $\psi$ and with $\phi$ and they emerge with a coupling strength, $\sqrt{2} gq$, which is completely fixed by gauge invariance. Also notice that the auxiliary field can, as usual, easily be integrated out producing another contribution to the scalar potential called ``$D$-term potential''. It reads
\beq\ds
V_D(\phi)=\frac12q^2g^2|\phi|^2\,.
\eeq
Once again, like the Yukawa's previously mentioned, its coefficient is completely specified in terms of the representation of the gauge group in which the field lives (i.e., the charge $q$ in our example) and by the gauge coupling $g$ of the theory.

\subsection{Why SUSY is Great: a Tale from the 80's}\label{tale}

The possibility of SUSY being the right tool to construct realistic extensions of the SM below or at the TeV scale (and not ``just'' a tool to build string theories of quantum gravity and to study deep theoretical aspects of Quantum Field Theory) is supported by a number of surprising phenomenological properties SUSY theories happen to possess (see \cite{Drees:1996ca,Martin:1997ns} for a complete discussion). These properties were discovered in the early 80's and produced enormous excitement in the theory community. The virtues of SUSY, which I will describe in the present section, of course are still there today. However they are now accompanied by a set of issues, related with negative searches of super-particles at different experimental facilities and with the determination of the Higgs boson mass, as I will explain in sect.~\ref{after}. None of these experimental issues were of course known in the 80's, and thus the great excitement about SUSY was fully justified. The situation is different now. SUSY might still be waiting to be discovered at the TeV scale, but apparently not in the simple ``vanilla'' form theorists imagined in the 80's.

The main reason why SUSY should be relevant for TeV scale physics is that SUSY models can solve the Naturalness Problem, as first pointed out by several authors in '81, among which S.~Dimopoulos, H.~Georgi and E.~Witten. In order to see how this works, let us recall the Naturalness Argument, as formulated in sect.~\ref{natarg}, for the Higgs mass parameter $m_H^2$. The problem has to do with a contribution that comes (in whatever new physics model is ultimately responsible for the microscopic origin of the Higgs mass) from low-energies, below the SM cutoff $\Lambda_{\textrm{SM}}$, i.e. at energies where physics is known and is provided by the SM.  We focus on the largest contribution, the one from the top quark loop in eq.~(\ref{deltamh})
\beq\ds\label{dmh}
\delta_{\textrm{SM}}m_H^2=
\frac{3 y_t^2}{4\pi^2}\Lambda_{\textrm{SM}}^2\,.
\eeq
It can be interpreted, poorly speaking, as a divergent contribution to the Higgs mass. The Naturalness Problem is that this term becomes much larger than the actual value of $m_H^2$, obliging us to a cancellation, if $\Lambda_{\textrm{SM}}$ is much above the TeV, as eq.~(\ref{deltatuning}) shows.

The problem emerges because the Higgs mass has two properties, which have to be simultaneously verified for the Naturalness Problem to arise. These are the fact that the Higgs mass term is a parameter with positive energy dimension and the fact that it is not protected by any symmetry, namely no new symmetry emerges in the SM Lagrangian if $m_H^2$ is taken to vanish. Parameters that violate both conditions are instead, for instance, the SM Yukawa couplings. Take for simplicity one single fermion, with its Yukawa coupling $y_f$ to the Higgs field, and repeat for $y_f$ the considerations that led us to the Naturalness Problem in sect.~\ref{natarg}. We can still split the integral expression for it as in eq.~(\ref{splitint}), but now if we compute the $E<\Lambda_{\textrm{SM}}$ contribution we find
\beq\ds\label{yukNNP}
\delta_{\textrm{SM}}y_f\sim \frac{y_f g^2}{16\pi^2}\log(\Lambda_{\textrm{SM}}/M_{\textrm{EW}})\,.
\eeq
In the expression, $M_{\textrm{EW}}$ denotes the EW scale and $g$ is one SM coupling which, depending on the diagram, could be either $y_f$ itself or one of the gauge couplings. The result contains no power-like divergence, for the very simple reasons is that all the SM couplings are dimensionless. It is thus impossible to have a polynomial divergence on a dimensionless quantity, only logarithmic divergencies are allowed. Clearly a logarithm is much less dangerous. Even for $\Lambda_{\textrm{SM}}=M_P$, the log is around $40$ and hardly compensates the $g^2/16\pi^2$ loop factor. The contribution to $y_f$ is thus of order $y_f$ or smaller and no cancellation is required. The fact that $\delta_{\textrm{SM}}y_f$ contains at least one power of $y_f$ is instead less trivial and has to do with the fact that a symmetry (chiral symmetry, i.e. two independent phase transformations acting on the two chirality components) is recovered at $y_f=0$. Then if $y_f$ was really vanishing, loop corrections could not generate it. A diagram contributing to it must thus contain at least one insertion of the $y_f$ vertex. As mentioned in sect.~\ref{natarg}, this symmetry argument can be extended in order to deal with all the three SM families, ensuring that the correction of each Yukawa coupling is proportional to itself. This avoids, for instance, relatively large contribution to the  Yukawa coupling of the up induced by the much larger coupling of the top quark.

A less simple case is the one of a massive fermion. Of course we don't have one in the SM since no fermion masses (but only Yukawa's) are present in the SM above the EW scale. Consider however a toy model in which a massive fermion is included in the theory, coupled through a set of dimensionless couplings ``$g$'' to the SM fields. Its mass $m_F$ has of course positive energy dimension like $m_H$, but still the low-energy contribution to it is only logarithmically divergent \footnote{From now on, since the low-energy (IR) theory we are considering to compute the low energy contribution is not anymore the SM, I will substitute ``$\delta_{\textrm{SM}}$'' with ``$\delta_{\textrm{IR}}$'' and ``$\Lambda_{\textrm{SM}}$'' with $\Lambda$, representing  the cutoff scale of the IR theory.}
\beq\ds
\delta_{\textrm{IR}}m_F\sim \frac{g^2}{16\pi^2}m_F\log(\Lambda/M_{\textrm{IR}})\,.
\eeq
Unlike the Yukawa's, $m_F$ has positive dimension but, exactly like the Yukawa's, it is protected by the chiral symmetry which is recovered in the theory if $m_F=0$. Thus loop corrections are proportional to $m_F$ itself and are not large. As such, $m_F$ does not suffer of a Naturalness Problem.

\begin{figure}[t]
  \centering
  \includegraphics[height=0.13\textwidth]{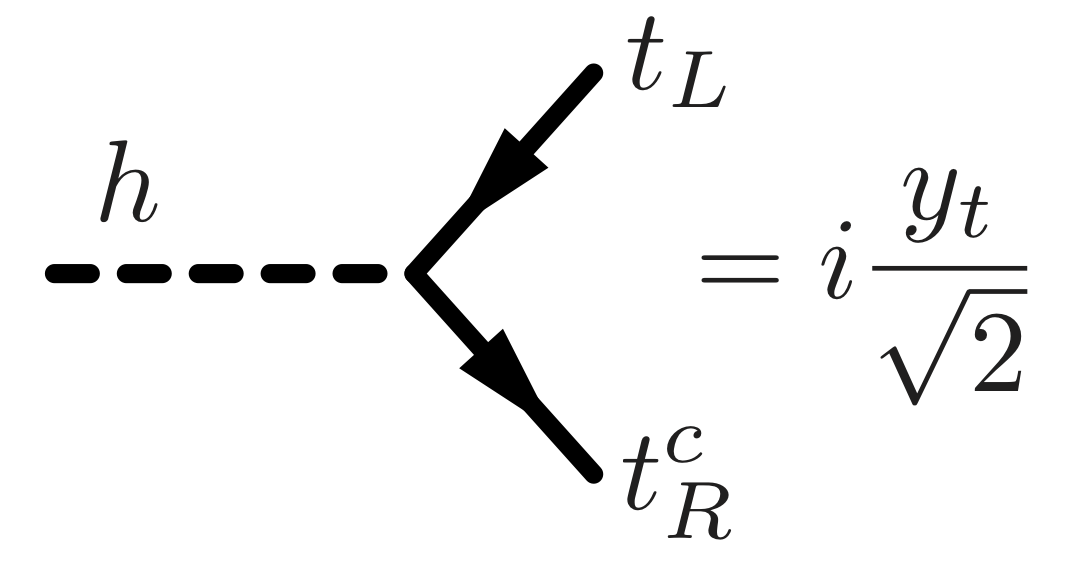}\hspace{40pt}
    \includegraphics[height=0.13\textwidth]{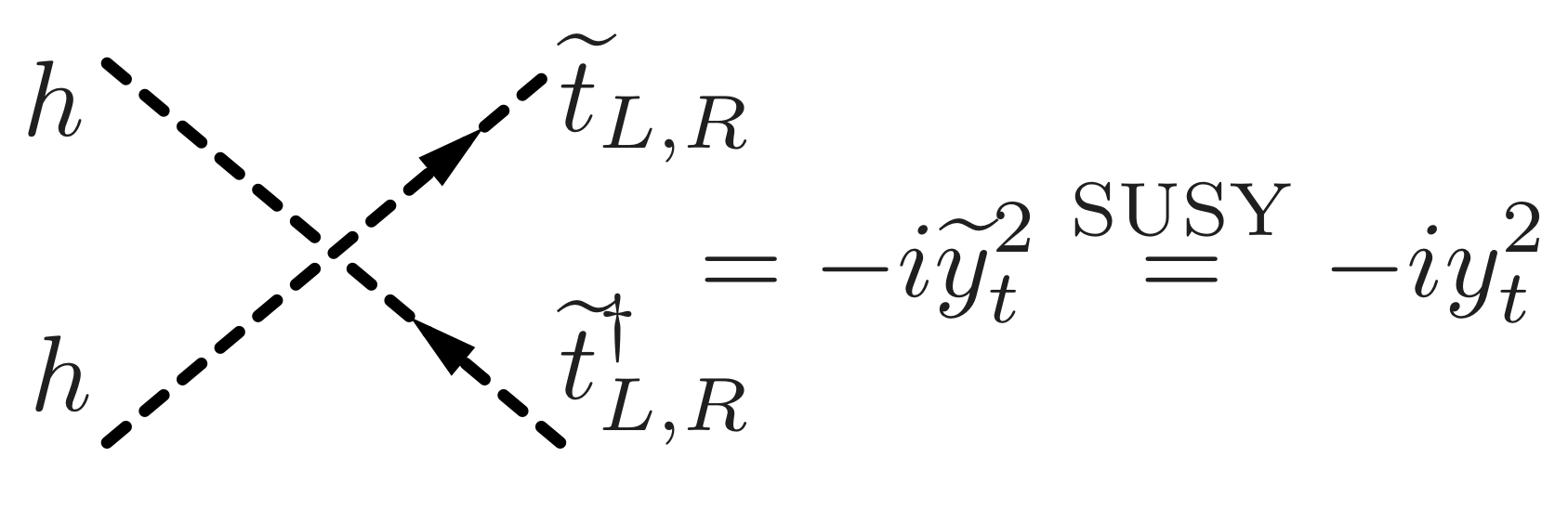}
    \caption{The ordinary Yukawa coupling (left) with its SUSY counterpart (right).\label{FD}}
\end{figure}

We just discovered that a fermion, unlike a scalar boson, can be ``Naturally'' light, even if the cutoff $\Lambda$ of the theory it is part of is extremely large. It is thus now clear why SUSY, which obliges the mass of the scalar Higgs boson to be equal to the one of its fermionic higgsino partner, can help us with the Naturalness Problem. If the former is ``Naturally'' light, the latter must be ``Natural'' as well in a SUSY model. In order to illustrate how this works, let us only consider the Higgs boson, the top quark and the Yukawa interaction between them, which is responsible for the largest contribution to $m_H^2$ in eq.~(\ref{dmh}). I will even ignore the bottom quark, as well as the other components of the Higgs doublet, and I will just focus on the neutral Higgs field component $h$ coupled to $t_L$ and $t_R$ through the Yukawa coupling. In order to construct a supersymmetric version of this theory, three chiral super-fields need to be introduced: $\Phi_h$, $\Phi_{t_L}$ and $\Phi_{t_R}$. After integrating out the auxiliary fields, they lead respectively to the fields $\{h,\widetilde{h}\}$, $\{t_L,\widetilde{t}_L\}$ and $\{t_R^c,\widetilde{t}_R^\dagger\}$, where $\widetilde{h}$ is the higgsino, $\widetilde{t}_L$ is the left-handed stop and $\widetilde{t}_R$ is the right-handed stop. Notice that what appears in $\Phi_{t_R}$ is the conjugate of the right-handed top, $t_R^c$, which is a left-handed field and as such can appear in the chiral super-field. Correspondingly, the right-handed stop is defined with a conjugate such that it has the same quantum numbers of the (not conjugate) SM $t_R$. Introducing the SM Yukawa in the theory requires us to put a trilinear term in the super-potential
\beq\ds
W=\frac{y_t}{\sqrt{2}}\Phi_h\Phi_{t_L}\Phi_{t_R}\;\longrightarrow\;
\left\{
\begin{array}{l}\ds
{\textrm{SM Yukawa (from eq.~(\ref{lesssimp})):}}\;\;\;\;\; -\frac{y_t}{\sqrt{2}} h \overline{t} t_R\,, \\[10pt]
\ds
F{\textrm{-term\ potential (from eq.~(\ref{fterm})):}} -\frac{y_t^2}2h^2[|\widetilde{t}_L|^2+|\widetilde{t}_R|^2]\,.
\end{array}
\right.
\eeq
Therefore in SUSY the Yukawa coupling, diagrammatically represented on the left panel of  fig.~\ref{FD}, is necessarily associated with a quartic $h^2\widetilde{t}_{L,R}^2$ vertex with the stops, reported on the right. Both couplings must be included in the calculation of $\delta_{\rm{IR}}m_H^2$, and the stop loop cancel exactly the one of the top\\\vspace{-20pt}
\begin{figure}[h]
  \centering
  \includegraphics[width=0.5\textwidth]{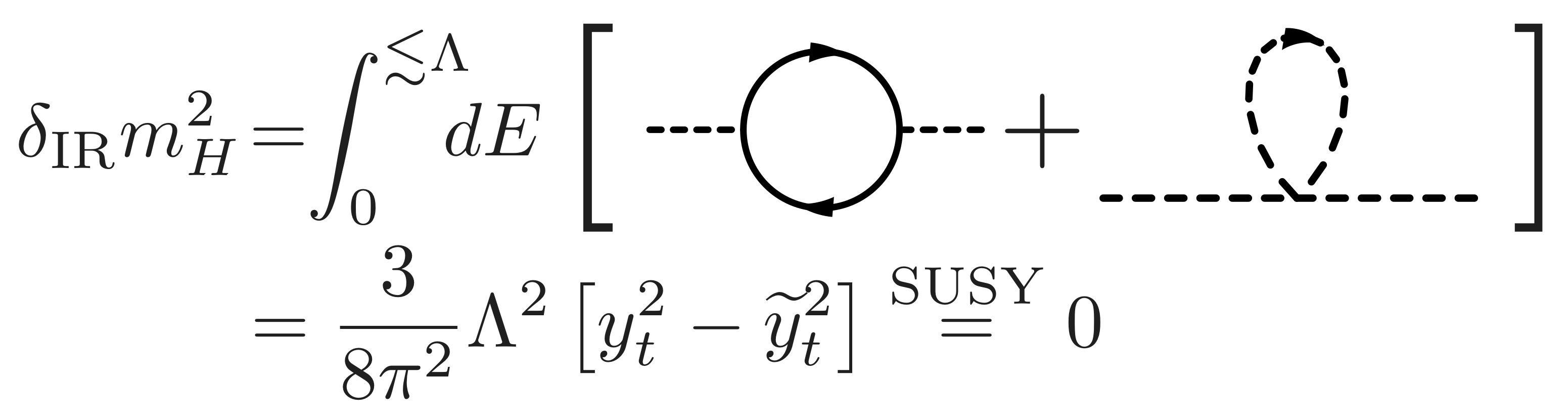}
\end{figure}

\vspace{-15pt}
\noindent{}The result is that the Naturalness Problem is solved, as expected, in a supersymmetric theory.

Obviously in oder to exploit the solution to the Naturalness Problem offered by supersymmetry we cannot just replace the SM with its SUSY version. This would be in sharp contrast with observations given that the particles we know about, their spectrum and interactions, do not respect SUSY. What one has to do is to first extend the SM to its (possibly minimal, but not necessarily so) SUSY version, and then include extra terms in the Lagrangian that break supersymmetry and reconcile the model with observations. Very importantly, it turns out that it is possible to do this without spoiling the SUSY solution to the Naturalness Problem, by introducing a special set of SUSY-breaking terms called ``soft terms''. Equally importantly, explicit microscopic models exist where SUSY is exact at very high scale, gets spontaneously broken and produces only soft breaking terms at low energy. Soft SUSY-breaking terms, namely terms that break SUSY but preserve Naturalness, include (see e.g. \cite{Drees:1996ca,Martin:1997ns}) mass, bilinear and trilinear terms for the scalar fields and gaugino mass terms. Including them in the Lagrangian happens to be sufficient to make all the SUSY partners of the SM particles heavy, explaining why we have not yet seen them. SUSY models addressing the Naturalness Problem can thus be made fully realistic, as a result of a fortunate series of ``coincidences'' related with a bunch of non-trivial properties of SUSY.

\begin{figure}[t]
  \centering
  \includegraphics[height=0.3\textwidth]{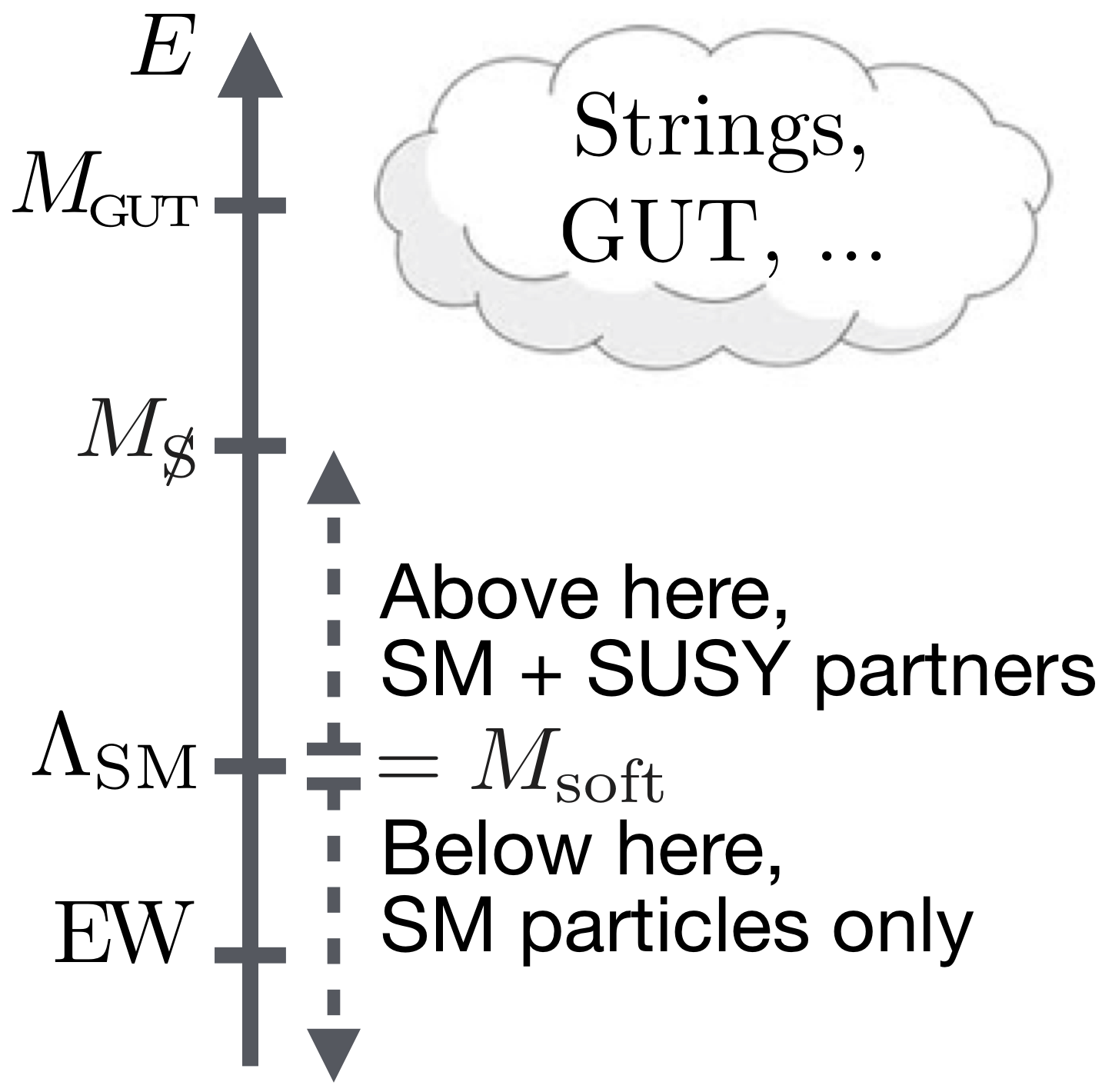}
    \caption{The SUSY picture of high-energy physics.\label{SUSYPIC}}
\end{figure}

The SUSY picture of high-energy physics is thus the one of fig.~\ref{SUSYPIC}. Starting from above, the theory is exactly supersymmetric at very high energies, until the scale $M_\slashed{\rm{S}}$ where SUSY is broken producing a set of soft terms. The typical mass-scale $M_{\rm{soft}}$ of the soft terms generated by the breaking, among which we have the mass of the supersymmetric particles, needs however not to be of the order of $M_\slashed{\rm{S}}$. It can be of that size in specific SUSY breaking scenarios, but it can also easily be much smaller than that, $M_{\rm{soft}}\ll M_\slashed{\rm{S}}$, as in the framework of ``gravity-mediated'' SUSY breaking (which used to be very popular in the 80's). Below $M_\slashed{\rm{S}}$, the theory reduces to a supersymmetric extensions of the SM containing both the SM particles and the SUSY partners as propagating degrees of freedom, the latter ones with a mass of order $M_{\rm{soft}}$, larger than the EW scale. Below $M_{\rm{soft}}$, SUSY partners decouple from the theory and one is left with the SM. Seen from below, $M_{\rm{soft}}$ is the scale at which BSM particles appear and thus it provides the SM cutoff $\Lambda_{\rm{SM}}$.

In view of the identification $M_{\rm{soft}}\sim\Lambda_{\rm{SM}}$, it is clear that the SUSY partners cannot be arbitrarily heavy if we really want to solve the Naturalness Problem, because of eq.~(\ref{dmh}). This is readily checked by giving a mass $M_{\widetilde{t}}\simeq M_{\rm{soft}}$ to the stops and repeating the calculation of $\delta_{\rm{IR}}m_H^2$. It is rather obvious by dimensional analysis that we are going to obtain
\beq\ds\label{dmhSUSY}
\delta_{\rm{IR}}m_H^2=\frac{3y_t^2}{8\pi^2}M_{\widetilde{t}}\log(M_\slashed{\rm{S}}/M_{\widetilde{t}})\,.
\eeq
Up to the log, which can just worsen the situation, we get the same expression as in eq.~(\ref{dmh}) with $\Lambda_{\rm{SM}}$ replaced by $M_{\widetilde{t}}\simeq M_{\rm{soft}}$. Consequently we get a large fine-tuning $\Delta$, as in eq.~(\ref{deltatuning}), if SUSY particles are not at the TeV scale or below.

\begin{figure}[t]
  \centering
  \includegraphics[width=1\textwidth]{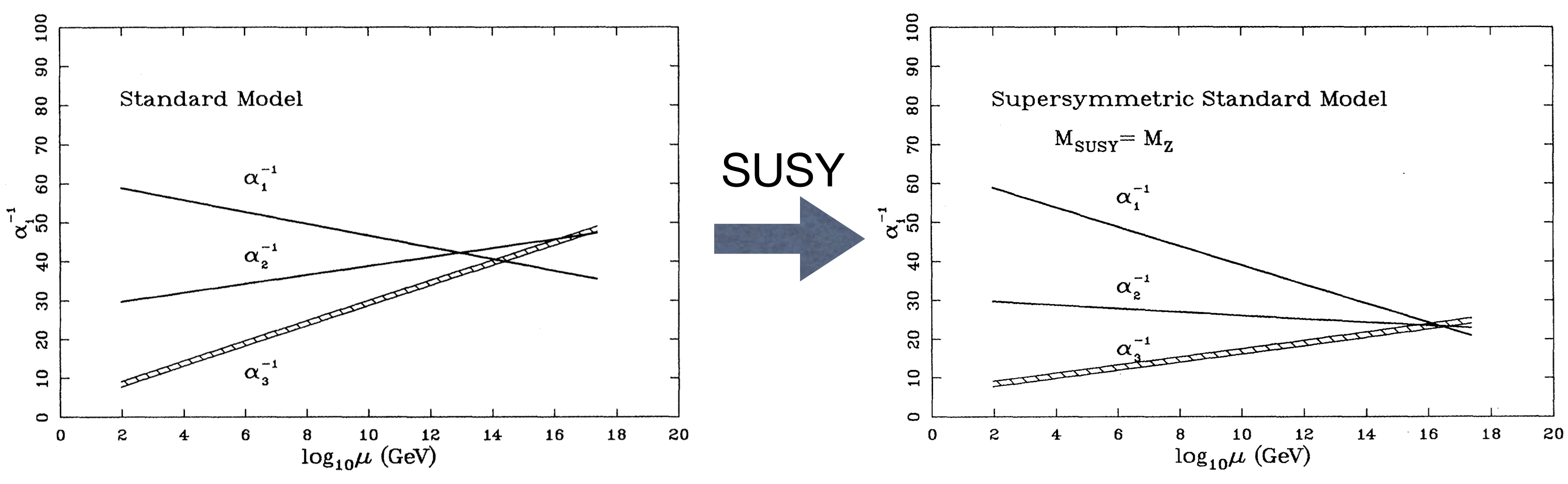} 
\caption{\label{GUT}The \mbox{SU$(3)$}, \mbox{SU$(2)$} and  \mbox{U$(1)$} inverse structure constant ($\alpha_i^{-1}=4\pi/g_i^2$) renormalisation group running in the SM (left) and in its minimal supersymmetric extension, the MSSM (right).} 
\end{figure}

SUSY having to show up before the TeV was of course not at all an issue in the 80', when this scale was far to be directly probed experimentally. It was actually a reason for excitement having all these new particles close enough to be discovered in the future. More reasons for excitement came from two more arguments, seemingly unrelated with SUSY: coupling unification and Dark Matter. Coupling unification (see \cite{Langacker:1980js,Raby:2006sk} for a review) is the idea that the three SM gauge forces might have a common origin at very high scales, where they are all described by a single simple unified gauge group (e.g., \mbox{SU$(5)$} or \mbox{SO$(10)$}), characterised by a single gauge coupling. This is supported, in the first place, by the fact that the SM matter fermion content fills, for no obvious reason, complete multiplets of the unified group (see \cite{Pomarol:2012sb} for a concise discussion). These multiplets contain at the same time quarks and leptons. GUT models are also supported by the fact that the running of the three SM gauge couplings makes them approach each other at high scale. As shown in fig.~\ref{GUT}, this more or less happens (but not very accurately) in the SM at a scale $M_{\rm{GUT}}\sim10^{14}$~GeV. At this scale, the full unified theory should show up. In particular, new massive gauge bosons should appear, with interactions connecting leptons and quarks that sit in the same GUT multiplets as previously mentioned. These interactions make the proton decay at an unacceptably large rate if $M_{\rm{GUT}}\sim10^{14}$~GeV. The situation is much better in supersymmetric extensions of the SM, as shown in the right panel of fig.~\ref{GUT}. First, the couplings unify  more accurately, simply due to the effect of the super-partners on the running, which happen to go in the right direction for no obvious reason. Second, unification is postponed to $M_{\rm{GUT}}\sim10^{16}$~GeV and proton decay experiments are not sensitive to such a high suppression scale. All this of course happens only provided $M_{\rm{soft}}$ is small enough for the super-partners starting to contribute to the running early enough, $M_{\rm{soft}}\sim100$~GeV is assumed in the plot. Clearly the running is logarithmically slow, so that $M_{\rm{soft}}=10$~TeV or even more would not change the situation radically.  However it is clear that also coupling unification, as well as the Naturalness Argument, point towards low-energy supersymmetry. The positive interplay between low-energy SUSY and unification is a very strong argument in favour of SUSY and of unification as well.

The interplay between SUSY and DM is equally impressive. It originates from a serious phenomenological problem of SUSY and of its solution, which consists in imposing a discrete symmetry called ``$R$-parity''. I stressed in sect.~\ref{SMONLY} the phenomenological importance of Baryon and Lepton number as accidental symmetry in the SM and how much non-trivial it is that these symmetries emerge at $d=4$ without being imposed in the construction of the theory. I also argued that BSM scenarios will in general not possess accidental Baryon and Lepton number and that those symmetries will have to be imposed in some way. This is the case also in SUSY. Indeed, when trying to construct the minimal supersymmetric extension of the SM (the Minimal Supersymmetric Standard Model, MSSM), one immediately encounters terms in the super-potential, allowed by the gauge symmetries, that violate both the Baryon and the Lepton number. For instance, Baryon number is violated by (see e.g. \cite{Martin:1997ns} for more details)
\beq\ds\label{dw}
\Delta W_{\Delta B=1}=\lambda'' \varepsilon_{\alpha\beta\gamma}\Phi_{u_R}^\alpha \Phi_{d_R}^\beta \Phi_{d_R}^\gamma\,,
\eeq 
where $\alpha,\beta,\gamma$ are QCD color indices while the flavour indices are understood. Adding those terms in the super-potential produces SUSY-invariant $d=4$ interactions that violate Baryon and Lepton number, in sharp contrast with observations. However all these dangerous terms, and all the soft SUSY-breaking ones which also violate Baryon and Lepton number, are avoided by imposing only one discrete symmetry, $R$-parity. $R$-parity consists in the sign-flip $\theta\rightarrow-\theta$ and $\overline\theta\rightarrow-\overline\theta$ of the super-space coordinates, times an additional overall minus sign for all the matter fermions (quarks and leptons) super-fields. A quick look at fig.~\ref{SF} immediately reveals that with this assignment all the SM fields (quarks, leptons, gauge and Higgs) are even and all the super-partners (or s-particles) are odd. The super-potential in eq.~(\ref{dw}) is obviously odd under $R$-parity and it is thus forbidden, together with all the other  Baryon and Lepton number-violating terms, if $R$-parity is imposed as a symmetry of the MSSM.

Since they are odd under $R$-parity, s-particles cannot decay to SM particles only, at least one  s-particle must be present in the final state. In particular this means that the lightest of the s-particles (the LSP) cannot decay at all and it is absolutely stable. If it happens to be electrically and QCD neutral, it is potentially a viable DM canditate. Moreover, the LSP mass will be of the order of $M_{\rm{soft}}$, which we argued above to be likely of the $100$~GeV to TeV order. Furthermore, the LSP will typically couple to SM through EW gauge interactions. A particle with these properties is called a Weakly-Interacting Massive Particle (WIMP) and it can perfectly account for the observed DM component of the Universe through the mechanism of thermal freeze-out (see \cite{dslect} for a review). This is the so-called ``WIMP miracle'', which automatically emerges as a byproduct of SUSY model-building.

\subsection{SUSY after LEP, Tevatron and LHC run-$1$}\label{after}

Naturalness, coupling unification and Dark Matter are extremely strong arguments in favour of low-scale SUSY, and all the enthusiasm they triggered towards SUSY is perfectly justified. However this enthusiasm cooled considerably after 30 years of negative experimental s-particle searches. LEP was the collider at which SUSY had its first chance to be discovered, in spite of the fact that LEP energy was far below $450$~GeV (see eq.~(\ref{deltatuning})), which is what we nowadays consider to be threshold for ``Natural'' BSM physics. This is because in the fine-tuning estimate we should not forget the logarithmic term we found in eq.~(\ref{dmhSUSY}), and we should remember that a high SUSY-breaking scale was expected in the 80's. With this expectation, taking for definiteness $M_\slashed{\rm{S}}=10^{15}$~GeV, the log is around $30$ and the Naturalness threshold moves down to $450/\sqrt{30}=82$~GeV. Even taking into account that s-particles must be produced in pairs because of $R$-parity, the LEP collider (in the LEP-II stage) could have had enough energy to produce them. Of course $M_\slashed{\rm{S}}$ needs not to be that high, viable SUSY-breaking scenarios exist where $M_\slashed{\rm{S}}$ is not far from the TeV scale and the log is small. Still, negative LEP search were the first evidence against the ``vanilla'' SUSY picture described in the previous section.

The search for s-particles continued at Tevatron and at the LHC run-$1$, with negative results.\footnote{And at the LHC run-$2$, however here I stick to the run-$1$ results, the only ones that were available when I gave the lectures.} Current limits on certain SUSY particles (light squarks and gluinos) are as high as $1.7$~TeV signalling, if taken at face value, that SUSY is a quite ``Un-Natural'' theory.  One should however be more careful, because the s-particles needs not to be all degenerate and a bound on few of them cannot be directly translated into a bound on $M_{\rm{soft}}$. Furthermore, not all the s-particles are equally important as far as fine-tuning is concerned because the way in which they contribute to the Higgs mass is very different. For instance, the stops are those that give the largest radiative contribution, in eq.~(\ref{dmhSUSY}), because their coupling to the Higgs is the largest one. The $450$~GeV threshold only applies to the stops, and the limit on their mass is only $700$~GeV or less, still compatible with Naturalness.\footnote{Tree-level contributions to $m_H^2$ emerge from higgsinos, and thus the Naturalness threshold on these particles is extremely low. However there no tension with the experimental bounds, which are too weak.} The strong limit on the light squarks is instead irrelevant for Naturalness, given that the squark contribution to $m_H^2$ is extremely suppressed by the small Yukawa couplings. The partners of the EW gauge bosons (EWinos) give the second largest radiative contribution (see eq.~(\ref{deltamh})), which is proportional to the Weak coupling square rather than to $y_t$. The Naturalness threshold for the EWinos is thus around the TeV, much above the limits. The gluinos are also relevant for Naturalness. In spite of the fact that their contribution to $m_H^2$ arises at two loops, the strong QCD coupling and certain color multiplicity factors produce a Naturalness threshold for gluinos around the TeV, which is comparable with the run-$1$ limit. The overall picture that emerges from this kind of considerations (the so-called ``Natural SUSY'' approach) is that the LHC run-$1$ started probing the ``Natural'' parameter space of SUSY, but no conclusive statement can be made. For an extensive presentation of this viewpoint and a quantitative discussion of run-$1$ searches the reader is referred to the lecture notes in Ref.~\cite{Craig:2013cxa}. 

The very last topic of these lectures is the structure of the Higgs potential in supersymmetry. This topic is relevant by itself, as it constitutes the starting point for SUSY Higgs phenomenology, extensively discussed in \cite{Djouadi:2005gj}. However it is also relevant in order to assess the current status of SUSY because it will allow us to understand and to qualify the often-heard statement that the LEP bound on the Higgs mass (and its measurement at the LHC) is problematic for SUSY. The first important point is that any SUSY extension of the SM requires us to introduce at least two Higgs chiral super-fields: $\Phi_{\rm{u}}$ and $\Phi_{\rm{d}}$. This follows from the fact in order to generate the Yukawa couplings in the up and in the down sector two Higgs doublets are needed, with respectively Hypercharge equal to $1/2$ and $-1/2$. Only one doublet is introduced in the SM because the other one can be obtained by complex conjugation, but this is impossible in SUSY since the conjugate of a chiral super-field cannot appear in the super-potential. Two chiral super-fields are thus needed,\footnote{The cancellation of gauge anomalies also requires two Higgs super-fields.} with super-potential terms
\beq\ds
W_{\rm{u}}=y_{\rm{u}}\Phi_{q_L}\Phi_{\rm{u}}\Phi_{u_R}\,,\;\;\;\;\;
W_{\rm{d}}=y_{\rm{d}}\Phi_{q_L}\Phi_{\rm{d}}\Phi_{d_R}\,.
\eeq
Therefore two scalar Higgs doublets $H_{\rm{u}}$ and $H_{\rm{d}}$ are present in SUSY, coupled to up- and down-type quarks respectively. After EWSB, three of these $8$ real degrees of freedom are eaten by the EW bosons becoming massive, one provides the neutral SM Higgs boson and the remain four are extra scalars which are absent in the SM. The extra scalars in SUSY are one heavy neutral CP-even state $H_0$, one charged $H_\pm$ and a neutral CP-odd $A$. Searching for these particles directly, or indirectly by studying their effects (through mixing) on the couplings of the SM-like Higgs, is one way to test supersymmetry.

The scalar potential for the $H_{\rm{u}}$ and $H_{\rm{d}}$ doublets consists of three terms \footnote{The contraction with the $\varepsilon_{ij}$ tensor is understood in last term of the equation that follows.}
\bea
\ds V(H_{\rm{u}},H_{\rm{d}})&=&\ds \mu^2(|H_{\rm{u}}|^2+|H_{\rm{d}}|^2)\nonumber\\
&&\ds+\frac{g^2+g^{\prime2}}8(|H_{\rm{u}}|^2-|H_{\rm{d}}|^2)^2+\frac{g^2}2 |H_{\rm{u}}^\dagger|H_{\rm{d}}|^2\nonumber\\
&&\ds + m_{\rm{u}}^2 |H_{\rm{u}}|^2+m_{\rm{d}}^2 |H_{\rm{d}}|^2 + B (H_{\rm{u}} H_{\rm{d}}+H_{\rm{u}}^* H_{\rm{d}}^*)\,.\label{Higgspot}
\eea
The one on the first line is an $F$-term, originating from the $\mu$-term $\mu\Phi_{\rm{u}}\Phi_{\rm{d}}$ in the super-potential. The one on the second line is a $D$-term, dictated by the gauge quantum numbers of the Higgs doublets. Notice that it is the only one that contains quartic couplings, which are thus completely fixed in terms of the SM gauge couplings $g$ and $g'$. The soft SUSY-breaking terms are displayed in the last line. The potential in eq.~(\ref{Higgspot}) allows, with the appropriate choice of its parameters, EWSB to occur. Also, it allows (or better, generically requires) both doublets to get a VEV
\beq\ds
\langle|H_{\rm{u}}|^2\rangle=\frac{v_{\rm{u}}^2}2\,,\;\;\;\;\;\langle|H_{\rm{d}}|^2\rangle=\frac{v_{\rm{d}}^2}2\,.
\eeq 
The sum of the square of the two VEVs is fixed to $v_{\rm{u}}^2+v_{\rm{d}}^2=v^2$, where $v\simeq246$~GeV, but the ratio between them is a free parameter, which is typically traded for the tangent of the ``$\beta$'' angle
\beq
\tan\beta\equiv\frac{v_{\rm{u}}}{v_{\rm{d}}}\,.
\eeq
Notice that both Higgses taking VEV is necessary in order to generate quark masses since, as we discussed, the up- and down-type Yukawa couplings are only present for $H_{\rm{u}}$ and for $H_{\rm{d}}$, respectively.

With the knowledge of the 80's, the potential in eq.~(\ref{Higgspot}) is quite successful. It produces realistic EWSB and fermion masses, and an extended Higgs sector which was perfectly plausible at that times, in which almost no experimental information was available on Higgs physics. After LEP could not discover the Higgs boson and set a lower bound $m_H>115$~GeV, the potential (\ref{Higgspot}) started being in tension with observations. Indeed, it is possible to show that the structure of the potential is such that the Higgs mass is unavoidably smaller than the one of the $Z$ boson. More precisely, it turns out that for any choice of the free parameters one has
\beq\ds\label{mhSUSY}
m_H\leq |\cos2\beta| m_Z\leq m_Z\,.
\eeq
The relation follows from the fact that the quartic terms in the potential are not free parameters, but instead they are uniquely dictated, through supersymmetry, by the gauge coupling. In order to see how this works, consider a simplified limit, the so-called ``decoupling limit'', in which the soft mass of the $H_{\rm{d}}$, $m_{\rm{d}}^2$, is taken to be large. In the limit, $H_{\rm{d}}$ decouples and it can be just ignored (i.e., set to zero) in eq.~(\ref{Higgspot}), obtaining a SM-like potential
\beq
V=\mu_{\rm{SM}}^2|H_{\rm{u}}|^2+\lambda_{\rm{SM}}|H_{\rm{u}}|^4\,,
\eeq
with $\mu_{\rm{SM}}^2=\mu^2+m_{\rm{u}}^2$ and $\lambda_{\rm{SM}}=(g^2+g^{\prime2})/8$. The habitual SM formula $m_H=\sqrt{2\lambda} v$ thus tells us that $m_H=m_Z$ in the decoupling limit. This matches eq.~(\ref{mhSUSY}) because in the decoupling limit one finds that $\tan\beta\rightarrow\infty$, i.e. $\beta\rightarrow\pi/2$.

Since the mass relation in eq.~(\ref{mhSUSY}) is violated experimentally, we might wander if it excludes SUSY as a realistic theory of Nature. Of course it does not, because of two reasons, but it has important implications. The first point is that eq.~(\ref{mhSUSY}) is only valid at the tree-level order, radiative corrections violate it. For instance top and stop loops contribute to the quartic by an amount
\beq\ds
\delta\lambda\sim \frac{3 y_t^2}{8\pi^2}\log\frac{M_{\widetilde{t}}}{m_t}\,,
\eeq
so that by making stops heavy one can get a large enough quartic and a large enough Higgs mass. Working in the decoupling limit for simplicity (and because it is the most favourable one), the shift we need on $\lambda$ is
\beq\ds
\delta\lambda=\frac{m_H^2-m_Z^2}{2 v^2}\simeq0.06\,,\;\;\;\Rightarrow\;\;\;
M_{\widetilde{t}}\simeq 1.3\, {\textrm{TeV}}\,.
\eeq
That heavy stops cost quite a lot of fine-tuning, definitely above ten. More refined estimates \cite{Hall:2011aa}, taking into account the need of a separation between $M_\slashed{\rm{S}}$ and the weak scale (i.e., of some log enhancement in the tuning), reveal that the tuning needed to accomodate the $125$~GeV Higgs mass is at least $100$ .

The second reason why eq.~(\ref{mhSUSY}) cannot disprove supersymmetry is that it only holds in the MSSM, thus it is not a robust property of SUSY models. It can be violated in SUSY scenarios like $\lambda$SUSY \cite{Barbieri:2006bg} (A.K.A. NMSSM), in which an extra singlet chiral super-field $\Phi_S$ is added to the theory with a $\lambda \Phi_S \Phi_{\rm{u}} \Phi_{\rm{d}}$ term in the super-potential.  This contributes to the quartic Higgs coupling and leads to a heavy enough Higgs boson if $\lambda$ is sufficiently large. The main drawback of this construction is that $\lambda$ needs to be relatively big, therefore its RG-running is very fast and reaches a Landau pole not much above the $10$~TeV scale. The alternative to a fine-tuned scenario seems thus to be a model which cannot be extended far above the TeV scale. This clearly seems very different from the basic SUSY picture we had in mind in fig.~\ref{SUSYPIC}. However there might be caveats, new model-building ways out, or space for a ``partially Un-Natural'', but still true, SUSY model at the TeV scale. Let us wait and see, the LHC run-$2$ will tell us more about SUSY.

\section{Conclusions and Outlook}

My purpose, when giving these lectures, was to outline that BSM physics is not (only) a collection of models, but rather a set of structural questions on fundamental physics and of possible answers to be checked against data. The microscopic origin of the Higgs mass, in connection with Naturalness (or Un-Naturalness), is only one of such questions. However it is the one about which decisive experimental progress will be made at the LHC, this is why I built the lectures around it. Several other relevant questions and ideas were encountered during the lectures, among which GUT, DM, neutrino masses and vacuum stability, each of which deserves a separate course. Some of these courses were given at this School \cite{pklect,dslect}. For what is missing, the lectures in Ref.~\cite{Pomarol:2012sb} are a valid starting point. The course aimed at providing a pedagogical introduction to BSM physics, for this reason basic material was presented and many recent developments were left out from the discussion. This should not obscure the fact that ``Natural'' BSM model-building is an active research area. Approaches related with the ``Twin Higgs'' mechanism \cite{Chacko:2005pe} are worth mentioning in this context.

Concerning the future of BSM physics, there is not much I can add to what discussed in sect.~\ref{wf}. There is not guarantee that the ongoing LHC program will produce a new physics discovery, but it is sure that it will improve our comprehension of fundamental interactions. This is more than enough to work on LHC physics to the best of our abilities. On a longer timescale, the future is impossible to predict. We will definitely keep asking structural questions on fundamental physics, however it is unclear if high-energy collider experiments will continue being the optimal investigation tools to search for answers. My viewpoint is well summarised by a famous sentence\\

\centerline{\emph{``Learn from yesterday, live for today, hope for tomorrow.}}
\centerline{\emph{The important thing is not to stop questioning.''}}

\hfill -- Albert Einstein \hspace{40pt}

\end{document}